# A Microgrid Deployment Framework to Support Drayage Electrification

Joseph N. E. Lucero,[1,]* Ruixiao Sun,[2] Brandon A. Miller,[3] Simona Onori,[4] and Vivek A. Sujan[2,5,]**

[1]Department of Chemistry, Stanford University, Stanford, CA 94305, USA

[2]Buildings and Transportation Science Division, Oak Ridge National Laboratory, Oak Ridge, TN 37831, USA

[3]Computational Sciences and Engineering Division, Oak Ridge National Laboratory, Oak Ridge, TN 37831, USA

[4]Department of Energy Science and Engineering, Stanford University, Stanford, CA 94305, USA

[5]Lead Contact

*Correspondence: jlucero@stanford.edu

**Correspondence: sujanva@ornl.gov

## SUMMARY

**The electrification of heavy-duty commercial vehicles (HDCVs) is pivotal in reducing greenhouse gas emissions and urban air pollution; however, this transition poses significant challenges for the existing electric grid, which is not designed to meet the high electricity demands of HDCVs. This can lead to a less effective reduction in freight transportation's carbon intensity despite significant electrification efforts. Deploying renewable energy sources, such as photovoltaics, alongside energy storage solutions, is essential to address these challenges. This paper examines the current grid limitations and explores the critical role of microgrid deployment, integrating solar and battery energy storage systems, in supporting the electrification of HDCVs. We propose an integrated framework that is designed to enhance regional grid capacity and decrease carbon intensity by identifying viable sites where a microgrid can be deployed and provide estimates for the deployment cost. Furthermore, using this framework, we quantify the maximal impact of microgrid deployment in reducing $CO_2$ emissions when we optimize the use of the available power. As a demonstration, we apply our framework to the region of the Port of Savannah, GA USA.**

## Context & Scale

The electrification of heavy-duty commercial vehicles operating at maritime ports is essential for reducing greenhouse gas emissions; however, electrification serves as a formidable challenge to the existing grid's capacity. Deploying renewable energy sources and battery storage through microgrids can help mitigate excess $CO_2$ emissions by reducing reliance on the grid, particularly in regions like the Port of Savannah. A framework for identifying viable sites for solar and battery deployment is presented and, through site and cost optimization, we show that excess $CO_2$ emissions can be reduced by more than 80% while simultaneously achieving a 7% reduction in vehicle costs. This integrated framework can straightforwardly be applied to other intermodal hubs around the US and may be modified to incorporate both front-of-the-meter, as well as behind-the-meter, microgrids.

## INTRODUCTION

The transportation sector is undergoing a significant transformation with the electrification of heavy-duty commercial vehicles (HDCVs), particularly in drayage applications around intermodal hubs such as maritime ports[1,2]. These diesel-powered vehicles, essential for "first-mile" delivery between ports and distribution centers, contribute substantially to greenhouse gas emissions and urban air pollution. As countries strive to meet climate goals, transitioning to electric drayage vehicles is critical[3].

However, the widespread adoption of electric HDCVs faces challenges, related to electricity demand and grid capacity. The current grid, designed for residential and light commercial needs, may struggle to support the high energy requirements of charging infrastructure for large vehicles. Microgrids, which integrate renewable energy sources and energy storage systems, can alleviate this strain by providing additional capacity, reducing the carbon footprint, and avoiding costly grid upgrades. These localized solutions enhance grid resilience and sustainability, supporting the transition to a cleaner transportation system.

This work focuses on the deployment of a microgrid incorporating solar power and lithium-ion battery (LIB) energy storage to meet the energy demands of an electrified HDCV fleet. Solar energy, combined with battery storage, offers a renewable power source that reduces reliance on fossil fuels and stabilizes energy supply during peak demand periods. The use of LIB is highlighted for their efficiency, longevity, and decreasing costs, making them ideal for grid applications.

The concept of HDCV electrification, particularly in drayage applications, has been explored extensively in recent years. Studies such as those by Moultak et al.[4] and Kotz et al.[5] have highlighted the potential of electric trucks to reduce emissions significantly, especially in urban and port areas where air quality is a major concern. This research indicates that electrification of drayage trucks operating at the port can lead to substantial reductions in both greenhouse gases and local pollutants. The integration of clean energy technologies into the grid has been examined in multiple contexts. For instance, the works of Rosales-Asensio et al.[6] and Lund et al.[7] discusses the advantages of incorporating diverse clean energy sources to enhance energy resilience and sustainability in various contexts. These works emphasize that integrating these renewable technologies can effectively reduce reliance on fossil fuels and provide stability to the grid. Furthermore, the importance of energy storage in supporting the integration of clean energy is underscored by research from Yang et al.[8]. Their findings suggest that advanced battery storage systems can mitigate the variability of clean energy, ensuring a steady supply of electricity for high-demand applications such as electric drayage vehicles. Moreover, studies focusing on the cost-benefit aspects of electrifying port operations[5,9–11], indicate that while initial investments in infrastructure and vehicles are high, the long-term benefits in terms of fuel savings, maintenance costs, and environmental impact are substantial.

Although existing research consistently affirms the feasibility and benefits of electrifying HDCVs, the capacity of the current electricity grid to support such electrification remains uncertain. In this paper, we demonstrate that, indeed, the current grid infrastructure is insufficient to meet the additional electricity demand that would result from electrification of HDCV fleets. Moreover, we quantify the impact of this increased demand on $CO_2$ emissions, revealing that without grid enhancements the expected reduction in emissions from electrification is significantly diminished. To address both of these challenges, we propose a integrated framework with three major and novel components: (1) a pipeline that synthesizes the capabilities of Oak Ridge National Laboratory's (ORNL) "Oak Ridge Siting Analysis for Generation Expansion" (OR-SAGE) tool with the "Renewable Energy Potential" (ReV) model from the National Renewable Energy Laboratory (NREL) to enable both the identification and evaluation of sites for microgrid deployment, (2) a power distribution optimization framework that yields the maximum amount of $CO_2$ emissions reduction that each of the identified sites is capable of, and (3) a total cost of ownership analysis that offers cost estimates of microgrid deployment allowing for the determination of the most cost-effective microgrid architecture and site from the perspectives of either the utility provider or the fleet operator, two primary stakeholders in the freight transportation sector. To demonstrate this framework, we apply it to the Port of Savannah, GA USA and the surrounding region as a representative case study.

## TWO PROBLEMS WITH FLEET ELECTRIFICATION

We discuss how the excess (electricity) load demand for a fleet of electrified HDCVs is estimated, as well as the corresponding increase in $CO_2$ emissions from the grid. We examine how the excess load demand poses a problem for the current grid infrastructure and how this may undermine $CO_2$ emissions reductions, one of the primary benefits of fleet electrification.

*Estimating the excess load demand and associated $CO_2$ emissions of fleet electrification*

To estimate the excess load demand due to HDCV electrification, we leverage a recently introduced framework known as the "Optimal Regional Architecture Generation for Electrified National Transport" (OR-AGENT) from ORNL, developed in collaboration with the Ohio State University and Stanford University[12]. This framework uses a bottom-up approach to generate sustainable roadmaps for HDCV electrification. To this end, OR-AGENT uses a comprehensive electric HDCV powertrain simulator. The forward-looking simulator accounts for various factors such as driver behavior, aerodynamic load variation based on truck configuration, tire thermal dynamics, rolling resistance, and power consumption of auxiliary components such as the cabin HVAC compressor and battery thermal management system while the vehicle operates in the region.

For each month of the year, OR-AGENT determines the different routes, the number of trucks in operation on those routes, and the trucks' associated weight/class statistics for a given U.S. region by integrating a variety of data sources. From the vehicle statistics, information about the routes, in conjunction with the estimate of energy used by a single truck operating on each of the routes, an estimate for the hourly energy required by the full fleet of electrified trucks is generated (Fig. S1).

Associated with this excess load demand is an amount of emitted $CO_2$. Transitioning the HDCV fleet operating at the Port of Savannah from a diesel to an electrified powertrain without building any additional energy resources, implies the current grid infrastructure must bear the entirety of the excess load demand. As the current grid infrastructure is not carbon-neutral, drawing energy from it will incur additional $CO_2$ emissions. We refer to this additional $CO_2$ emitted, over and above what the current grid is producing, as *excess $CO_2$ emissions.* The excess $CO_2$ emissions can be estimated by an approach first proposed in Sujan et al.[12]. Briefly, this approach integrates historical spatiotemporal data on the carbon intensity and the load demand of the grid for all the counties in the U.S. and performs forecasting for the $CO_2$ emissions associated with the excess load demand. Thus, given a county of interest, the county's historical electric load profiles, the historical energy generation profiles for that region, as well as the excess load profile, the excess $CO_2$ emissions can be estimated for every hour of the year (Fig. S2).

*Identifying grid capacity gaps and mitigated carbon emissions reduction*

To make all of the routes in the Port of Savannah region viable for electrification either in-route stationary charging or electrified road systems must be implemented to extend the range of HDCVs[13]. Consider the scenario where stationary charging is allowed at both at the Port of Savannah as well as at the destinations of the HDCVs. We find that a certain number of chargers must be installed at each location to meet the demand, as seen in Figure 1a. Moreover, we can also estimate the peak power of each location in any given hour (Figure 1b).

To understand the capabilities of the current grid to support this energy demand, all substations within a 15-mile radius of the major routes in which trucks operate are identified. Using ORNL proprietary data and models, the capacity limit of the grid is determined by assessing the peak loads during the year at these substations and the amount of additional load that can be put on each grid bus that connects substations before a bus in the network observes a reduction in capability. Comparing this capacity limit to the energy requirements of the electrified HDCV fleet (Figure 1c) we find a mismatch: At certain locations, the hourly electricity demand is more than what the current grid can provide. To address this, additional energy resources must be built.

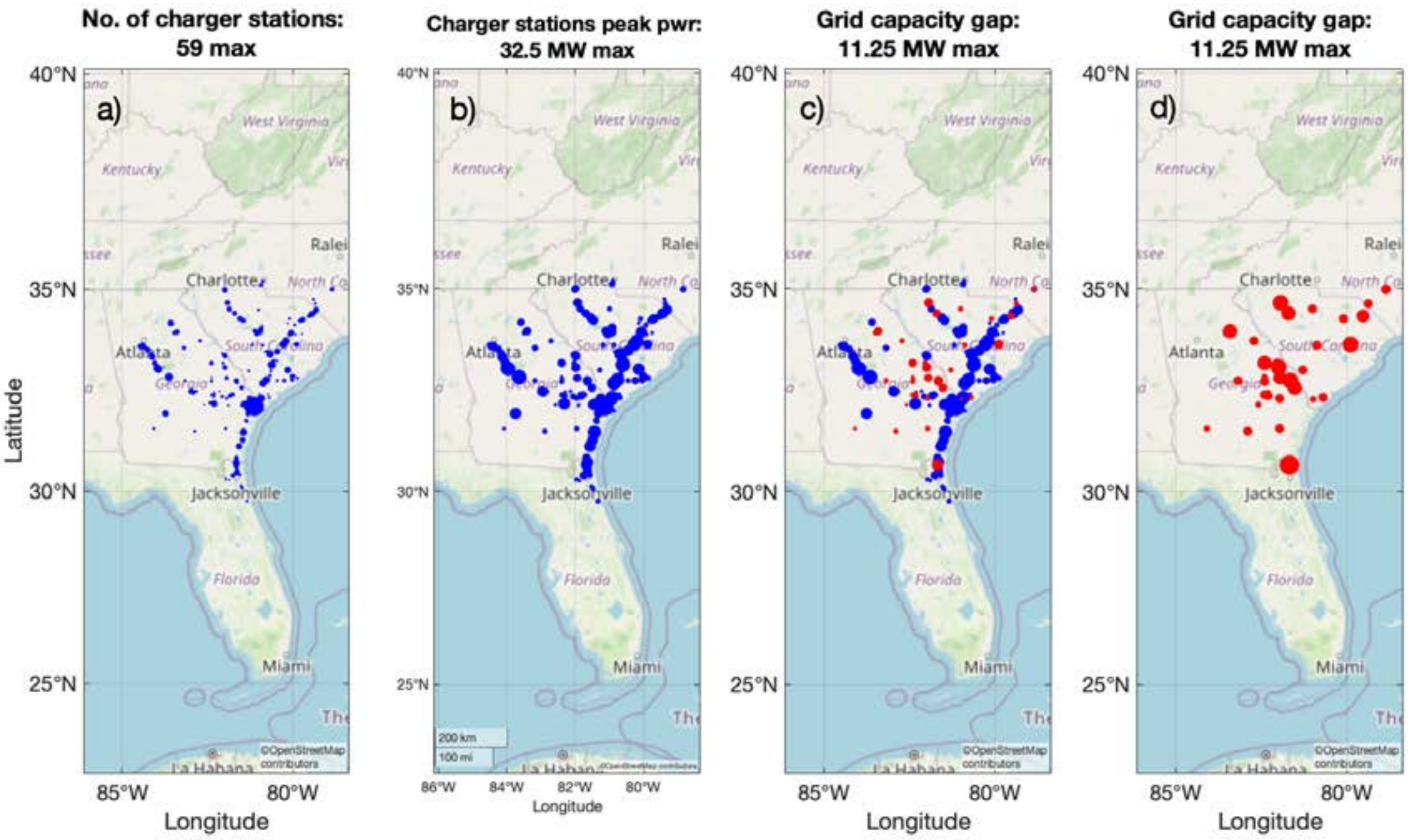

*Figure 1: Charger infrastructure and energy demands for the Port of Savannah region. a) Number of chargers at each station needed at destinations as well as the Port of Savannah to meet demand. Largest point corresponds to max number of 59 chargers. b) Peak power demand experienced at each charging station. Largest point corresponds to peak power of 32.5 MW. c) Grid capacity gap observed at red-marked locations. d) Charger locations where peak power would violate current grid capacity. Largest point indicates an excess of 11.25 MW above grid capacity.*

Deploying new energy resources as carbon intensive as the current grid infrastructure can significantly reduce the $CO_2$ emissions benefits of vehicle electrification. In the Port of Savannah region, electrifying ~80% of the fleet with trucks using 800 kWh batteries---with the remainder of the fleet continuing to be diesel trucks---would ensure continued full coverage of all the routes; however, if carbon-intensive technologies are used to power the excess load demand, this would only reduce overall emissions by about 37.6% compared to a fully diesel-based fleet (Fig. S3). While electrification lowers vehicle $CO_2$ emissions, some emissions are transferred from the vehicle tailpipe to the energy generation plant. Thus, deployment of renewable, carbon-neutral energy resources is necessary to realize greater $CO_2$ emissions reductions.

## USING A MICROGRID TO MEET INCREASED ELECTRICITY DEMANDS

### *The renewable energy network*

We propose a *renewable energy network* that integrates a microgrid with solar and battery power to supplement the available grid power in meeting the excess load demand. In this context, we consider both existing and future non-carbon-neutral resources as part of the 'grid.' Figure 2 provides a schematic representation of this network, with arrows indicating the flow of energy between the network's components[14,15]. The variables of the energy network along with their interpretations are summarized in Table 1.

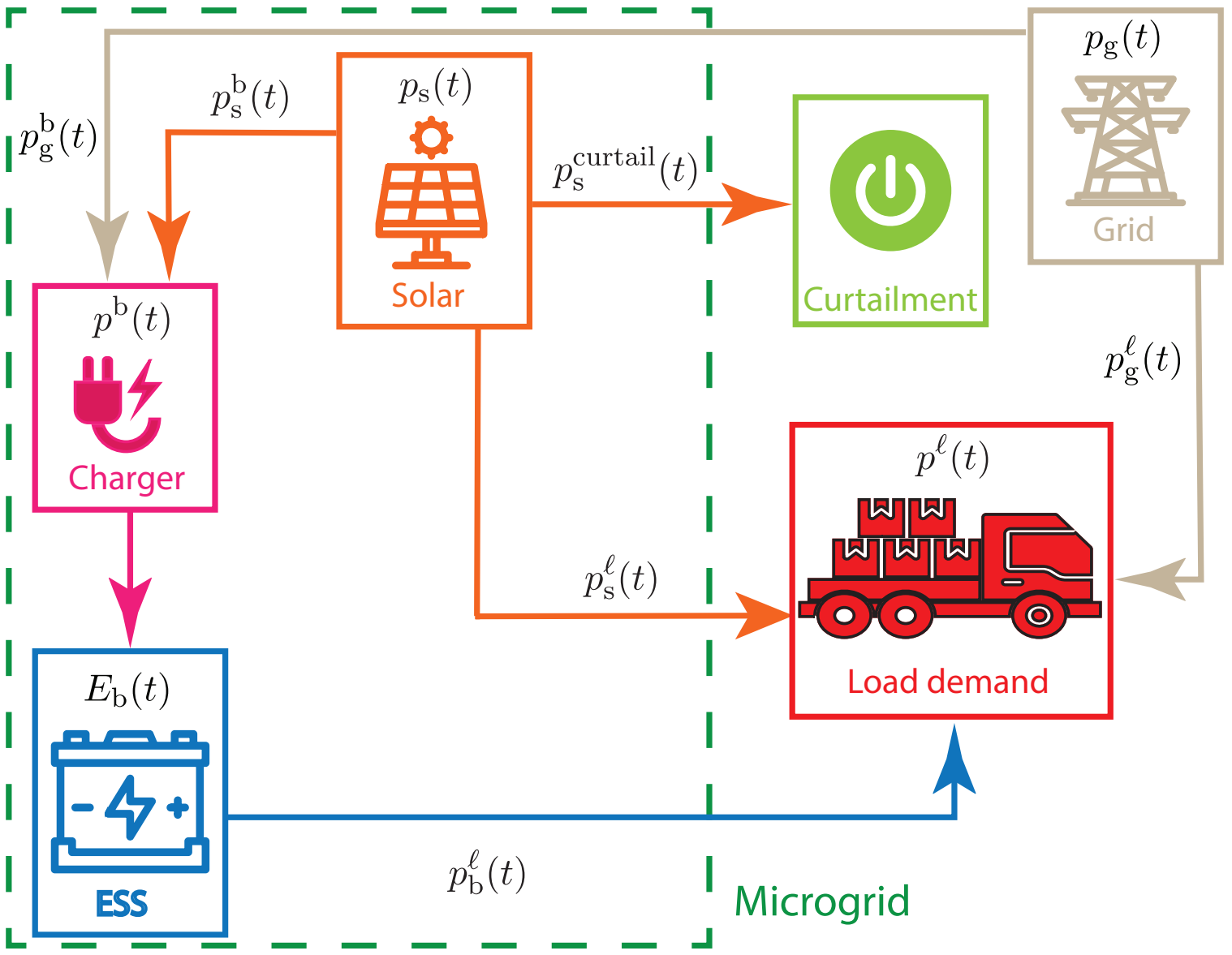

*Figure 2: Schematic of energy network. Colors correspond to origin of power: orange lines from solar, blue lines from the battery, and beige lines from the grid. Curtailment acts as an energy sink for solar.*

*Table 1: Energy network variables and their interpretation.*

| Variable [Unit] | Interpretation |
|---|---|
| $p^{\ell}(t)$ [MW] | Excess load demand |
| $p_{\mathrm{s}}(t)$ [MW] | Available solar power available |
| $p_{\mathrm{g}}^{\ell}$(t) [MW] | Power provided to load by grid |
| $p_{\mathrm{s}}^{\ell}$(t) [MW] | Power provided to load by solar |
| $p_{\mathrm{b}}^{\ell}$(t) [MW] | Power provided to load by battery |
| $p_{\mathrm{s}}^{\mathrm{b}}$(t) [MW] | Power provided by solar to charge the battery |
| $p_{\mathrm{s}}^{\mathrm{curtail}}$(t) [MW] | Curtailed solar power |
| $p_{\mathrm{g}}$(t) [MW] | Total power provided by the grid |
| $p^{\mathrm{b}}$(t) [MW] | Charging power to the battery |
| $p_{\mathrm{g}}^{\mathrm{b}}$(t) [MW] | Power provided by grid to charge battery |
| $E_{\mathrm{b}}(t)$ [MWh] | Battery energy remaining |

In our notation, for a power quantity (units of MW) denoted as $p_i^j(t)$, the subscript $i \in \{s, b, g\}$ denotes the network element origin of the power, that can be the solar, the battery, or the grid and $j \in \{b, \ell\}$ denotes the target network element, which can be the battery or the load. The variable $t$ denotes time. We direct the reader to the Supplemental Information – Section 2 for a mathematical formulation of the energy network.

*Maximally reducing excess $CO_2$ emissions by orchestrating resources*

We consider an energy network that relies solely on dispatching existing, non-carbon-neutral, grid resources to meet the load demand as a *baseline network* with the corresponding baseline emissions being calculated as the sum of the estimated hourly excess $CO_2$ emissions. In contrast, the renewable energy network uses a mix of solar, battery, and grid power to meet the excess load demand. As such, its emissions are calculated based on the proportion of grid power dispatched relative to the baseline. Given the hourly available solar capacity $p_s(t)$ and the excess load demand profile $p^{\ell}(t)$, we formulate a mixed-integer linear program, which determines the optimal power distribution from solar, battery, and the grid for each hour of the year, that maximizes the percentage reduction in $CO_2$ emissions for the renewable network compared to the baseline network, while meeting excess load demand. We direct the reader to Supplemental Information – Section 2 for a formal presentation of this optimization problem.

## MICROGRID SITING FRAMEWORK

We describe in this section, how the potential for siting and generation of solar resources is modeled by integrating two existing frameworks: the OR-SAGE tool from ORNL and the ReV model from NREL[16].

*Regional solar viability assessment*

The assessment of solar viability is performed using OR-SAGE[17]. OR-SAGE makes use of high-resolution geospatial data layers that integrates existing restrictions on land-use based on factors such as federal guidelines, environmental/human impacts, and existing infrastructure. Each layer divides the region under consideration into 100x100 m grid cells, which is upsampled to a 90 m resolution. Some layers, such as wetlands and open water, are known as decision layers: The grid values directly correspond to whether the region they describe is suitable for siting a particular technology. In contrast, other layers, such as layers describing variations in slope, must be converted into a decision layer by using a filtering threshold on the data. Further details on the technology parameters and threshold values used for filtering in this analysis may be found in Supplemental Information – Section 3.

Within each decision layer, a value of 0 indicates that a grid square is suitable for siting, and a value of 1 indicates a violation of the constraint specific to that layer. These data layers can then be overlayed to create a composite siting map, where the value of a grid square corresponds to the number of constituent data layers that conflict with a siting decision at the given location (Fig. S4). For this analysis, we focus on the region surrounding Chatham County, GA, where the Port of Savannah is located. Specifically, the three counties surrounding Chatham are also included in this analysis to encompass all available areas within the immediate surrounding region of the Port (Figs. S5-S6). The section of the solar viability map corresponding to this region can be seen in the red starred binary map in Figure 3.

*Regional solar capacity availability*

The calculations of technical potential and average capacity for the Port of Savannah region are performed using the various modules available in NREL's ReV tool[16], integrating the viability map created in the previous section. A schematic overview of these modules and how they are integrated with OR-SAGE is displayed in Figure 3.

The pipeline begins with the selection of a region of interest (ROI). In ReV, the ROI is defined by a series of grid-points that are associated with traditional geographic coordinates. In the case of solar technical capacity, these grid points correspond to spatial data points within the National Solar Radiation Database (NSRDB), which measures three forms of solar radiation alongside a series of atmospheric and surface conditions every hour at an approximately 2 km resolution[18]. Next, the generation module of ReV is used to calculate capacity factor values---the ratio between the actual generation capacity of a site and its nameplate capacity---for a solar farm placed at each selected grid point in the NSRDB dataset. This is done for every hour in the chosen year, which results in a set of temporal capacity factor profiles for each farm.

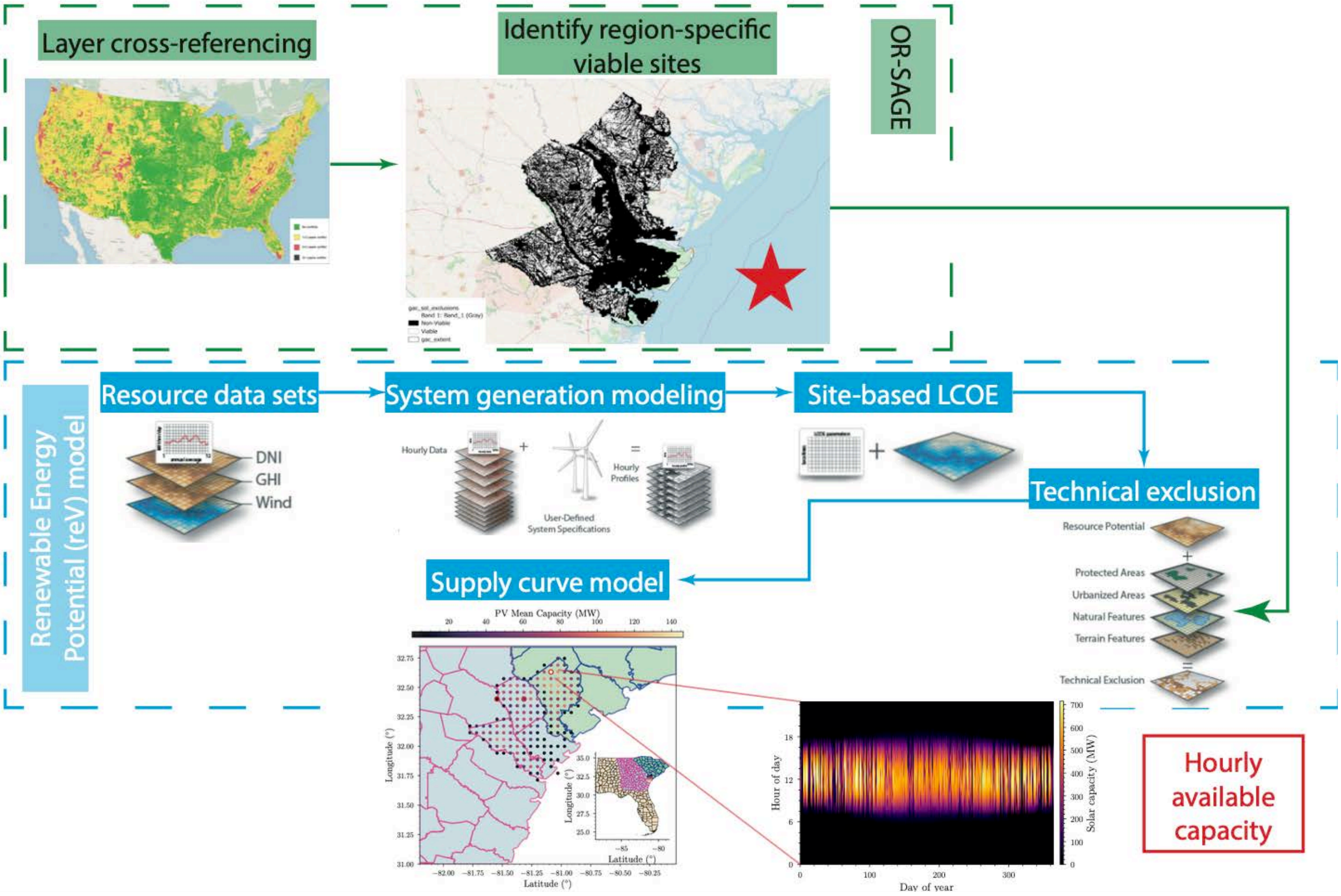

*Figure 3: Schematic of solar siting and hourly available capacity estimation framework. Plot marked with red star shows the binary map of non-viable (marked black) sub-regions within the greater Port of Savannah region considered.*

Subsequently, two processes are initiated in the ReV model: (1) supply curve aggregation and (2) representative profile generation. Supply curve aggregation maps the low-resolution NSRDB grid points to regions within the high resolution (90 m) site suitability data output from OR-SAGE. Following this, the model re-aggregates the high-resolution data and calculates the total amount of viable land and associated capacity factors for the resulting ~33 $km^2$ land parcels. The land values are multiplied by a scalar power density value of 36 $MW/km^2$ to calculate the technical potential generation capacity for each region. The resulting value is the nameplate capacity for a solar farm occupying all available land in the parcel. For the Port of Savannah region considered, this aggregation identifies a set of 171 *viable solar sites*. The geographic locations of these sites and the nameplate solar capacity of each site is shown in Figure 4. By convention, the sites are numbered in order of increasing nameplate solar capacity.

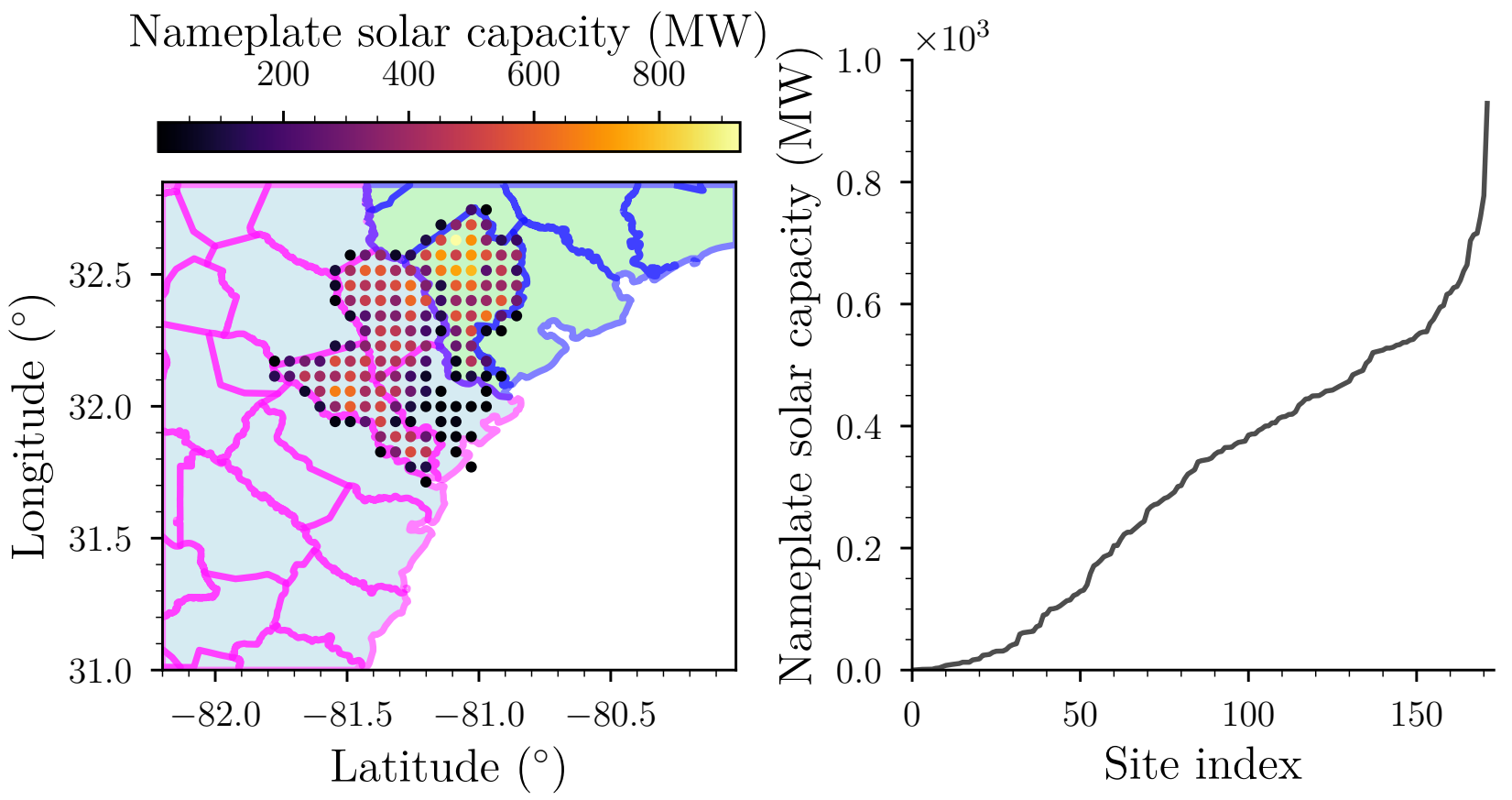

*Figure 4: Possible solar sites in the region around Port of Savannah. (Left plot) Geographic locations of the viable sites colored by their nameplate capacity. (Right plot) Nameplate solar capacity vs. site index.*

Representative profile generation aggregates the capacity factor profiles created in the generation module to the same resolution as the parcels in the supply curve aggregation module. Post-processing is performed on the ReV output to calculate the hourly available capacity profiles $p_s(t)$ from representative capacity factors and nameplate generation capacities for each viable site (Fig. S7).

## ESTIMATING THE COSTS OF MICROGRID DEPLOYMENT

Having established the set of viable sites and the framework which determines how each element of the energy network is used if a microgrid is deployed at a given site, this section formulates the cost of deployment for the microgrid. We then propose a set of cost metrics for the utility provider Here, we consider the microgrid to be installed as a front-of-the-meter resource. Thus, the capital costs of deployment are borne by the utility. The changes to the total cost of electricity due to the microgrid deployment, however, will affect both front-of-the-meter (utility) and behind-the-meter (fleet operator) stakeholders. A mathematical presentation of the costs and cost metrics described in this section is available in Supplemental Information – Section 4.

### *Capital cost expenses for photovoltaic farms*

To assess the capital cost of a utility-scale solar system, a comprehensive capital cost calculator has been developed based on the U.S. solar PV system benchmarks reported by NREL and their corresponding "PV System Cost Model" (PVSCM) model[19].

A 100-$MW_{DC}$ utility-scale PV system with single-axis tracking, a central inverter with an inverter load ratio of 1.34 $kW_{DC}/kW_{AC}$, and crystalline silicon (c-Si) solar cell modules with an efficiency of 0.203 $m^2/kW_{DC}$ is used as the representative system for collecting values from the PVSCM model. Our capital cost calculator integrates these components and values, adjusting for the nameplate capacity of the system, to provide a detailed estimate of the total investment required for deploying a solar PV system of a given nameplate capacity. The use of up-to-date benchmarks ensures that the calculations reflect current market conditions and technological advancements.

*Capital cost expenses for battery energy storage*

To estimate the capital cost variation of different LIB systems, we leverage the capabilities of the Battery Performance and Cost (BatPaC Version 5.1) tool, developed by Argonne National Lab[20]. Using this tool, we estimate the capital costs for building a battery system using the leading three chemistries[1] currently used in the industry: NMC (811), NCA, and LFP[21].

To obtain a capital cost estimate for the battery pack using a given chemistry, BatPaC requires the user to specify certain details of the battery pack. Here, we consider a battery pack with a duration of 4 hours, obtained by appropriately specifying the target power and the rated energy of the pack. As utility-scale storage is a long-duration application, we consider only energy-type cells in our analysis.

The BatPaC analysis generally yields numbers that are significantly less than the average price across the literature, as seen in the review by Mauler et al.[21]. To better align with literature estimates and predictions, while retaining the variation of cost between chemistries, we take the average cost, over a range of rated durations, of a pack rated at $1MW_{DC}$ power produced by BatPaC and find a scale factor to the average price of that chemistry reported in the Mauler et al. report[21]. This scale factor is then applied uniformly all higher power ratings and durations for that chemistry.

In addition to the cost of the battery storage unit, we also account for the costs associated with installation and integration of the battery storage unit in obtaining the overall cost of the battery system.

*The total cost of ownership*

To quantify the techno-economic benefits and costs of a microgrid with solar and battery systems, the total cost of ownership (TCO) is calculated as the sum of initial costs (IC) and annual operating costs (AOC) over the years of operation (YrsOp). The total electricity cost (TEC) is key for all stakeholders and depends on two metrics: the levelized cost of photovoltaic recharge (LCOPR) and the levelized cost of storage (LCOS)[22]. LCOPR reflects the cost of solar energy dispatch based on the capital cost and total energy output over the solar plant's life. LCOS measures the battery system’s discharge cost, based on capital cost and the total energy discharged throughout the battery’s life, informed by the number of charge-discharge cycles that can be realized until the end of life. In addition, the TEC also reflects the average utilization of each resource to meet the excess load demand.

*Table 2: Levelized cost-of-storage for averaged for each battery chemistry.*

| Battery chemistry | LCOS ($/MWh) |
|---|---|
| NMC | 147.91 |
| NCA | 267.58 |
| LFP | 69.20 |

[1] For the LIB we consider, we assume that the negative electrode material is always lithiated graphite. As such, here “chemistry” in this work refers to a choice of positive electrode material. Other choices could be made for the negative electrode which could constitute a different “chemistry”.

The LCOS value, averaged over a range of battery sizes between 10 to 100 MWh, for different chemistry choices are shown in Table 2. Due to its high expected lifetime, the LCOS of LFP is lower than the other chemistries despite its high capital cost. In contrast, NCA has the highest LCOS of all 3 chemistries due to its high capital cost and low expected lifetime.

*Utility cost metric*

We propose the cost metric for the utility provider to be the TCO per kilogram of excess CO2 emissions removed. This metric measures the success of a utility to meet the power demands of electrified fleets while minimizing $CO_2$ emissions, preventing reliance on non-carbon-neutral resources. The utility's TCO includes the overnight capital costs for deploying the microgrid and the annual operating cost, calculated by multiplying the total excess load demand by the TEC. The microgrid system's operational lifespan YrsOp is the minimum of 30 years for solar or the battery's lifespan, determined by the usage profile of the battery and the cycle life of the chosen chemistry. The utility's cost metric is thus the ratio of its TCO, given by the sum of the above costs, to the weight of $CO_2$ emissions removed, calculated by the difference in $CO_2$ emissions between the renewable and baseline energy networks.

*Fleet operator cost metric*

The fleet operator's cost metric is the TCO per mile for a fleet HDCVs using the renewable energy network. Here, the TCO is calculated for a fleet of 1,825 vehicles. The initial cost for each vehicle in the fleet includes the Manufacturer's Suggested Retail Price and the registration cost but offset slightly by any existing government subsidies. At the end of the fleet's 5-year operational lifetime, vehicles retain a residual value based on a percentage of the MSRP. The fleet operator's AOC includes insurance costs per vehicle, vehicle maintenance costs, and the cost of electricity used by the fleet. The TCO for the full fleet is calculated by summing these costs. Dividing this TCO by the total vehicle miles travelled by the fleet over the 5 years of operation yields the fleet operator's cost metric. We additionally account for a penalty to account for the increased cost of dwell time to charge, as well as the payload capacity loss for electrified HDCVs which are generally heavier due to the on-board battery [23].

As a point of comparison, the TCO/mile of a diesel-based HDCV fleet os evaluated using the same framework above, but with diesel rather than electricity as the fuel source, and we find that the cost of operating a diesel-based fleet in the Port of Savannah region is 1.68 $/mile. This serves as a benchmark for which we can compare the costs of an electrified fleet.

## CAPABILITY AND TECHNO-ECONOMIC ANALYSIS OF VIABLE MICROGRID DEPLOYMENT SITES

In this section, we examine the capabilities of each of the 171 viable sites identified by our siting framework, in reducing excess $CO_2$ emissions if a microgrid with solar and battery energy storage are deployed there. We then examine the costs associated with deployment and investigate the tradeoffs that manifest for different transportation stakeholders.

*Site-specific maximum excess $CO_2$ emissions reduction potential*

To perform the resource dispatch optimization for each viable site, we use the estimated hourly available solar capacity profile for that site. We note that while this analysis yields an estimate for the maximum excess $CO_2$ emissions reduction potential of a given site, reduction values less than the maximum can always be realized by building a solar farm with a lower nameplate solar capacity than the site's nameplate solar capacity. An extended analysis on how variation of solar and battery system size affects the capabilities and deployment costs at a given site, is performed in Supplemental Information – Section 6. Hereafter, we will consider two cases for each viable site: (1) a microgrid is deployed with a solar system alone, and (2) a microgrid is deployed with both solar and a 100 MWh battery system.

We find (Figure 5, right plot) that a maximum excess $CO_2$ emissions reduction ~90% is achievable by the largest nameplate capacity site with the addition of solar resources alone to the energy network. Tantalizingly, the addition of a 100 MWh battery at the same site increases this reduction level to near 100%. Thus, with both solar and batteries deployed at the site with the highest nameplate capacity in this region, the full load demand of the electrified HDCV fleet in the Port of Savannah region may be provided entirely by carbon-neutral resources, thereby eliminating the operational carbon emissions of this fleet.

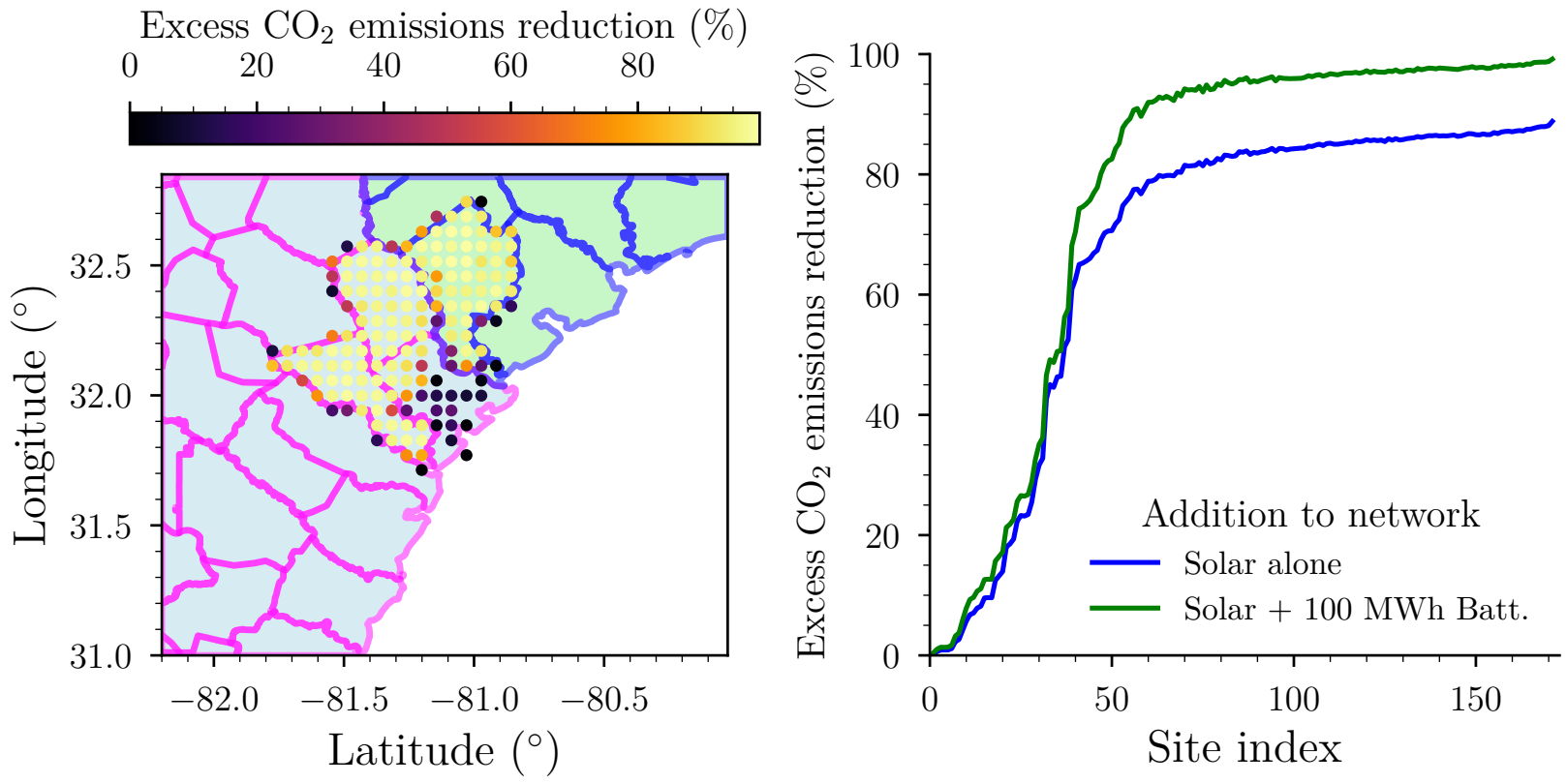

*Figure 5: Excess CO2 emissions reduction potential of the viable sites in the Port of Savannah region. (Left plot) Geographic locations of viable sites. Point colors: maximum total excess CO2 emissions reduction, assuming deployment of solar and a 100 MWh battery system. (Right plot) Maximum total excess $CO_2$ emissions reduction for every site. Blue curve: only solar system deployed, no battery. Green curve: Solar system and 100 MWh battery deployed.*

### *Site-specific total cost of electricity*

We estimate the TEC for all viable solar sites in the Port of Savannah region in Figure 6. The details of the overnight capital costs, site-specific LCOPR, LCOS, and average resource utilization that result in this TEC estimate may be found in Supplemental Information – Section 5. The TEC of the baseline network is equivalent to the grid electricity price (GEP) of 160 $/MWh. As the LCOPR of ~26 $/MWh is significantly lower than the GEP in this region (Fig. S16), we find that deploying solar alone generally improves the TEC over the baseline network. As sites with increasing nameplate solar capacity are considered, the TEC decreases further as there is less utilization of the grid.

Deploying a microgrid with an additional 100 MWh battery can increase or decrease the TEC, depending on the choice of battery chemistry. We find that choosing LFP as the battery chemistry, across all the viable solar sites, results in a TEC lower than that of a microgrid with solar alone. This is due to the LCOS of LFP (Fig. S12) being significantly lower than the GEP: Using the battery to provide power to the load is considerably cheaper than using the grid. In contrast, deploying an NMC battery increases the cost relative to just deploying solar alone. As the LCOS of NMC is only marginally smaller than the GEP, using an NMC battery in this region has a mitigated improvement on the cost; however, the TEC associated with NMC is still better than the baseline for all except the lowest capacity sites. Due to its high LCOS relative to the GEP, there are only a few select sites in which the TEC of a NCA battery system deployed alongside a solar farm is less than that of the baseline network. Thus, we find that LFP is the best choice of battery chemistry for this application, in this region.

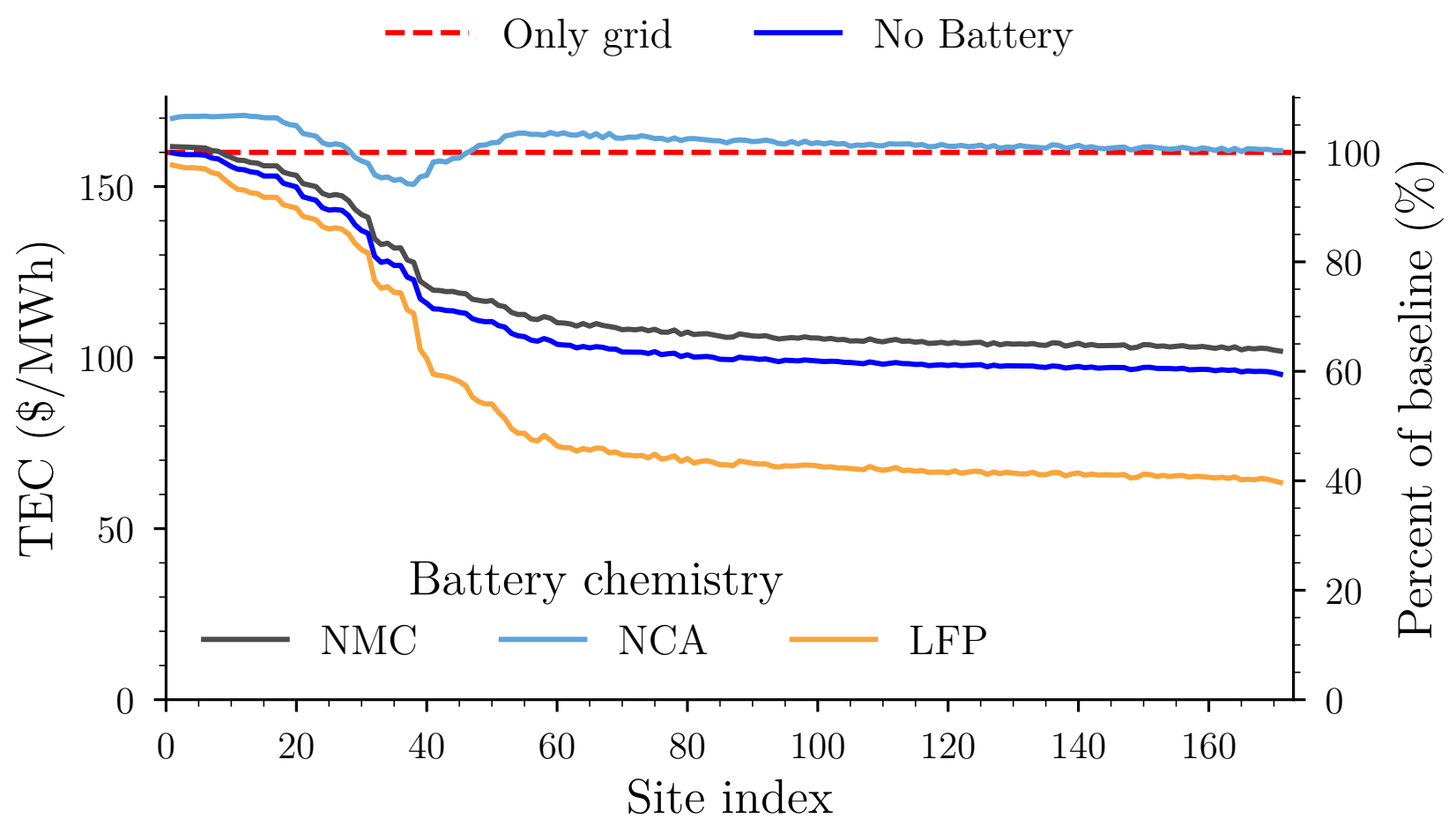

*Figure 6: Total electricity cost (TEC) for viable solar sites in Port of Savannah region. Only grid (red dashed line): TEC of baseline network where only grid power is used to meet excess load demand. No battery (blue solid curve): microgrid deployed with only solar system. Other colors denote different choices of chemistry when 100 MWh battery is deployed along with the solar system.*

### *Site-specific utility cost metric*

We show the estimated cost metric for the utility in Figure 7. Notably, we find that there is a minimum in the cost metric at a site with an intermediate nameplate solar capacity. On one hand, low nameplate capacity sites have a low ability to reduce excess $CO_2$ emissions but still must incur the capital costs of deploying solar and battery resources, leading to a relatively high value of the utility cost metric. On the other hand, due to the diminishing ability for high nameplate capacity sites to reduce excess $CO_2$ emissions, the costs of deploying at sites with increasingly larger nameplate capacity eventually becomes dominant. The competition between these two effects leads to a clear minimum in the cost for the utility.

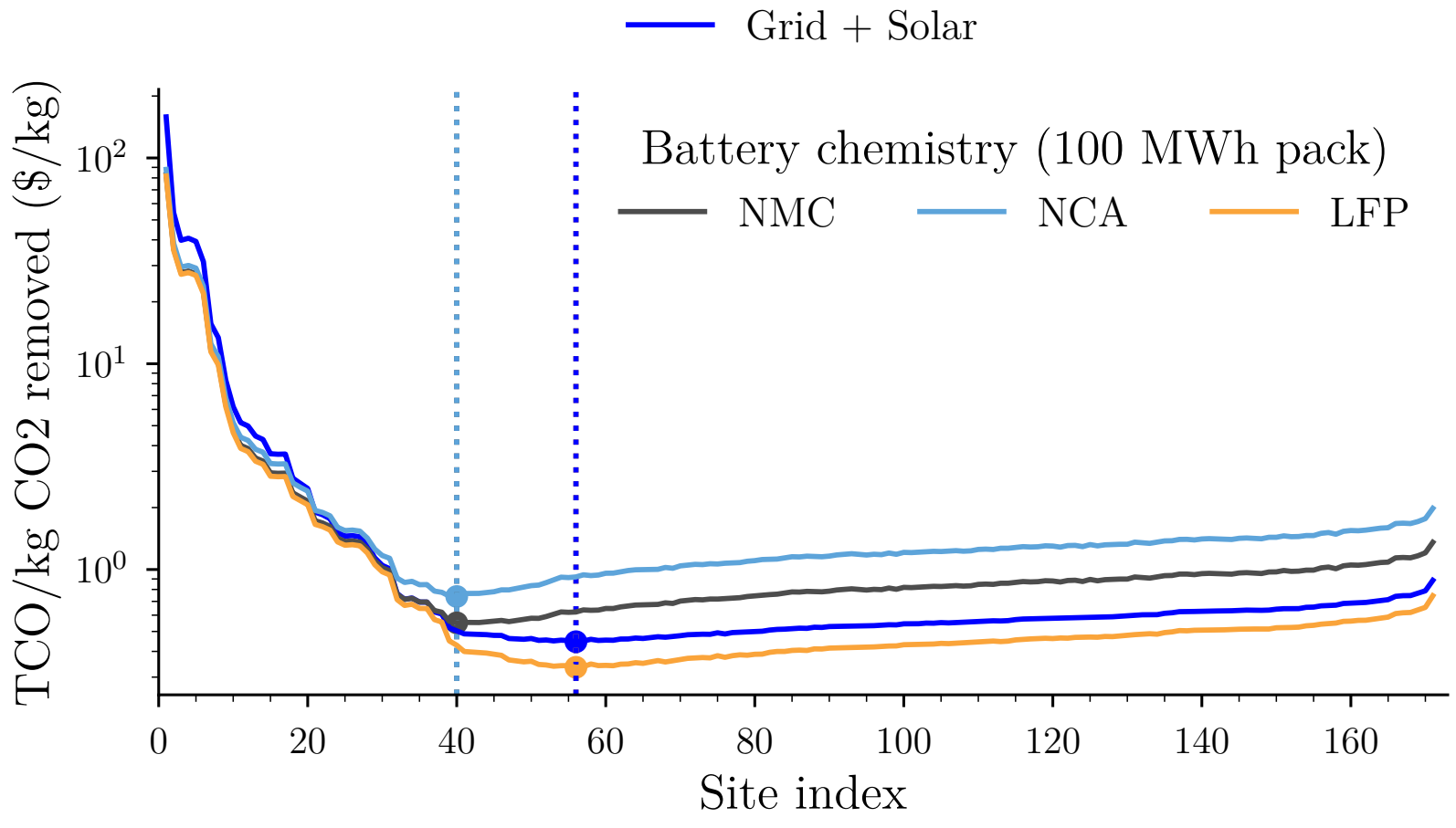

*Figure 7: Utility cost metric. Grid + Solar: Microgrid deployment with only solar alone. Other colors denote different choices of chemistry when 100 MWh battery is deployed along with the solar system.*

The addition of a 100 MWh LFP battery system tends to decrease costs further, relative to when only solar is installed, across all viable sites. In contrast, the addition of a 100 MWh NMC or NCA battery system is only beneficial sites with lower nameplate solar capacities. For higher nameplate capacity sites, the additional cost of introducing an NMC or NCA battery system results in a cost metric that is higher than that of deploying solar alone.

*Site-specific fleet operator cost metric*

Turning to the fleet operator, we estimate the TCO/mile for every viable site in Figure 8. We first consider the case of full electrification: The fleet transitions the maximum number of trucks to electrified powertrains while maintaining coverage of all current routes in the region. We observe (Figure 8, left plot) that without deployment of a microgrid in this case, leads to a TCO/mile to the operator of ~2.68 $/mile. Therefore, electrification amounts to a ~68% increase in total cost to the fleet operator per mile. We find that the increase in cost can be mitigated with the deployment of a microgrid. In the best-case scenario, deploying at the largest nameplate capacity site and using an LFP battery, the cost can be lowered to ~2.44 $/mile, or equivalently a ~53% increase in price relative to a pure diesel-based fleet. Deploying only solar alone at this site increases the price or 58% increase in cost metric relative to diesel.

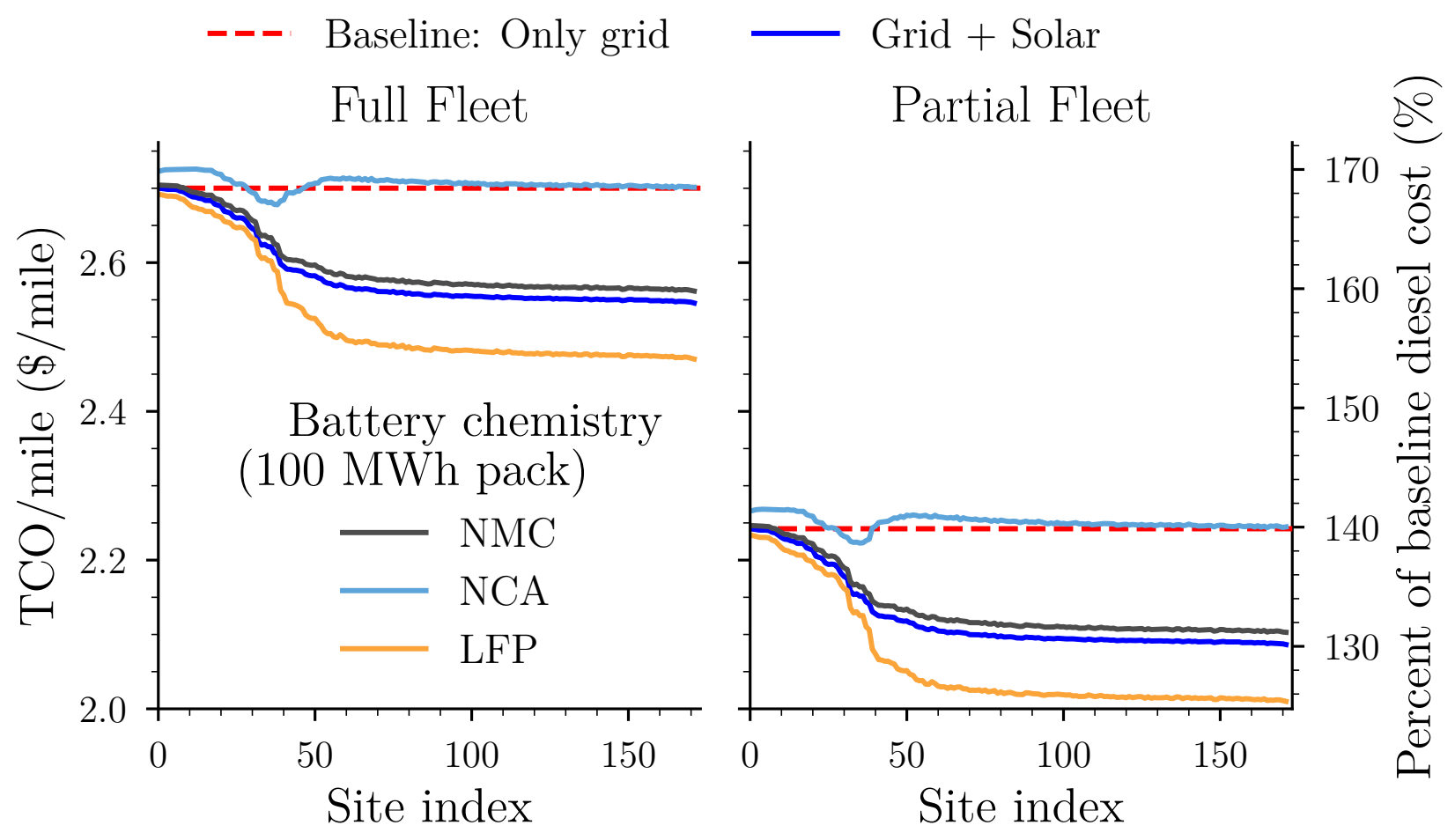

*Figure 8: Cost per mile to fleet operator with a fully electrified fleet (left plot) and a partially electrified fleet (right plot). Only grid (red dashed line): TEC of baseline network where only grid power is used to meet excess load demand. No battery (blue solid curve): microgrid deployed with only solar system. Other colors denote different choices of chemistry when 100 MWh battery is deployed along with the solar system.*

As the full fleet scenario results in significant cost increases to the fleet operator, a more conservative scenario sees fleet operators only partially electrifying their fleets. Specifically, we consider a scenario wherein only trucks that travel a significant number of miles, with annual VMTs of 60,000 miles or greater, are electrified. This amounts to $N_{trucks} = 1087$ and the average annual VMT per truck is $VMT_{tot} = 84{,}932$. Performing the same analysis for this conservative scenario sees the TCO/mile lower significantly (Figure 8, right plot), with the cost increase only amounting to a 40% increase in TCO/mile relative to diesel in the no microgrid case.

Interestingly, the sites where the minimum in the utility cost metric occurs correspond to sites before diminishing returns are observed in the fleet's cost metric. While quantitatively the optimal site for the fleet operator is the largest nameplate capacity site, as it leads to the smallest TCO/mile, deploying resources at the site optimal for the utility results in only a 1% increase to the fleet operator's cost metric for either electrification scenario. Thus, we find an alignment in the optimal site corresponding to the (near) lowest cost metrics for both stakeholders.

## CONCLUSION

The electrification of drayage HDCVs at intermodal hubs like ports is crucial for reducing greenhouse gas emissions and achieving U.S. climate goals; however, this transition increases electricity demand and challenges the existing grid infrastructure. Without integrating renewable energy, electrification may shift some $CO_2$ emissions from the vehicle tailpipe to the energy generation plants.

We developed an integrated framework that integrates a pipeline using ORNL's OR-SAGE tool and NREL's ReV model to identify viable sites for microgrid deployment and evaluate each site's ability to provide energy, an optimization framework that allows for the estimation of the maximal $CO_2$ emissions reduction impact that a microgrid deployment at a given site may have and a cost analysis that reveals how microgrid deployment and usage changes the TCO changes for various stakeholders. Applying this to the region around the Port of Savannah, GA USA, we identify and evaluate the capabilities of 171 viable sites for microgrid deployment. Our findings suggest that deploying a microgrid at sites with intermediate nameplate solar capacity in this region maximizes $CO_2$ reduction at the lowest cost, aligning the interests of both utilities and fleet operators.

While we have shown the capability of our framework to evaluate the potential impact and cost of deploying a microgrid in a single region, it may be straightforwardly extended in various ways. Firstly, it would be interesting to consider the siting of multiple microgrids, installed in either a front-of-the-meter or behind-the-meter fashion, and investigate how this changes the costs for both the utilities and the fleet operators. Secondly, these deployed microgrids may consist of distributed clean energy resources such as solar, but also wind, hydro, or nuclear power, as well as other long-duration energy storage systems beyond LIBs. Finally, these microgrids would connect to one or more vehicle charging networks, which could serve private or public fleets. The energy storage capabilities of the vehicles in these fleets could be used for bidirectional energy transfer, managed by permissions and state management controls. A network with these additional elements is schematically shown in Figure 9.

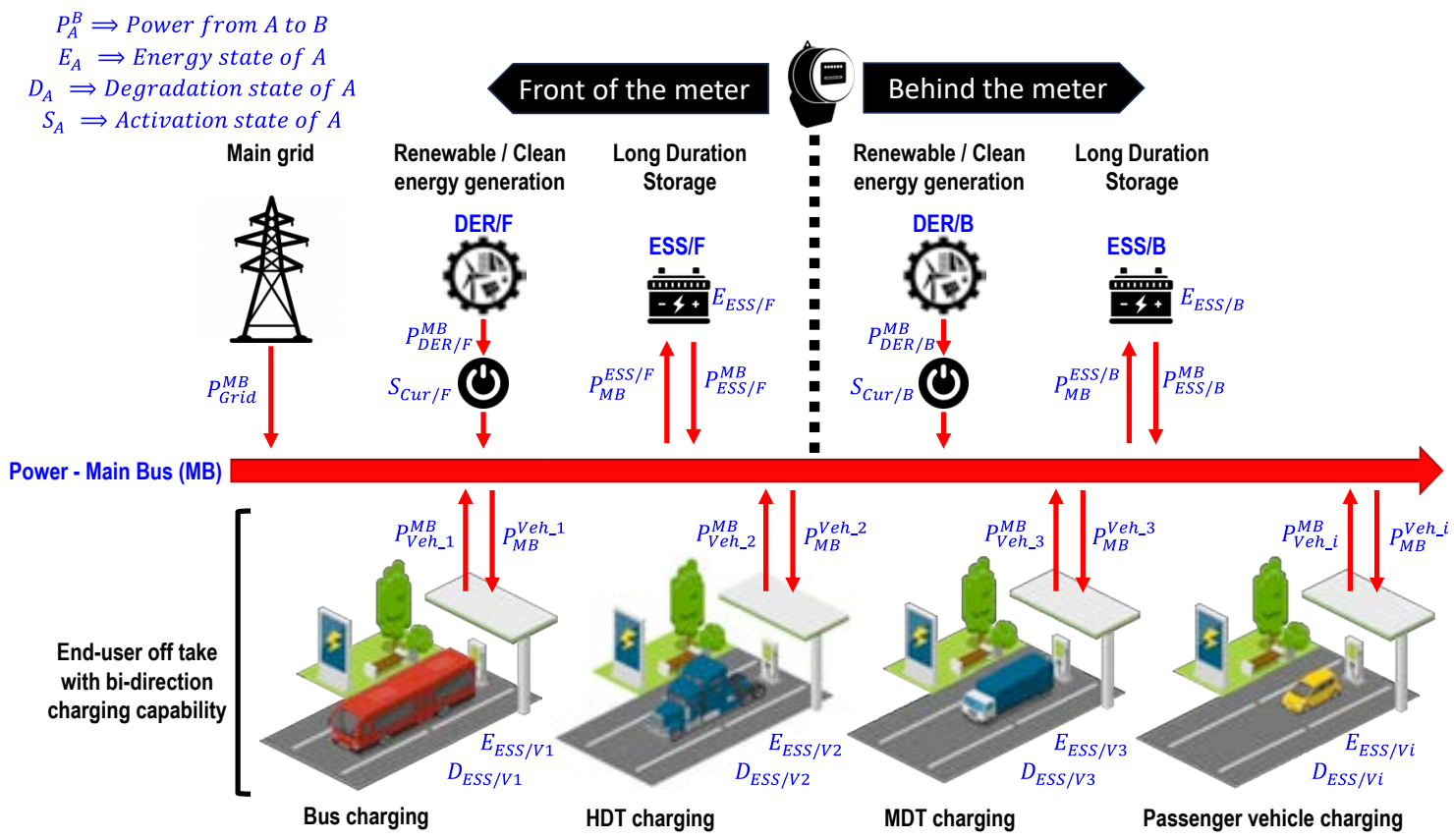

*Figure 9: Power management/control system of i2MG.*

Our analysis, focusing on the Port of Savannah, demonstrates the critical role of microgrids, comprised of renewable energy and utility-scale storage, in reducing excess $CO_2$ emissions originating from electrification: Integrating solar and battery resources via a microgrid can significantly lower or even eliminate these excess emissions. Regional differences in grid composition and renewable energy availability however, mean that the impact of microgrid deployment will vary across the U.S. In regions with lower grid costs and emissions, the benefits of renewables could be less pronounced. Expanding the application of this framework to the whole U.S. would allow the quantitative assessment of microgrid deployment's potential impact and feasibility in regions around different intermodal hubs across the nation. Understanding where microgrid deployment can realize the largest $CO_2$ emissions reduction, as well as the associated costs, will help both policymakers and industry stakeholders that are aiming to mitigate the climate impact of the freight transportation sector in making maximally impactful and cost-effective decisions.

## EXPERIMENTAL PROCEDURES

### Resource Availability

*Lead Contact*

Further information and requests for resources should be directed to and will be fulfilled by the Lead Contact, Vivek A. Sujan (sujanva@ornl.gov).

*Materials Availability*

This study did not generate new materials.

*Data and Code Availability*

The generated datasets, example optimization implementation, and associated analysis scripts developed in this work can be obtained from DOI: 10.5281/zenodo.13509426. The microgrid siting pipeline is based on previously published tools of OR-SAGE (https://doi.org/10.2172/1032036) and ReV (https://doi.org/10.2172/1563140). The TCO calculators incorporate protected information and thus cannot be released publicly. Access to these may be obtained by contacting the Lead Contact, Vivek A. Sujan (sujanva@ornl.gov).

## SUPPLEMENTAL INFORMATION

Document S1. Supplemental Numerical Procedures and Data, Figures S1–S25, and Table S1-S6.

## ACKNOWLEDGMENTS

This research was funded by the US Department of Energy (DOE) Office of Energy Efficiency and Renewable Energy, Vehicle Technologies Office under Contract No DE-AC05-00OR22725. This research was supported in part by an appoint to the Oak Ridge National Laboratory GRO Program, sponsored by the U.S. Department of Energy and administered by the Oak Ridge Institute for Science and Education. This research used resources of the Compute and Data Environment for Science (CADES) at the Oak Ridge National Laboratory, which is supported by the Office of Science of the U.S. Department of Energy under Contract No. DE-AC05-00OR22725. Some of the computing for this project was performed on the Sherlock cluster. We would like to thank Stanford University and the Stanford Research Computing Center for providing computational resources and support that contributed to these research results.

## AUTHOR CONTRIBUTIONS

Conceptualization, J.N.E.L. and V.A.S.; methodology, J.N.E.L., R.S., B.A.M., V.A.S.; data curation, J.N.E.L., B.A.M.; software, J.N.E.L., B.A.M. V.A.S.; visualization, J.N.E.L., B.A.M.; writing – original draft, J.N.E.L.; Writing – review and editing,

J.N.E.L., R.S., B.A.M., S.O., V.A.S.; supervision, S.O., V.A.S.; project administration and funding acquisition, V.A.S.

## DECLARATION OF INTERESTS

The authors declare no competing interests.

# Supplemental Information

## 1. Quantifying the two problems with fleet electrification

This section provides an expanded discussion on how the excess load demand of an electrified HDCV fleet operating in the Port of Savannah region is estimated, as well as the anticipated increase in $CO_2$ emissions.

As described in the main text, using the OR-agent framework, an estimate of the hourly energy required by the full fleet of electrified trucks, shown in *Figure S1*, is generated.

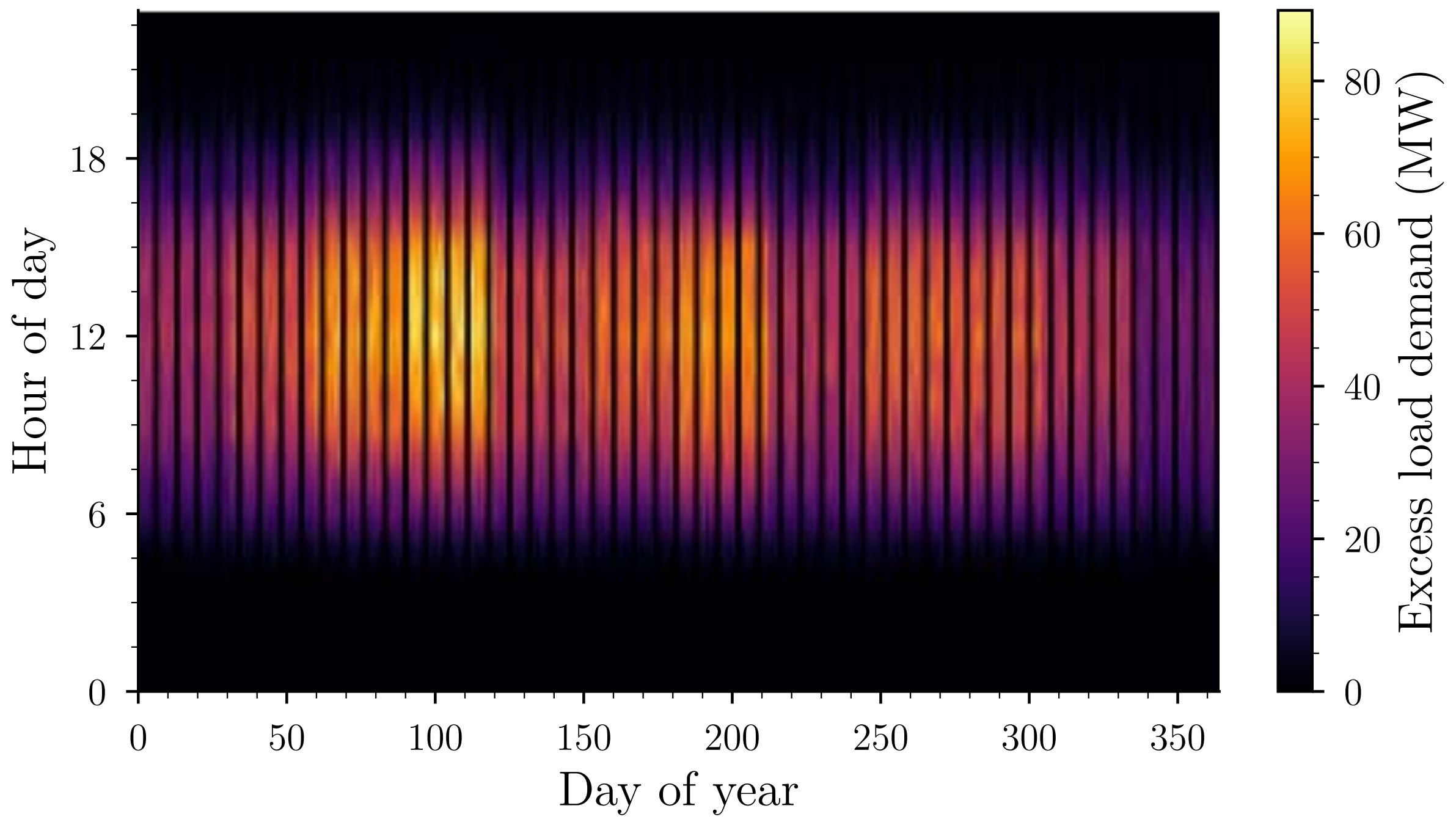

*Figure S1: Excess load demand from a fleet of electrified HDCVs operating in the region around Port of Savannah.*

Associated with this excess load demand is an amount of emitted $CO_2$. The excess $CO_2$ emissions can be estimated by an approach first proposed in Sujan et al.[1]. Briefly, this approach integrates historical spatiotemporal data on the carbon intensity and the load demand of the grid for all the counties in the U.S. and performs forecasting for the $CO_2$ emissions associated with the excess load demand. Thus, given a county of interest, the county's historical electric load profiles, the historical energy generation profiles for that region, as well as the excess load profile, the excess $CO_2$ emissions can be estimated for every hour of the year. A schematic of the inputs and output of this pipeline is shown in *Figure S2*. Additionally, by construction, when the excess load profile is zero---as it is on the weekend hours for example---the excess CO2 emissions will also be zero.

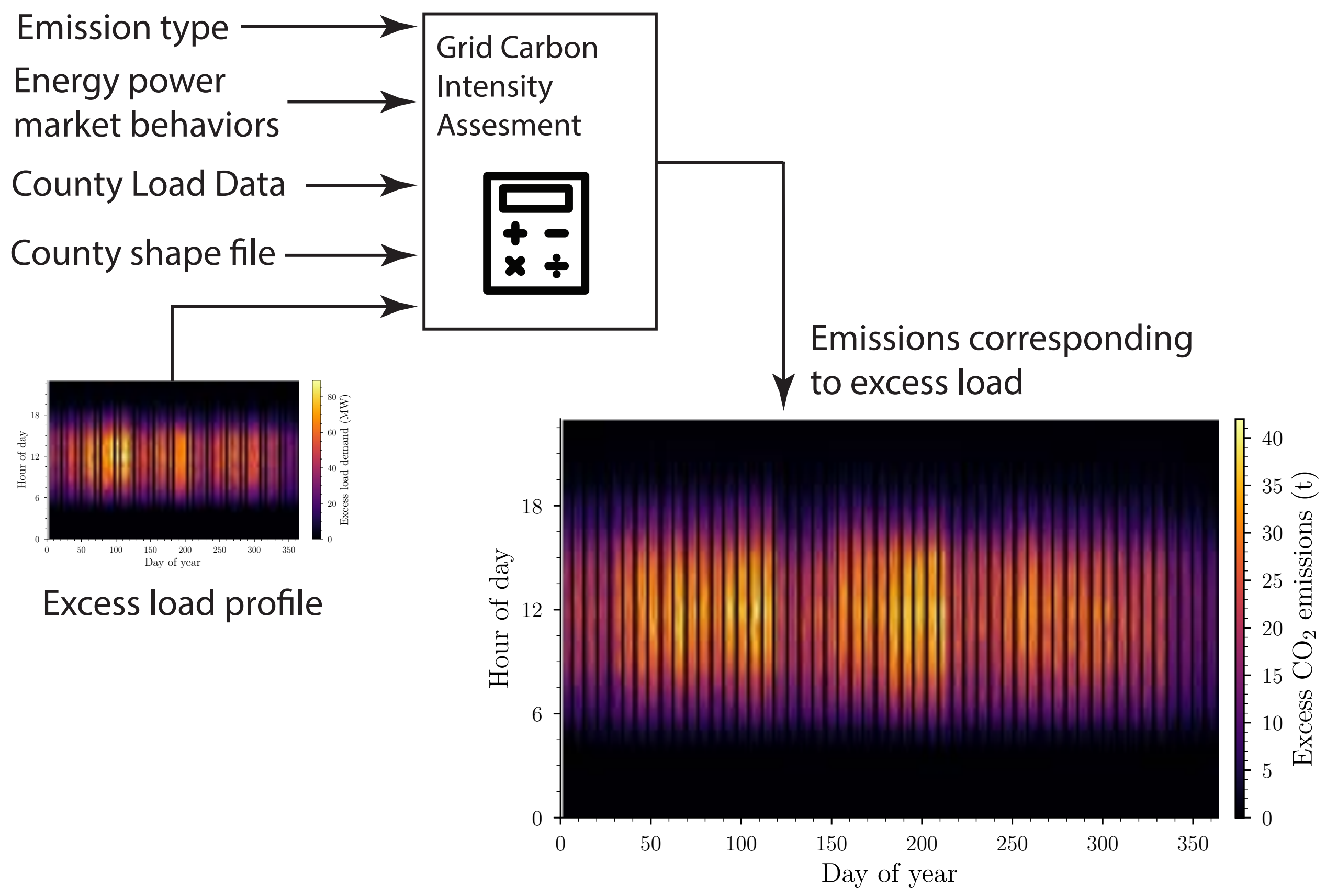

*Figure S2: Schematic of excess CO2 emissions estimation pipeline.*

If additional energy resources are built that are as carbon intensive as the current grid infrastructure, the electrification of HDCVs may yield limited $CO_2$ emissions reduction. For the Port of Savannah region, electrifying about 80% of the fleet with trucks that have an 800 kWh battery pack ensures full coverage of all the current routes; however, we estimate that this electrification will reduce emissions by only about 37.6% (see *Figure S3*) compared to a fully diesel-based fleet if the same carbon-intensive energy generation sources used in the grid today are used to power the excess load demand.

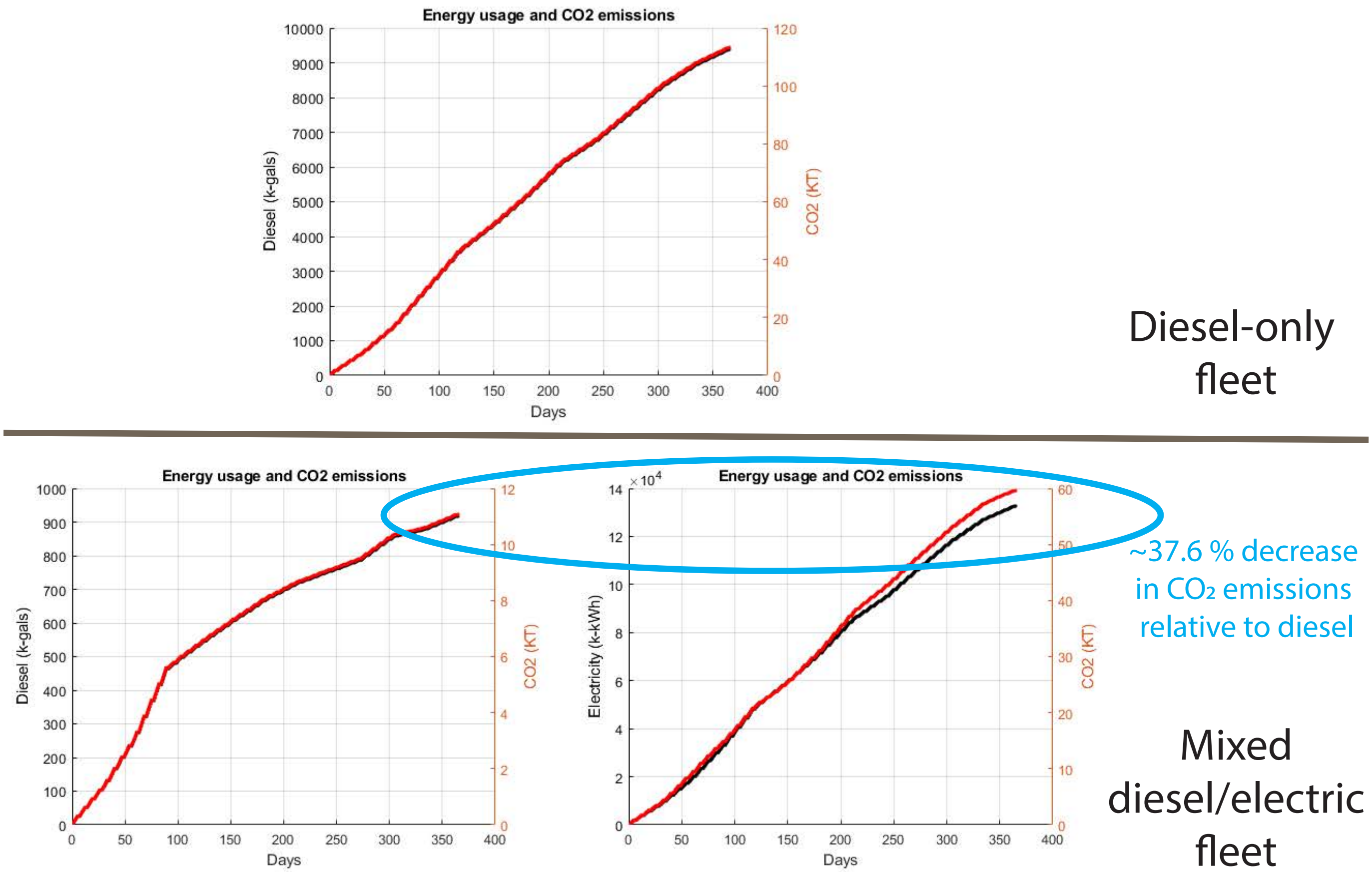

*Figure S3: Energy usage and CO2 emissions for diesel-based and mixed HDCV fleet operating in the Port of Savannah region. (Top plot) Diesel-based HDCV fleet. (Bottom plots) Mixed HDCV fleet. For each plot, the left-axis is the fuel/electric consumption, and the right axis is CO2 emissions.*

**2. Integration of utility-scale solar and battery power with the grid**

We propose a renewable energy network consisting of solar and battery power to supplement grid resources shown in Figure 5 of the main text. In this section, we formally introduce the modeling framework which captures the dynamics of the different elements in the network[2].

*2.1 Mathematical Formulation of the Renewable Energy Network*

As mentioned in the main text, in our notation, for a power quantity denoted $p_i^j(t)$, the subscript $i$ denotes the network element origin of the power and $j$ denotes the target network element. An integer variable $t \in [1, 8760]$ is used to index all 8760 hours of a single year. Here, we lump together both currently existing resources, as well as to-be-built resources, that are not carbon neutral as being part of the "grid".

We denote the excess load demand from the electrified HDCVs by $p^{\ell}(t)$. The hourly available solar capacity is denoted as $p_{\mathrm{s}}(t)$. The available solar capacity can be used to provide power to the load $p_{s}^{\ell}(t)$, provide power to the battery $p_{\mathrm{s}}^{\mathrm{b}}(t)$, or be curtailed $p_{\mathrm{s}}^{\mathrm{curtail}}(t)$ if too much solar capacity is available:

$$p_{\mathrm{s}}(t) = p_{\mathrm{s}}^{\ell}(t) + p_{\mathrm{s}}^{\mathrm{b}}(t) + p_{\mathrm{s}}^{\mathrm{curtail}}(t) \qquad \text{(Equation 1)}$$

Similarly, the battery can be used to provide power $p_{\mathrm{b}}^{\ell}(t)$to meet the excess load demand. It may be recharged (via a charger) either through solar power $p_{s}^{\mathrm{b}}(t)$ or grid power $p_{\mathrm{g}}^{\mathrm{b}}(t)$. Thus, the charger power is

$$p^{\mathrm{b}}(t) = p_{\mathrm{s}}^{\mathrm{b}}(t) + p_{\mathrm{g}}^{\mathrm{b}}(t) \qquad \text{(Equation 2)}$$

The remaining energy stored in the LIB at a given hour $\boldsymbol{t}$ is modeled as,

$$E_{\mathrm{b}}(t) = E_{\mathrm{b}}(t-1) + \mu \cdot p^{b}(t) - p_{\mathrm{b}}^{\ell}(t) \qquad \text{(Equation 3)}$$

The battery energy increases upon charging and decreases upon discharging. The roundtrip efficiency $\boldsymbol{\mu}$ characterizes the battery system's performance and encapsulates the whole set of system losses. To simplify the model, and in accordance with the National Renewable Energy Lab's (NREL) Annual Technology Baseline [3], we use a roundtrip efficiency of $\mu = 0.85$ as being characteristic for Li-ion battery technologies. The roundtrip efficiency generally varies between LIB chemistries/manufacturers; however, this variation is neglected here.

The existing grid resources can still be used as needed to meet the load demand and/or charge the battery. The power from the grid delivered to the load is denoted $p_{\mathrm{g}}^{\ell}(t)$ and the power from the grid used to charge the battery is $p_{\mathrm{g}}^{\mathrm{b}}(t)$. Thus, the total power generated by the grid is

$$p_{\mathrm{g}}(t) = p_{\mathrm{g}}^{\mathrm{b}}(t) + p_{\mathrm{g}}^{\ell}(t) \qquad \text{(Equation 4)}$$

The excess load demand in the renewable energy network is thus satisfied by a mix of solar, battery, and grid power according to

$$p^{\ell}(t) = p_{\mathrm{s}}^{\ell}(t) + p_{\mathrm{b}}^{\ell}(t) + p_{\mathrm{g}}^{\ell}(t) \qquad \text{(Equation 5)}$$

*2.2 Optimal dispatch framework to maximally reducing excess $CO_2$ emissions*

Assuming an estimate of the hourly available solar capacity $p_{\mathrm{s}}(t)$, and given the excess load demand profile $p^{\ell}(t)$, we now introduce the optimal dispatch framework used to determine how the grid and battery elements in the energy network are used. The dispatch is designed to maximize the yearly total excess $CO_2$ emissions reduction while ensuring that the excess load demand at every hour of the year.

To do this, we first consider an energy network where only the existing grid resources can be used to meet the excess load demand. This will be referred to as the *baseline network*. In the baseline network, the following relationship

$$p_{\mathrm{g}}(t) = p_{\mathrm{g}}^{\ell}(t) = p^{\ell}(t) \qquad \text{(Equation 6)}$$

between the total grid power and the excess load demand holds. The baseline network has an associated total excess $CO_2$ emissions of

$$C_{\mathrm{g}}^{\mathrm{base}} = \sum_{t=1}^{8760} C_{\mathrm{g}}(t) \qquad \text{(Equation 7)}$$

where $C_{\mathrm{g}}(t)$ denotes the hourly excess CO2 emissions.

In contrast, the renewable network has an associated total excess $CO_2$ emissions of

$$C_{\mathrm{g}}^{\mathrm{renew}} = \sum_{t=1, p^{\ell}(t)>0}^{8760} \frac{p_{\mathrm{g}}(t)}{p^{\ell}(t)} \cdot C_{\mathrm{g}}(t) \qquad \text{(Equation 8)}$$

where the sum only contains hours where there is nonzero excess load demand.

An optimization problem is formulated to obtain the optimal dispatch of the three energy sources in the network with the objective being to maximize the percentage of *total excess CO2 emissions reduction* of the renewable network,

$$J[\Theta] = 100 \times \left(1 - \frac{C_{\mathrm{g}}^{\mathrm{renew}}}{C_{\mathrm{g}}^{\mathrm{base}}}\right). \qquad \text{(Equation 9)}$$

This is accomplished by varying, within their respective bounds, the set of decision variables

$$\Theta = \left\{p_{\mathrm{g}}^{\ell}(t), p_{\mathrm{s}}^{\ell}(t), p_{\mathrm{b}}^{\ell}(t), p_{\mathrm{s}}^{\mathrm{b}}(t), p_{\mathrm{s}}^{\mathrm{curtail}}(t), p_{g}(t), p^{b}(t), p_{\mathrm{g}}^{\mathrm{b}}(t), E_{b}(t), \delta_{b}(t)\right\},$$

some of which were introduced in the main text, are recapitulated in *Table S1*.

*Table S1: Set of decision variables to minimize the objective and their interpretations*

| Variable [Unit] | Interpretation |
|---|---|
| $p_{\mathrm{g}}^{\ell}$(t) [MW] | Power provided to load by grid |
| $p_{\mathrm{s}}^{\ell}$(t) [MW] | Power provided to load by solar |
| $p_{\mathrm{b}}^{\ell}$(t) [MW] | Power provided to load by battery |
| $p_{\mathrm{s}}^{\mathrm{b}}$(t) [MW] | Power provided by solar to charge the battery |
| $p_{\mathrm{s}}^{\mathrm{curtail}}$(t) [MW] | Curtailed solar power |
| $p_{\mathrm{g}}$(t) [MW] | Total power pulled from the grid |
| $p^{\mathrm{b}}$(t) [MW] | Charging power to the battery |
| $p_{\mathrm{g}}^{\mathrm{b}}$(t) [MW] | Power from grid to charge battery |
| $E_{\mathrm{b}}(t)$ [MWh] | Energy of battery |
| $\delta_{\mathrm{b}}(t)$ [-] | Discharge indicator |

By convention, we set

$$p_{\mathrm{g}}^{\ell}(t), p_{\mathrm{s}}^{\ell}(t), p_{\mathrm{b}}^{\ell}(t), p_{\mathrm{s}}^{\mathrm{b}}(t), p_{\mathrm{s}}^{\mathrm{curtail}}(t), p_g(t), p^b(t), p_{\mathrm{g}}^{\mathrm{b}}(t) \geq 0\,. \qquad \text{(Equation 10)}$$

The battery system is assumed to start the year half full,

$$E(0) = 0.5 \cdot E_{\mathrm{b,rated}}\,, \qquad \text{(Equation 11)}$$

where the nominal energy of the battery system (which we will refer to as the *battery size*) is denoted $E_{\mathrm{b,rated}}$. To facilitate our analysis later, which investigates multiple years of system operation, we impose that

$$E(8760) = E(0) \qquad \text{(Equation 12)}$$

thereby, allowing us to straightforwardly extrapolate the same battery dispatch to subsequent years[1]. Additionally, to maximize lifetime the battery's energy is limited to the energy range

$$0.1\, E_{\mathrm{b,rated}} \leq E_{\mathrm{b}}(t) \leq 0.9\, E_{\mathrm{b,rated}} \qquad \text{(Equation 13)}$$

Moreover, the battery is unable to charge and discharge at the same time. When it does (dis)charge, it can only do so up to a specified maximum power set by the duration of the battery and the battery size

[1] This assumes that the excess load demand $p^{\ell}(t)$ and the available solar capacity profile $p_{\mathrm{s}}(t)$ also do not change in subsequent years.

$$\tau_{\mathrm{b}} = \frac{E_{\mathrm{b,rated}}}{p_{\mathrm{b,max}}} \quad \text{(Equation 14)}$$

In this work, we consider batteries with a 4-hour ($\tau_b = 4$) duration, as this duration currently serves as the de-facto limit for a large majority of deployed utility-scale battery systems [4]. These restrictions on the battery behavior can be summarized in the following relations:

$$p_{\mathrm{b}}^{\ell}(t) \leq \delta_{\mathrm{b}}(t) \cdot p_{\mathrm{b,max}} \quad \text{(Equation 15a)}$$

$$p^{\mathrm{b}}(t) \leq \left(1 - \delta_{\mathrm{b}}(t)\right) \cdot p_{\mathrm{b,max}} \quad \text{(Equation 15b)}$$

A binary decision variable $\delta_{\mathrm{b}}(t) \in \{0,1\}$ is introduced that indicates if the battery is discharging ($\delta_{\mathrm{b}}(t) = 1$) or charging ($\delta_{\mathrm{b}}(t) = 0$).

The grid power is restricted such that no power is pulled from the grid when excess load demand is zero,

$$p_{\mathrm{g}}(t) = 0 \ \ \forall\, t \mid p^{\ell}(t) = 0 \quad \text{(Equation 16)}$$

The optimal dispatch of the renewable network is thus the set of decision variables $\Theta^*$ that satisfy

$$\Theta^* = \underset{\Theta}{\operatorname{argmax}}\, J[\Theta] \quad \text{(Equation 17)}$$

subject to the Equations 1 to 5 and constraints Equations 10 to 17. This optimization is a mixed-integer linear program (MILP), consistent with optimal dispatch frameworks explored previously in literature[5–7]. The MILP is solved in MATLAB using the **intlinprog** function in MATLAB's Optimization Toolbox.

*2.3 Average resource utilization of resources to power excess load demand*

Having obtained the optimal set of powers $p_g^{\ell}(t)$ (grid), $p_s^{\ell}(t)$ (solar), and $p_b^{\ell}(t)$ (battery) that meets the excess load demand $p^{\ell}(t)$ from the optimization procedure outlined in Section 2.2, we are able to compute the corresponding average utilization of the grid, solar, and battery, respectively, by

$$f^{grid} = \frac{1}{N_{p^{\ell}(t)>0}} \sum_{t=1,p^{\ell}(t)>0}^{8760} \frac{p_g^{\ell}(t)}{p^{\ell}(t)} \quad \text{(Equation 18a)}$$

$$f^{solar} = \frac{1}{N_{p^{\ell}(t)>0}} \sum_{t=1,p^{\ell}(t)>0}^{8760} \frac{p_s^{\ell}(t)}{p^{\ell}(t)} \quad \text{(Equation 18b)}$$

$$f_{dchg}^{batt} = \frac{1}{N_{p^{\ell}(t)>0}} \sum_{t=1,p^{\ell}(t)>0}^{8760} \frac{p_b^{\ell}(t)}{p^{\ell}(t)} \quad \text{(Equation 18c)}$$

where $N_{p^{\ell}(t)>0}$ denotes the number of hours in a year where the excess load demand is nonzero. The average utilization of each resource to power the excess load demand sums to unity:

$$f^{grid} + f^{solar} + f_{dchg}^{batt} = 1 \quad \text{(Equation 19)}$$

*2.4 Fraction of charging resources*

Similar to calculations done in Section 1.3, we are also able to compute the average utilization of grid and solar that is used to charge the battery. These can be estimated by the following relations, respectively:

$$f_{chg}^{grid} = \frac{1}{N_{p_b(t)>0}} \sum_{t=1,p_b(t)>0}^{8760} \frac{p_g^b(t)}{p_b(t)} \quad \text{(Equation 20a)}$$

$$f_{chg}^{solar} = \frac{1}{N_{p_b(t)>0}} \sum_{t=1,p_b(t)>0}^{8760} \frac{p_s^b(t)}{p_b(t)} \quad \text{(Equation 20b)}$$

As, before, the average utilization of each resource for charging sums to unity:

$$f_{chg}^{grid} + f_{chg}^{solar} = 1 \quad \text{(Equation 21)}$$

## 3. Siting framework for photovoltaic resources

To reduce the overall reliance on fossil fuels and lower the risk of grid failure, it is necessary to supplement the existing power grid with renewable energy resources (RES). Modeling the potential for siting and generation of these resources is carried out here using two modeling technologies: the OR-SAGE tool from ORNL and the Renewable Energy Potential (ReV) model from the National Renewable Energy Laboratory (NREL)[19].

*3.1 Regional solar viability assessment*

The assessment of solar viability in this section is performed using a computational framework known as the Oak Ridge Siting Analysis for Generation Expansion (OR-SAGE) tool[20]. As discussed in the main text, OR-SAGE makes use of a myriad of high-resolution geospatial data layers that integrates existing restrictions on land use based on various factors. While some layers directly correspond to whether a given region is suitable for siting of a given technology, other layers, such as layers describing variations in slope, must be converted into a decision layer by using a filtering threshold on the data. These thresholds are generally determined from safety requirements or from constraints created by the technology under consideration. The set of data layers that are used in this work to assess the regional siting of solar energy, along with their associated thresholds and buffer zones, are detailed in *Table S2*.

*Table S2: Exclusion criteria, thresholds, and buffer distances for solar energy siting*

| **Criteria** | **Exclusion Threshold** | **Buffer** |
|---|---|---|
| Population Density | >500 people per sq mi | 20 miles |
| Protected Lands | None/No go | 3 km |
| Wetlands and Open Water | None/No go | None |
| 100-year Floodplains | None/No go | None |
| Landslide Hazards | None/No go | None |
| Slope | >5% | None |
| Major Roads | None/No go | None |
| Solar Radiation | <4.8 $kWh/m^2$/day | None |

Within each decision layer, a value of 0 indicates that a grid square is suitable for siting, and a value of 1 indicates a violation of the constraint specific to that layer. These data layers can then be overlayed to create a composite siting map, where the value of a grid square corresponds to the number of constituent data layers that conflict with a siting decision at the given location. The resulting composite map, obtained from cross-referencing the various solar energy siting layers, can be seen in *Figure S4*.

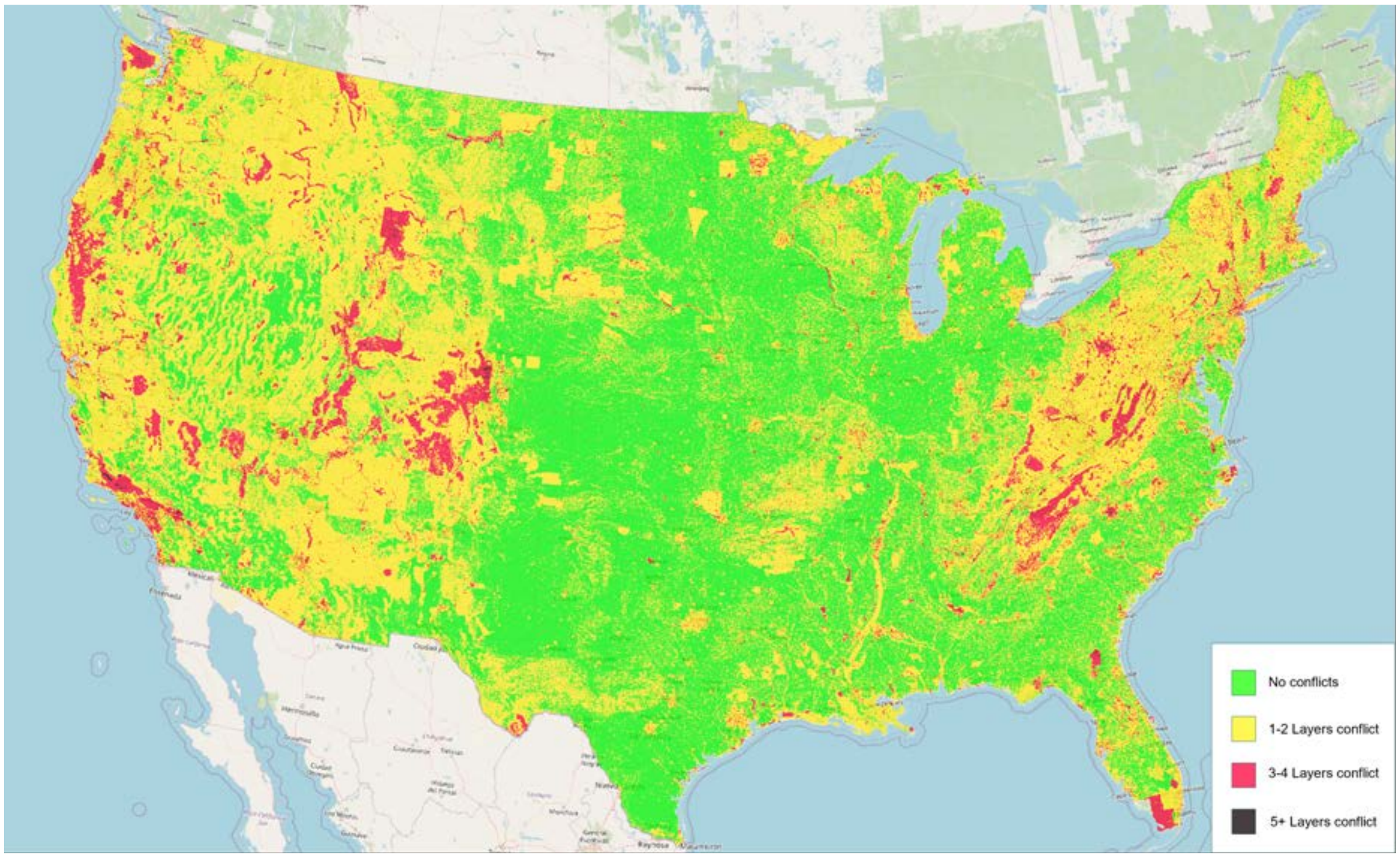

*Figure S4: Composite siting map depicting the number of conflicting decision layers for the conterminous US*.

For this analysis, the region of Chatham County, Georgia is selected for its proximity to the Port of Savannah. The three counties surrounding Chatham are also included in this analysis to encompass all available areas within the county. The section of the solar viability map corresponding to this region can be seen in *Figure S5*.

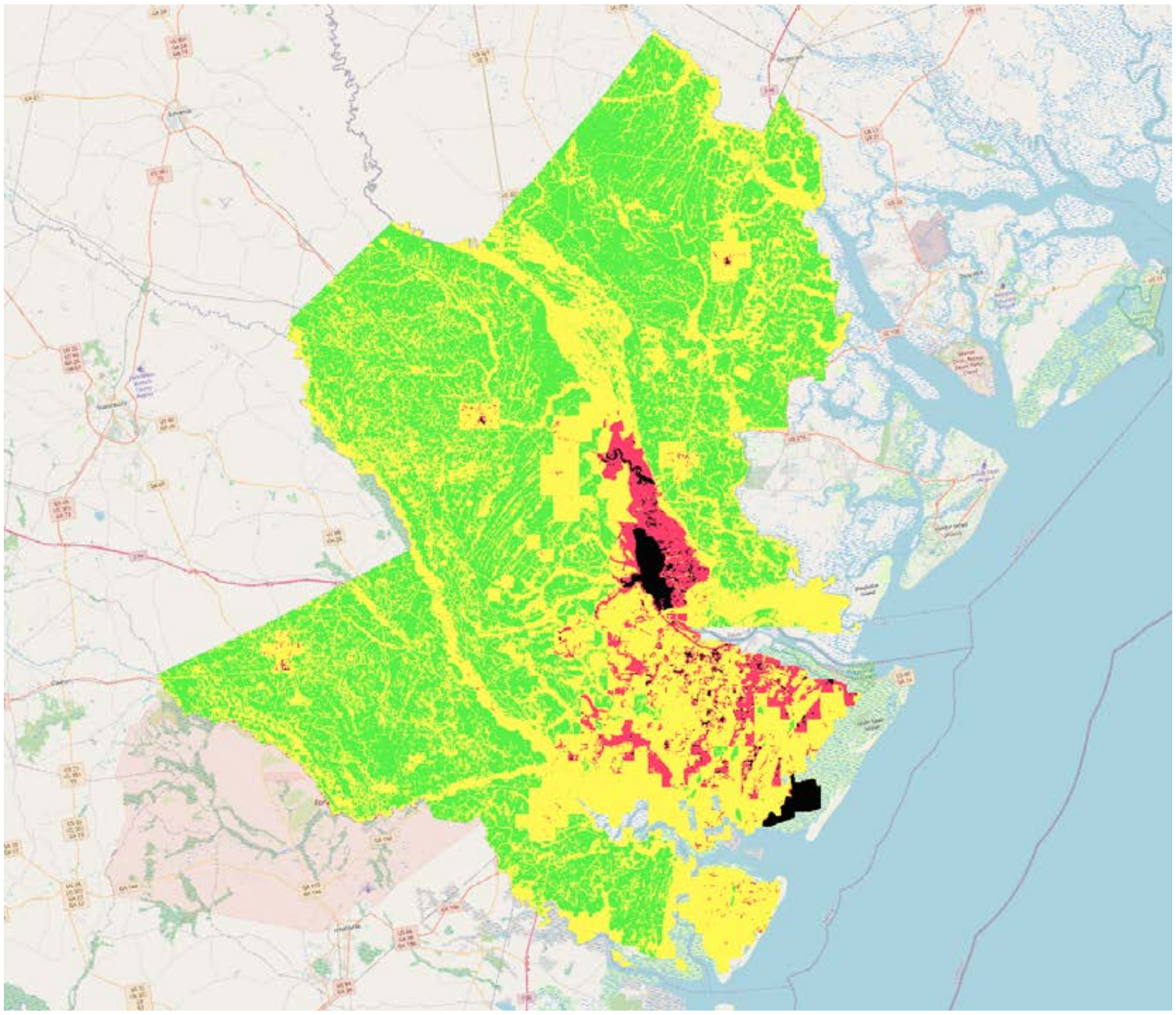

*Figure S5: Composite map of siting conflicts for Chatham County, Georgia, and surrounding counties. Green denotes regions viable to site a microgrid. Yellow, Red and Black indicate regions that have conflicts with land-use restrictions and thus are not viable for siting.*

*3.2 Hourly available solar capacity assessment per possible site*

For the analysis done in this paper, to obtain the available capacity of viable sites in a region, the 2019 National Solar Radiation Database (NSRDB) dataset from version 3 is chosen. Once the region of interest (ROI) is established, it is necessary to define the configurations and specifications of the solar technology under consideration. The system configuration parameters and values selected for this analysis can be seen in *Table S3*. These values reflect a relatively inexpensive configuration, utilizing standard crystalline silicon photovoltaic (PV) cells and a fixed-tilt mounting system.

*Table S3: Generation module input parameters*

| Variable | Value | Meaning |
|---|---|---|
| system_capacity | 1000 | Power rating (kW) for system at each grid point |
| losses | 14.07 | Percent power lost due to external/internal inefficiencies |
| array_type | 0 (Fixed tilt) | Designates whether to model solar tracking technology or rooftop mounted panels |
| azimuth | 180 | Module rotation from North (degrees) |
| dc_ac_ratio | 1.3 | Conversion factor between generated (DC) and distributed (AC) power ratings |
| gcr | 0.4 | Ratio between module surface area and land area occupied by the module |
| inv_eff | 96 | Inverter conversion efficiency between AC and DC power output |
| module_type | 0 (Standard crystalline silicon) | Module type/crystal material to be modeled |
| tilt | latitude | Degrees tilted from horizontal. Latitude designates that tilt should increase with distance from equator. |

Once assembled, these parameters are passed on to ReV's generation module along with the selected NSRDB points. The generation module is then used to calculate capacity factor values---the ratio between the actual generation capacity of a site and its nameplate capacity---for a theoretical solar farm placed at each selected grid-point in the NSRDB dataset. This is done for every hour in the chosen year, which results in a set of temporal capacity factor profiles for each farm. The average capacity factors for Chatham Georgia and the surrounding counties in 2019 can be seen in *Figure S6*.

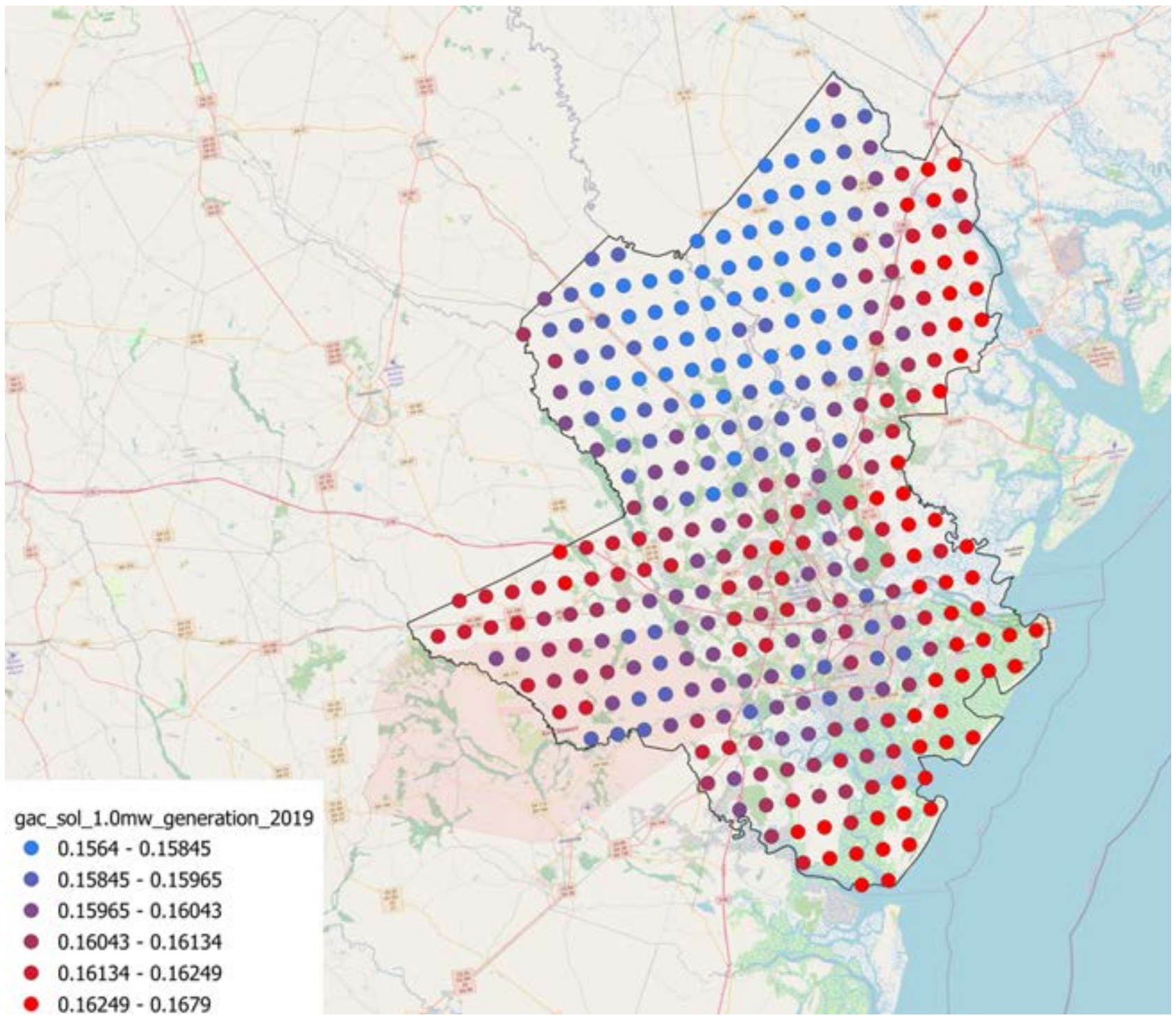

*Figure S6: Capacity factors for Chatham County and the surrounding counties.*

The representative profile module of ReV aggregates the capacity factor profiles created in the generation module to the same resolution as the parcels in the supply curve aggregation module. Here, we use the default 'meanoid' method in ReV to generate the representative profiles for each supply curve point. Post-processing is performed on the ReV output to calculate the hourly available capacity profiles $\boldsymbol{p_s(t)}$ from representative capacity factors and nameplate generation capacities. In *Figure S7*, the available solar capacity for three of the 171 viable sites, representing a site with low, median, and high nameplate solar capacity is shown.

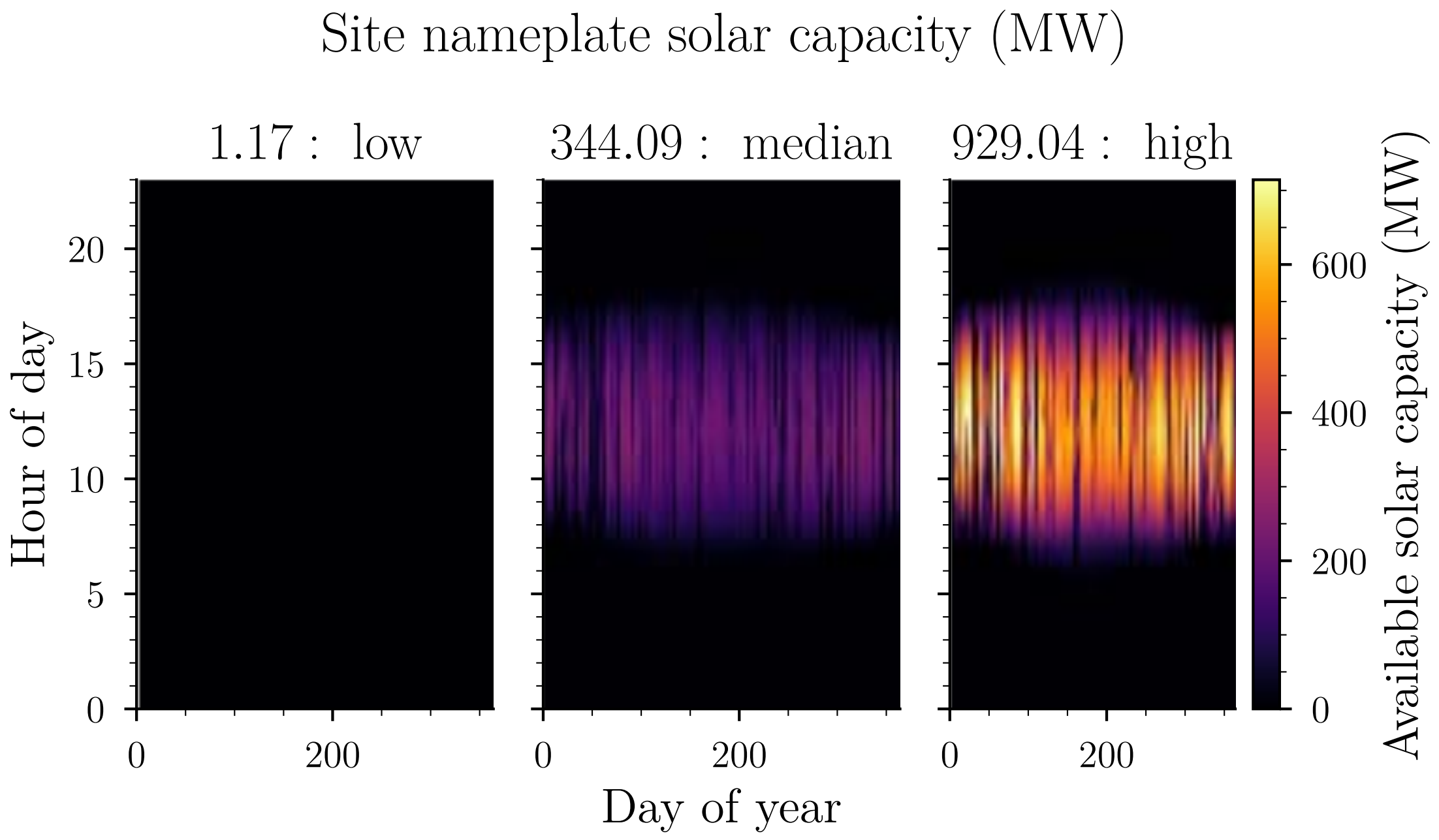

*Figure S7: Estimated hourly available solar capacity generated from the proposed framework for three different representative sites in the Port of Savannah area.*

**4. Estimating costs of deployment**

This section provides further details on the cost of deployment for a microgrid integrating solar and battery systems to different stakeholders in freight transportation. The capital cost estimates of the solar (Section 4.1) and battery systems (Section 4.2) are each estimated separately with additional costs being incorporated to capture these systems' integration and deployment. The trends in the total capital costs are discussed in Section 4.3. The total cost of ownership framework is then mathematically formulated in Section 4.4. Given this TCO framework, two separate cost metrics for deployment are assessed for the utility and the fleet operator, in Sections 4.5 and 4.6, respectively.

*4.1 Capital cost expenses for photovoltaic farms*

As described in the main text, we have developed a comprehensive capital cost calculator based on the U.S. solar PV system benchmarks reported by NREL in their PVSCM model[8]. The capital cost of the solar PV system is divided into two main categories: equipment cost and development cost.

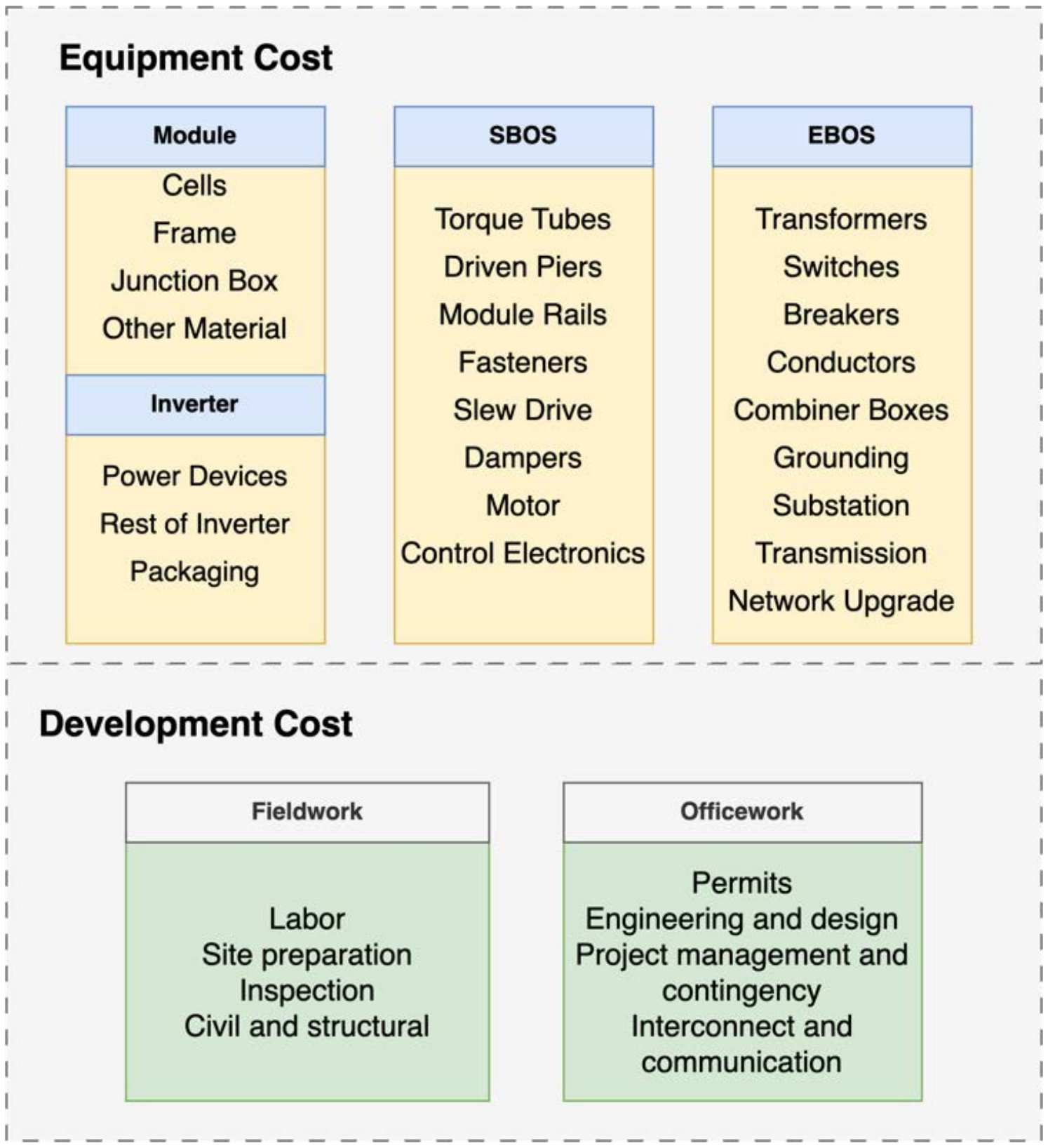

*Figure S8: Primary components of solar system cost*

The equipment cost comprises various components as shown in *Figure S8*:

- The module includes photovoltaic cells, which convert sunlight into electricity, and the frame, which provides structural support. Additional elements are the junction box, electrical connections, and assembly materials.
- Inverters include power devices that convert DC to AC power, the rest of the inverter system, and packaging materials for shipping and protection.
- The structural and balance of system (SBOS) includes torque tubes, driven piers, module rails, fasteners, slew drives for rotation, dampers to reduce vibrations, motors for tracking, and control electronics for PV operations.
- The electrical balance of system (EBOS) consists of transformers, switches, breakers, conductors, combiner boxes, grounding systems, substations, transmission lines, and network upgrades to enhance existing electrical infrastructure.

The development cost includes fieldwork and office work components. Fieldwork encompasses labor for installation, site preparation, inspection, and construction. Office work involves permitting, engineering and design, project management, contingency planning, and interconnect and communication.

Putting these costs together generates an estimate of the total overnight capital cost of the solar farm shown in *Figure S9*. Intuitively, the cost is monotonic with nameplate solar capacity; nevertheless, even the smallest nameplate solar capacity site corresponding to 1 MW incurs over 1 $MM to build and install while the largest sites that we consider will require about 1 $bn of investment.

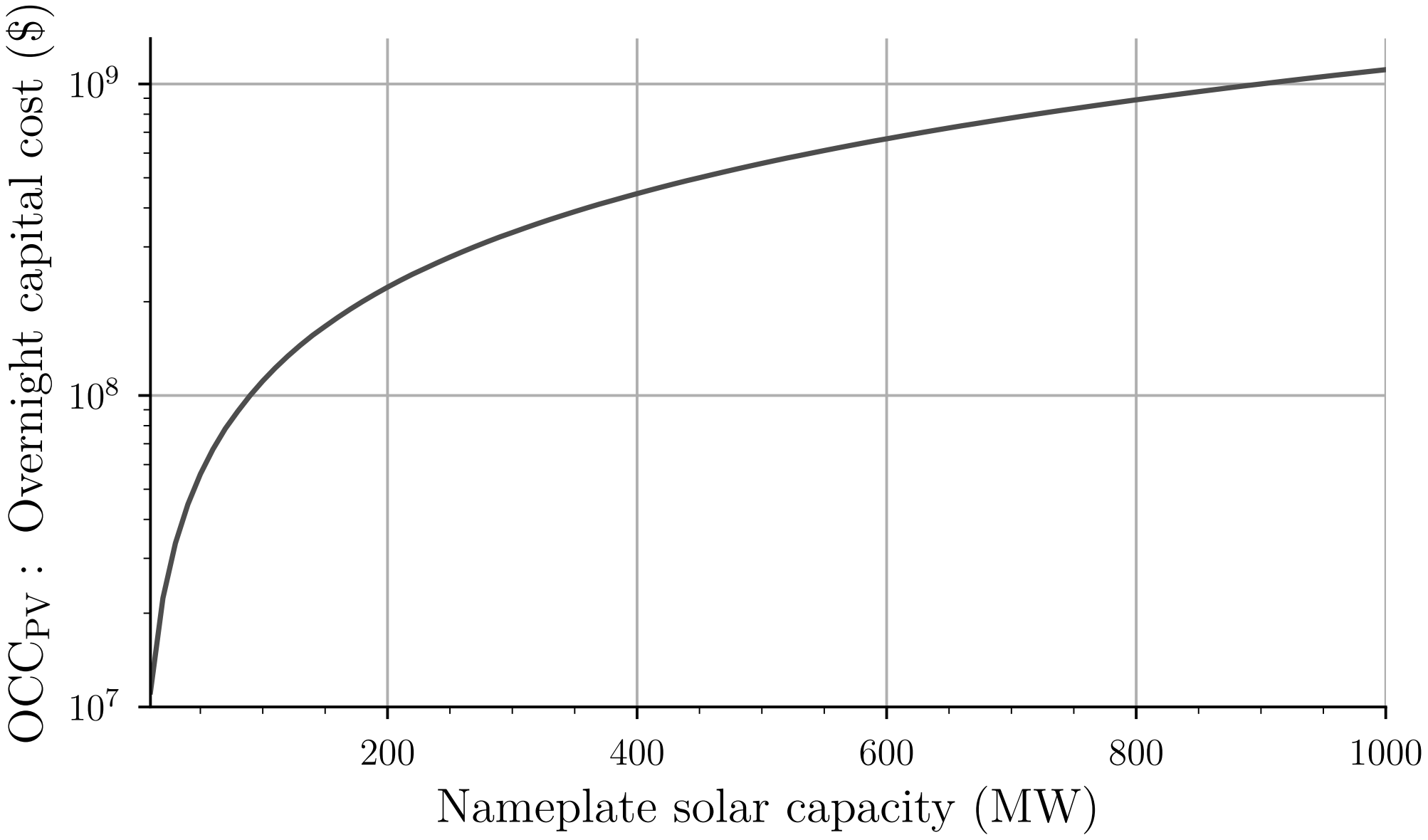

*Figure S9: Capital cost associated with deploying a solar farm of a given nameplate solar capacity.*

*4.2 Capital cost estimates for energy storage technologies*

To estimate the capital cost variation of different LIB systems, we leverage the capabilities of the Battery Performance and Cost (BatPaC Version 5.1) tool, developed by Argonne National Lab[9]. We estimate, in Section 4.2.1, the capital costs for building a battery pack using the leading three chemistries[2] currently used in the industry: NMC (811), NCA, and LFP[10]. Subsequently, the capital costs associated with installation and deployment are estimated in Sec. 4.2.2 leveraging a report from the Pacific Northwest National Laboratory (PNNL).

[2] For the lithium-ion batteries we consider, we assume that the negative electrode material is always lithiated graphite. As such, here "chemistry" in this work refers to a choice of positive electrode material. Other choices could be made for the negative electrode which could constitute a different "chemistry".

4.2.1 BatPaC capital cost estimation for battery pack

To obtain a capital cost estimate for the battery pack using a given chemistry, BatPaC requires the user to specify the details of the battery pack. We create battery pack in BatPaC with a rated duration of 4 hours, obtained by appropriately specifying the target power and the rated energy of the pack. As utility-scale storage is a long-duration application, we consider only energy-type cells in our analysis. The maximum positive electrode thickness is set to 80 micrometers while the maximum cell thickness is set to 25 millimeters. For the pack configuration, the number of modules in parallel[3] is set to 2, the number of modules in a row is set to 10, and the number of rows of modules is set to 4. Fixing these settings, the number of cells in parallel within a module is determined from

$$N_p = \frac{1}{2}\left\lceil \frac{E_{b,rated}}{V_{pack,target} \cdot Q_{cell,target}} \right\rceil \qquad \text{(Equation 22)}$$

where $V_{pack,target} = 800\ V$ and $Q_{cell,target} = 15\ Ah$ are the target pack voltage and cell capacity, respectively. The factor of ½ comes from the 2 modules that are connected in parallel. Correspondingly, the total number of cells in a module is computed by,

$$N_{total,module} = \frac{\left(\frac{V_{pack,target}}{V_{cell,nominal}}\right)}{\left(\frac{10 \cdot 4}{2}\right)} \cdot N_p \qquad \text{(Equation 23)}$$

We denote the nominal cell voltage (found after optimization in BatPaC) by $V_{cell,nominal}$. The denominator accounts for the number of modules connected in series. A summary of the characteristics for a 100 MWh battery pack for different chemistries is given in *Table S4*. We find that packs using NMC 811 or NCA battery cells are configured to the same configuration, owing to the similarity in their cell-level characteristics, particularly their nominal voltage. As the nominal voltage of each LFP cell is lower than that of an NMC 811 or NCA cell, the LFP battery pack requires more cells to meet the required target voltage.

*Table S4: Summary of a 4-hour duration, 100 MWh battery pack characteristics used in this work.*

| Battery Chemistry | Rated power (MW) | Battery size (MWh) | Battery system voltage (V) | $N_p$ | $N_{total,module}$ | Cell capacity (Ah) | Cell nominal voltage (V) |
|---|---|---|---|---|---|---|---|

[3] This is a limitation of BatPaC that it becomes more inaccurate in estimating the quantity of connecting equipment necessary if the number of modules connected in parallel is set to more than 2.

| | | | | | | | |
|---|---|---|---|---|---|---|---|
| NMC 811 | 25 | 100 | 811 | 4167 | 45837 | 14.80 | 3.68 |
| NCA | | | 808 | 4167 | 45837 | 14.86 | 3.67 |
| LFP | | | 794 | 4167 | 50004 | 15.11 | 3.31 |

BatPaC allows different assumptions related to the pack manufacturing. In our analysis, we make the following assumptions:

1. There is a single gigafactory dedicated to producing the grid-scale storage pack.
2. The energy throughput of this gigafactory is fixed to 35 GWh/year.
3. The cell yield is 95%.
4. Any additional costs in fitting the factory for a specific size of battery pack are neglected. (i.e., it is assumed that the factory can be retooled from producing 1 MWh battery packs to 50 MWh packs without incurring any additional costs)

These assumptions ultimately set a lower limit on the costs of the battery pack. Except for the thermal management system costs, which we remove from the BatPaC analysis (this cost is included in the storage balance of systems instead, see Section 4.2.2), default values included in BatPaC are generally used for this analysis. The cost of a 100 MWh battery pack for a given chemistry generated from this process is summarized in *Table S5*.

*Table S5: Battery pack cost estimate for a 100 MWh system for different choice of chemistries*

| Chemistry | Battery system cost ($) | Battery system cost ($/kWh) |
|---|---|---|
| NMC 811 | 15,097,974 | 150.98 |
| NCA | 19,403,415 | 194.03 |
| LFP | 21,616,938 | 216.17 |

4.2.2 ESS system installation and integration costs

Having estimated the cost of the battery pack, we now incorporate the costs associated with installation and integration of the battery pack to the grid to estimate the overall battery system cost. We leverage a report[11] from PNNL released in 2022, to obtain estimates for the associated costs of installation and integration of the battery pack. The additional cost items, their values, and descriptions are summarized in *Table S6*.

*Table S6: Extra capital costs for battery system. Items are evaluated in order from top to bottom.*

| Item | Cost | Notes |
|---|---|---|
| Storage Balance of System | + 23% of battery pack cost | This includes supporting cost components for the storage block with container, cabling, switchgear, flow battery pumps, and heating, ventilation, and air conditioning. |
| Power conversion system | 45 $/kW (rated) | This component includes bidirectional invertor, DC-DC converter, isolation protection, alternating current breakers, relays, communication interface, and software. This is the power conversion system for batteries. |
| Controls & Communication | 7.8 $/kW baseline cost for 10MW battery system. 50% reduction in cost from baseline going from 10 MW to 1 MW. 33% increase in cost from baseline going from 10 MW to 100 MW. Linear interpolation is used to obtain a cost for battery systems with power ratings between 10 and 100 MW. | This includes the energy management system for the entire battery system and is responsible for system operation. This may also include annual licensing costs for software. The cost is typically represented as a fixed cost scalable with respect to power and independent of duration. |
| System Integration | + 5% of all items above | Price charged by the system integrator to integrate sub-components of a battery pack into a single functional system. This includes procurement and shipment to the site of battery modules, racks with cables in place, containers, and power equipment. At the site, the modules and racks are containerized with HVAC and fire suppression installed and integrated with the power equipment to provide a turnkey system. |
| Engineering, Procurement & Construction | + 20% of all items above | This includes non-recurring engineering costs and construction equipment as well as shipping, siting and installation, and commissioning of the battery pack. |
| Project development | + 20% of all items above | This includes costs associated with permitting, power purchase agreements, interconnection agreements, site control, and financing. |
| Grid integration | + 1.5% of all items above | This includes the direct cost associated with connecting the ESS to the grid, including transformer cost, metering, and isolation breakers. For the last component, it could be a single disconnect breaker or a breaker bay for larger systems. |

The total battery system cost is shown in *Figure S10*. Intuitively, as the battery system size (i.e. the system nominal energy) increases, the greater the cost becomes. Interestingly, the cost of NCA and LFP battery systems is greater than that of the solar system, while NMC is relatively similar. Nevertheless, in raw capital cost, NMC is estimated to provide the lowest capital cost of all the chemistries considered.

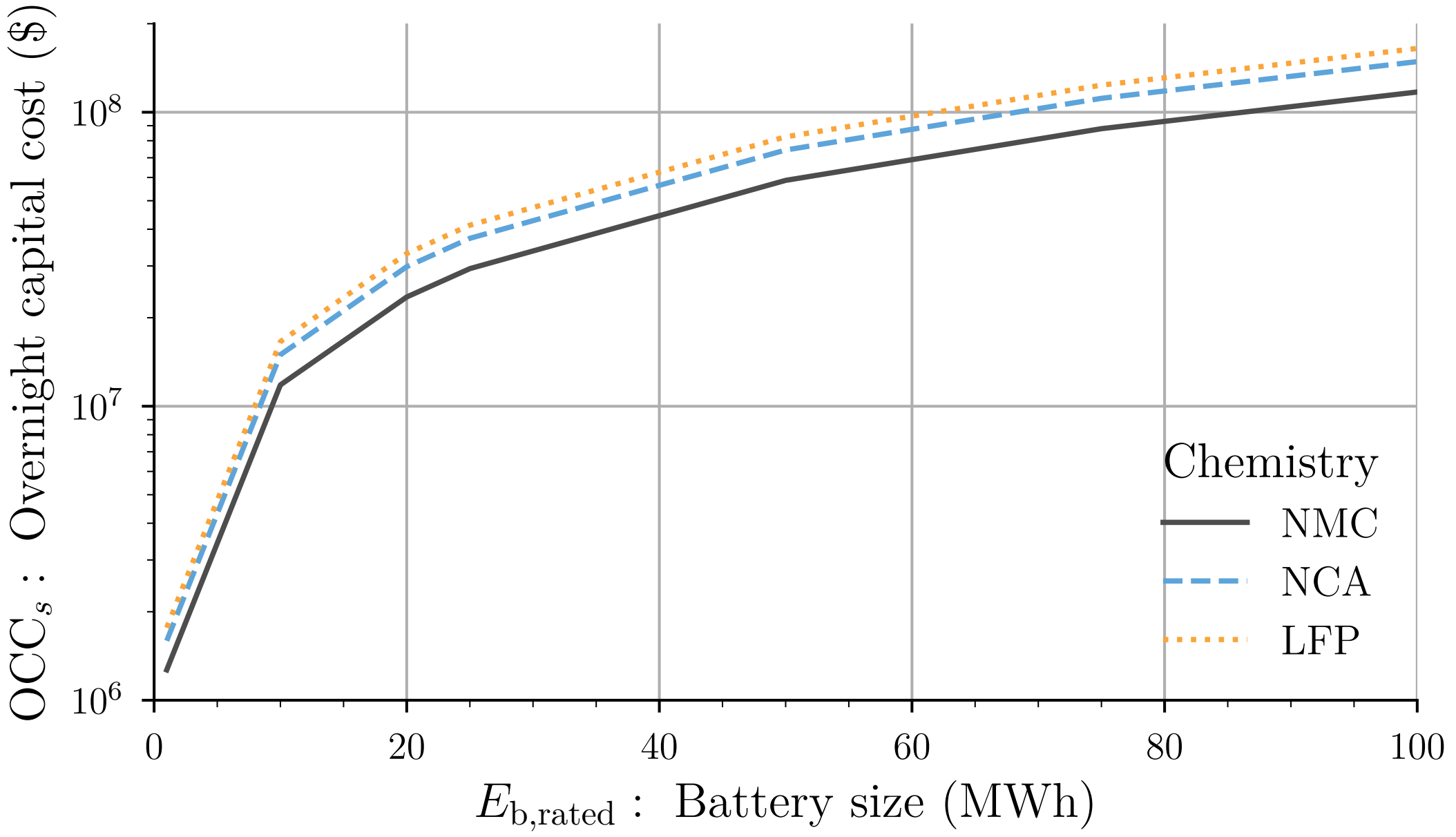

*Figure S10: Cost of deployment for battery systems of different sizes for different chemistries*

*4.3 Total overnight capital cost*

The total overnight capital cost of deployment for all different combinations of nameplate solar capacities and battery sizes, respectively, is shown in *Figure S11*. Recapitulating their separate trends, we find that the most expensive systems are those with the largest nameplate solar capacity site and the largest battery size.

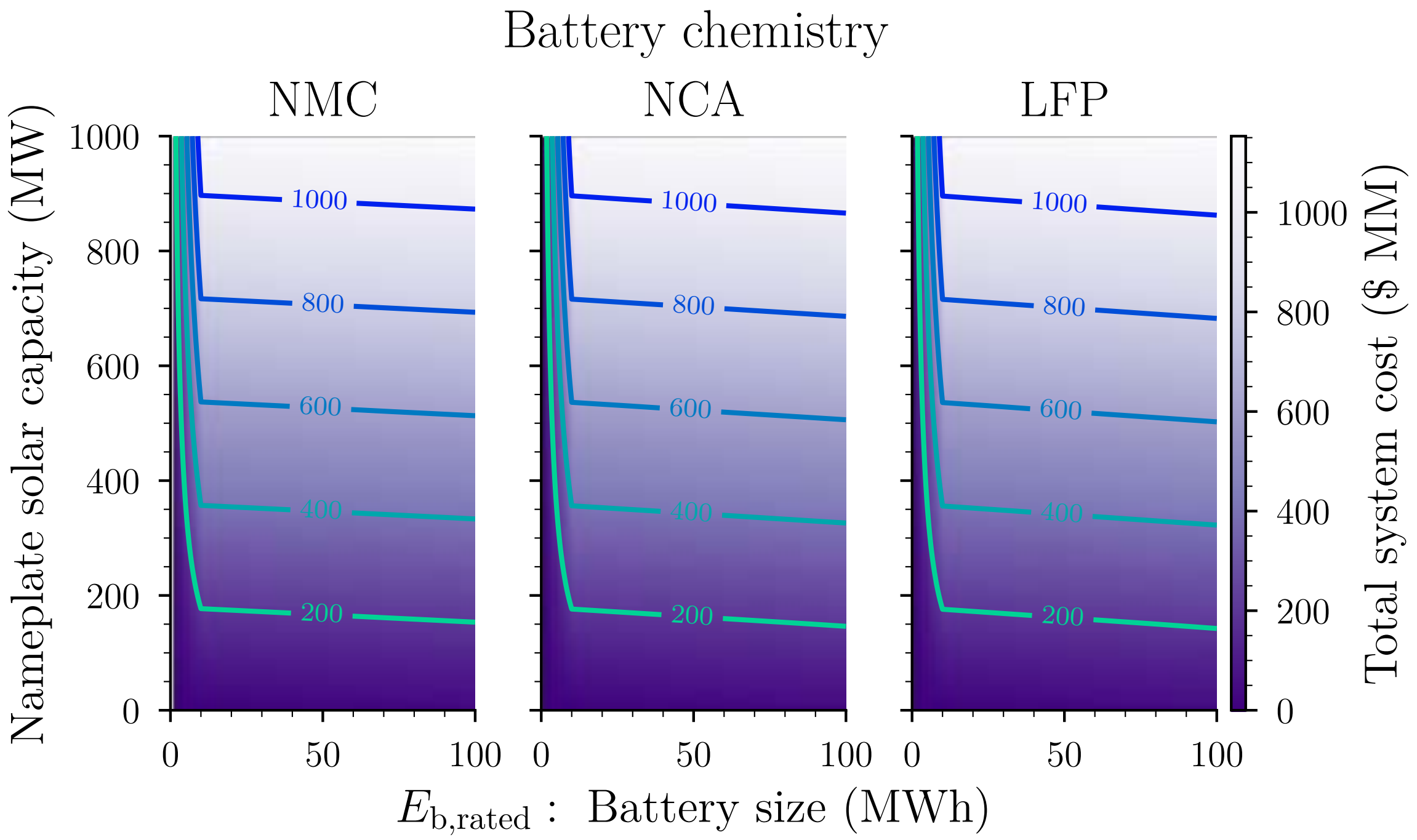

*Figure S11: Total overnight capital cost (in millions USD) of deploying solar with a given nameplate solar capacity and an attached battery system of a given size.*

*4.4 Total cost of ownership framework*

In this section, we mathematically formulate the total cost of ownership framework, described in the main text. Here, we take the total cost of ownership (TCO) to be,

$$TCO = IC + AOC * YrsOp \,, \qquad \text{(Equation 24)}$$

where $IC$ ($) is the initial costs, $AOC$ ($/year) is the annual operating cost, and $YrsOp$ is the number of years of operation considered. The costs comprising $IC$ vary depending on the stakeholder of interest. Aspects of the $AOC$ also depend on the stakeholder; however, a common quantity between stakeholders in the context of freight electrification is the total electricity cost, which we denote as TEC ($/MWh). To compute the TEC, two metrics must be defined: (1) the levelized cost of photovoltaic recharge, and (2) the levelized cost of storage. We now describe each of these quantities and their computation in turn.

The levelized cost of photovoltaic recharge, denoted as LCOPR ($/MWh), quantifies the cost of discharging a given solar resource [12] and is defined as

$$LCOPR = \frac{OCC_{PV}}{N_y^{op} \cdot E_y^{out}} \qquad \text{(Equation 25)}$$

where $OCC_{PV}$ ($) is the overnight capital cost of the solar installation, $N_y^{op}$ (year) is the number of years that the solar plant expects to operate, and $E_y^{out}$ (MWh/year) is the total energy output of a given solar installation/site over a single year

$$E_y^{out} = \sum_{t=1}^{8760} p_s(t) \quad \text{(Equation 26)}$$

The denominator of Eq. (24) is thus the total energy output of the solar farm throughout its life. Here, we assume that no degradation in the performance of the solar plant will occur during its lifetime. This quantity will be estimated for different scenarios in Sections 5 and 7.

The levelized cost of storage, denoted as LCOS ($/MWh), quantifies the cost of discharging the battery system and is defined by

$$LCOS = \frac{OCC_b}{N_{cycles}^{EOL} \cdot E_{\mathrm{b,rated}}} \quad \text{(Equation 27)}$$

where $OCC_b$ ($) is the overnight capital cost of the solar installation, $N_{cycles}^{EOL}$ (-) is the number of cycles that the battery can undergo before the end-of-life. The denominator of Equation 27 thus measures the total amount of discharged energy by the battery system throughout its life. The values used for the number of cycles until of end of life for each Li-ion battery chemistry are informed by [13] and listed in *Table 7*.

*Table 7: Assumptions on cycle life for different Li-ion battery chemistries*

| Battery chemistry | $\boldsymbol{N_{cycles}^{EOL}}$ |
|---|---|
| NMC | 2000 |
| NCA | 1400 |
| LFP | 6000 |

The LCOS for different battery sizes and for different chemistry choices are shown in *Figure S12*. The average value for each curve is reported in Table 2 of the main text. Due to its high expected lifetime, the LCOS of LFP is lower than the other chemistries despite its high capital cost. In contrast, NCA has the highest LCOS of all 3 chemistries due to its high capital cost and low expected lifetime.

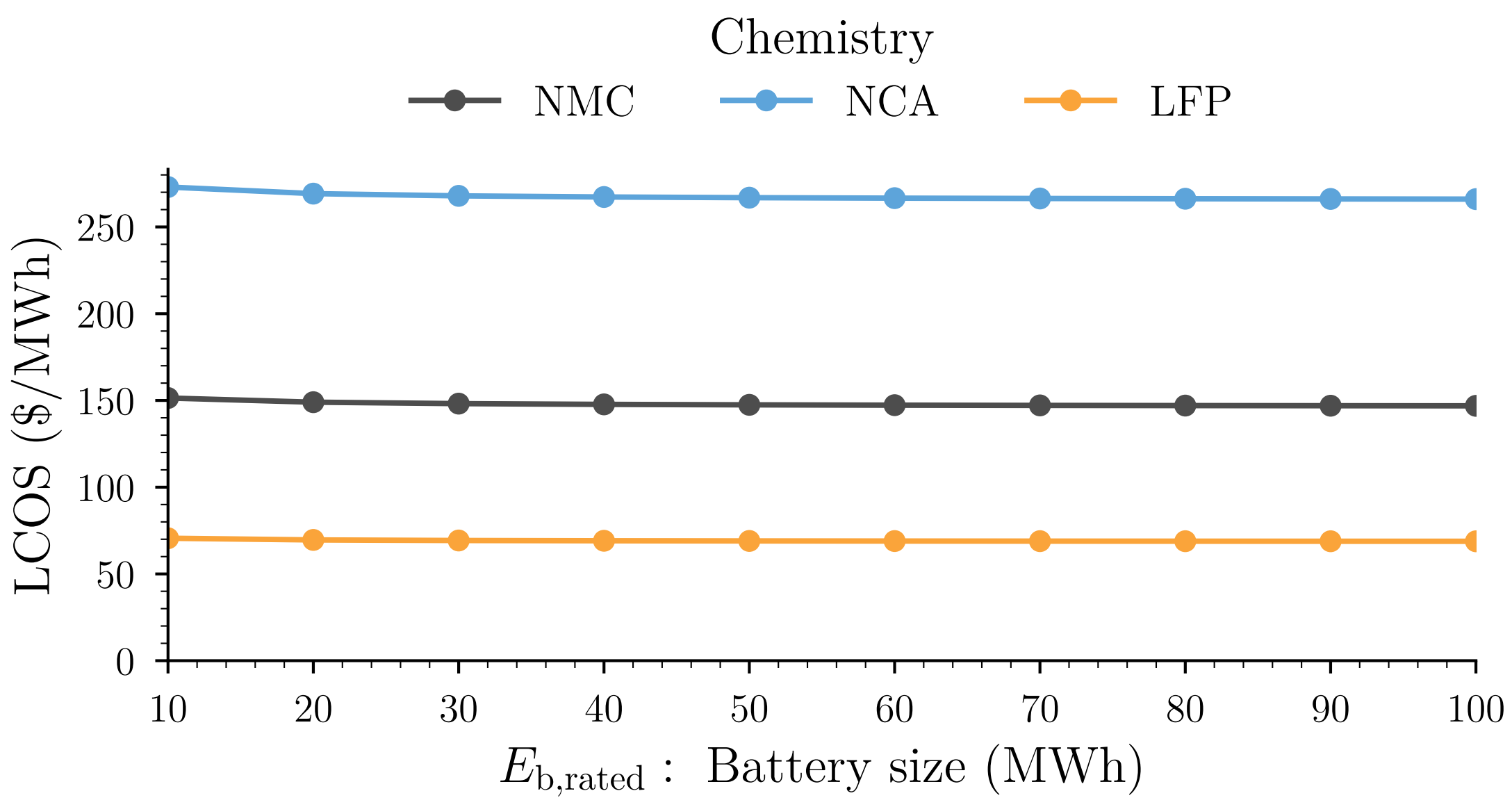

*Figure S12: Levelized cost of storage for different size battery systems. Different colors correspond to different battery chemistries.*

Using these metrics, and with the utilization of each resource used meet the excess load demand that was computed in Section 3.3, the TEC is formulated as

$$TEC = f^{solar} \cdot LCOPR + f^{batt}_{dchg} \cdot (COC + LCOS) + f^{grid} \cdot GEP \qquad \text{(Equation 28)}$$

Here, $GEP$ ($/MWh) represents the electricity price associated with the existing grid, which we take to be 160 $/MWh for the Port of Savannah region, aligned with prices reported by the Bureau of Labor Statistics for the Atlanta, GA region[14], and $COC$ ($/MWh) is the cost of charging that is computed by

$$COC = f^{solar}_{chg} \cdot LCOPR + f^{grid}_{chg} \cdot GEP \qquad \text{(Equation 29)}$$

where the average utilization of solar and grid for charging, $f^{solar}_{chg}$ and $f^{grid}_{chg}$, are defined in Section 3.4. Having defined the TEC, we now turn to defining the costs for two different stakeholders: the utility and the fleet operator

4.4.1 The cost to the utility

In this section, we introduce a cost metric for the utility: the TCO per kilogram of excess CO2 emissions removed. This metric is designed to incentivize utilities to supply the necessary power for the electrified fleet while minimizing excess CO2 emissions. Without such an incentive, utilities may resort to deploying non-carbon-neutral resources to meet the increased demand. Thus, this metric quantifies the balance for utilities between the financial cost of deployment and their effectiveness in utilizing renewable energy technologies. By doing so, it ensures that the environmental benefits of electrifying heavy-duty vehicles are fully realized.

To compute this metric, we first estimate the TCO of solar and battery system deployment, as relevant to the utility. For the utility, the initial costs ($IC_{utility}$, $) will the overnight capital costs of deployment for both solar, $OCC_{PV}$, and the battery systems, $OCC_b$:

$$IC_{utility} = OCC_{PV} + OCC_b \qquad \text{(Equation 30)}$$

Given the total excess load demand per year

$$E_{total}^{\ell} = \sum_{t=1}^{8760} p^{\ell}(t) \qquad \text{(Equation 31)}$$

the annual operating cost of the utility is the product of the of this total excess load demand and the total electricity cost

$$AOC_{utility} = TEC \cdot E_{total}^{\ell} \qquad \text{(Equation 32)}$$

The system's total number of years of operation, as seen from the eyes of the utility, are taken to be the minimum number of years before either the solar or the battery system reaches its end of life. For solar, the number of years of operation is set to be 30 years [15]. For the battery's lifetime, we compute the number of years required before the end-of-life number of cycles is reached. To do this, from the battery dispatch signal we estimate how many cycles $N_y^{cycles}$ the battery goes through each year of operation by the cycle counting algorithm detailed in [16]. Then, the number of years of operation for the battery will depend on chemistry according to

$$N_y^{operation} = \frac{N_{cycles}^{EOL}}{N_y^{cycles}} \qquad \text{(Equation 33)}$$

Thus, the overall lifetime of the system is

$$YrsOp = \min\left(30, N_y^{operation}\right) \qquad \text{(Equation 34)}$$

We note that the estimate of Eq. (33) will generally overestimate the lifetime of the battery system as it neglects any calendar aging effects . As such, this analysis gives a lower bound on the costs to the utility. Future work will seek to improve the life models of the batteries and improve the estimates for the TCO to the utility. The utility TCO for a solar + battery system is therefore

$$TCO_{utility} = OCC_{PV} + OCC_b + \left(TEC \cdot E_{total}^{\ell}\right) \cdot YrsOp \qquad \text{(Equation 35)}$$

The weight of excess CO2 emissions removed, measured in kilograms, is computed by the difference of the total excess carbon emissions of the renewable and the baseline network

$$W_{CO_2} = C_{\mathrm{g}}^{\mathrm{renew}} - C_{\mathrm{g}}^{\mathrm{base}} \qquad \text{(Equation 36)}$$

obtained from the optimization detailed in Section 2.2.

4.4.2 The cost per mile to the fleet operator

To quantify the cost of using the renewable energy network to the fleet operator, we use the TCO of an HDCV per mile that it is driven, computed for the full fleet. For each vehicle in the fleet, the initial cost is computed by the sum of the Manufacturer's Suggested Retail Price (MSRP, $/vehicle) and the registration cost (RC, $/vehicle) less the subsidies (SUB, $/vehicle) of the government. Assuming a uniform initial cost for all $N_{veh} = 1825$ vehicles in the fleet, the total initial cost of the fleet is

$$IC_{fleet} = N_{veh} \cdot (MSRP + RC - SUB) \qquad \text{(Equation 37)}$$

Like in the case of the utility in Section 4.4.1, the total cost of operating can be computed as the product of the annual operating costs multiplied by the number of years of operation. The annual operating costs of the fleet depends on the insurance costs per truck in the fleet (INS, $/vehicle), paid every year, the maintenance cost of the fleet (MAINT, $/mile), and the amount of electricity that the fleet uses

$$AOC_{fleet} = INS \cdot N_{veh} + MAINT \cdot VMT_{tot} + TEC \cdot E_{total}^{\ell} \qquad \text{(Equation 38)}$$

The amount of vehicle miles travelled by all the vehicles in the fleet per year is denoted by $VMT_{tot}$. The number of years of operation for the electrified HDCVs is taken to be 5 years, $YrsOp = 5$. Note the correspondence between the last term in the AOC of the fleet (Equation 38) and the AOC of the utility (Equation 32): The energy provided by the utility is used by the fleet. The residual value (RV, $) of the fleet is modeled here as a percentage of the vehicle's MSRP

$$RV = \eta \cdot MSRP \qquad \text{(Equation 39)}$$

The TCO of the full fleet is thus computed

$$TCO_{fleet} = N_{veh} \cdot \left((1-\eta) \cdot MSRP + RC - SUB + INS \cdot YrsOp + MAINT \cdot VMT_{tot} \cdot YrsOp\right) + \left(TEC \cdot E_{total}^{\ell}\right) \cdot YrsOp \qquad \text{(Equation 40)}$$

The assumed TCO model variable values for the electrified fleet are summarized in *Table S8*.

*Table S8: Model variable values for total cost of ownership calculation for electrified fleet.*

| Variable | Value |
| --- | --- |
| MSRP ($/vehicle) | 334, 313 |
| RC ($/vehicle) | 1,500 |
| SUB ($/vehicle) | 40,000 |
| INS ($/vehicle/year) | 11,700 |
| MAINT ($/mi) | 0.055 |
| $\boldsymbol{\eta}$ | 0.25 |

The cost to the fleet for deploying renewable energy sources is the ratio of the fleet TCO with the total vehicle miles travelled by the fleet over the years of operation,

$$\frac{TCO}{mile} = \frac{TCO}{VMT_{tot} \cdot YrsOp} + Penalty \qquad \text{(Equation 41)}$$

where we have added a $Penalty = 0.5$ $/mile accounting for the price of dwell time and payload capacity loss of electrified HDCVs[18].

For comparison, we will compare the TCO/mile of the electrified fleet with the TCO/mile of a diesel-based HDCV fleet using the same framework above but with diesel rather than electricity as the fuel source. Assuming a fuel cost of 0.68 $/mile, the other values listed in *Table S9*, and performing the same cost analysis as above, we find that the cost of operating a diesel-based fleet is 1.63 $/mile.

*Table S9: Model variables for total cost of ownership for diesel fleet.*

| Variable | Value |
| --- | --- |
| MSRP ($/vehicle) | 133,841 |
| RC ($/vehicle) | 1,500 |
| SUB ($/vehicle) | 0 |
| INS ($/vehicle/year) | 8,000 |
| MAINT ($/mi) | 0.086 |
| $\eta$ | 0.35 |

## 5. Excess $CO_2$ emissions reduction potential of viable sites in the Port of Savannah region

In this section, we examine the capabilities of the 171 viable sites identified by our proposed framework in Section 6 in reducing excess CO2 emissions if a solar and battery system are deployed there. We then examine the costs associated with deployment and investigate the tradeoffs that occur for different stakeholders.

### *5.1 Average utilization of resources*

The average utilization of each resource is shown in *Figure S13* for the cases where: 1) only solar resources are built at each viable site and 2) where solar resources and a 100 MWh battery pack are deployed at each viable site. Aligning with intuition, the average utilization of the grid decreases as sites with higher nameplate solar capacity are considered. As we saw in Section 5, the average utilization of solar generally remains the same even with the addition of a 100 MWh battery pack. Instead, the average utilization of the grid decreases, as the battery utilization increases. Interestingly, for the largest nameplate capacity site, we find that solar alone can only reduce the grid utilization to ~50%. While this still amounts to a significant excess CO2 emissions reduction, this illustrates the limitations of solar power in terms of efficiency and intermittency.

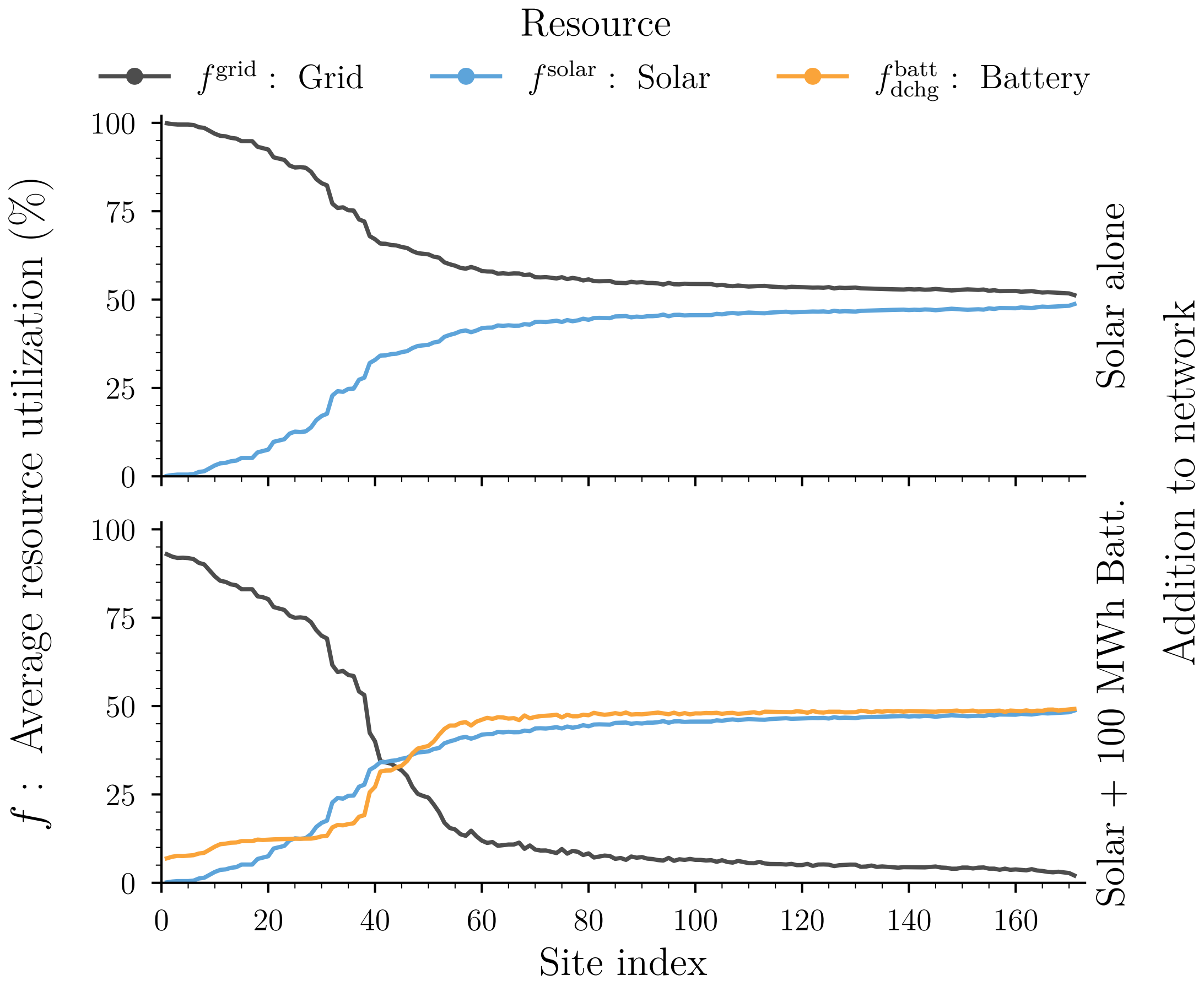

*Figure S13: Breakdown of resource usage for each site in the region to power the excess load demand. (Top plot) Only solar system deployed at site. (Bottom plot) Solar system and 100 MWh battery deployed at site.*

Additionally, in the case where a 100 MWh battery system is deployed, it is informative to examine the origin of energy used to charge the battery. Across all viable sites, greater than 80% of the time solar power is used to charge the battery; however, we observe that grid power is used to charge the battery for sites with lower nameplate capacities. This usage of the grid will incur a small amount of excess CO2 emissions, contributing to the limited the maximum excess CO2 emissions reduction for these lower nameplate capacity sites.

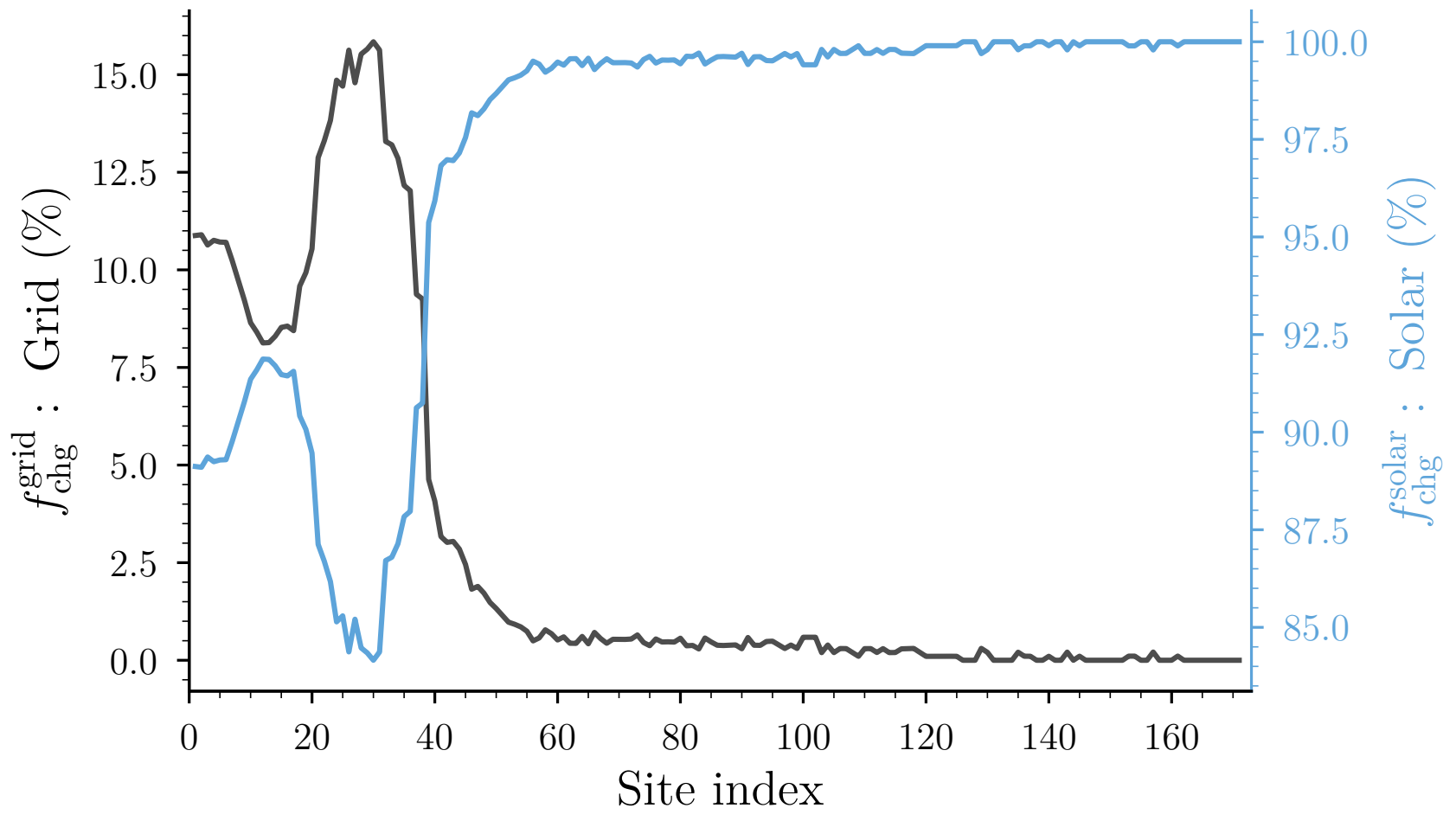

*Figure S14: Average resource utilization to charge a 100 MWh battery for each viable site in the region. (Left axis, black curve) Grid average utilization. (Right axis, blue curve) Solar average utilization.*

*5.2 Capital costs of deployment*

Using the capital costs estimates for both the solar and battery systems generated in Sections 7.1 and 7.2, we now estimate, for each viable site in the Port of Savannah, the cost of building a solar installation alone at the site, and a solar installation with a 100 MWh battery system collocated at that site. *Figure S15* illustrates that deploying a solar system at sites with higher nameplate capacity will generally cost more, aligning with intuition. Moreover, co-locating a battery system with the solar installation will further increase the costs. We again emphasize that, for a given site, a lower total cost than the maximum cost observed here can be realized if a smaller size battery system or a smaller nameplate capacity solar farm is built at the site; however, this would also lower the capabilities of this site to reduce excess CO2 emissions.

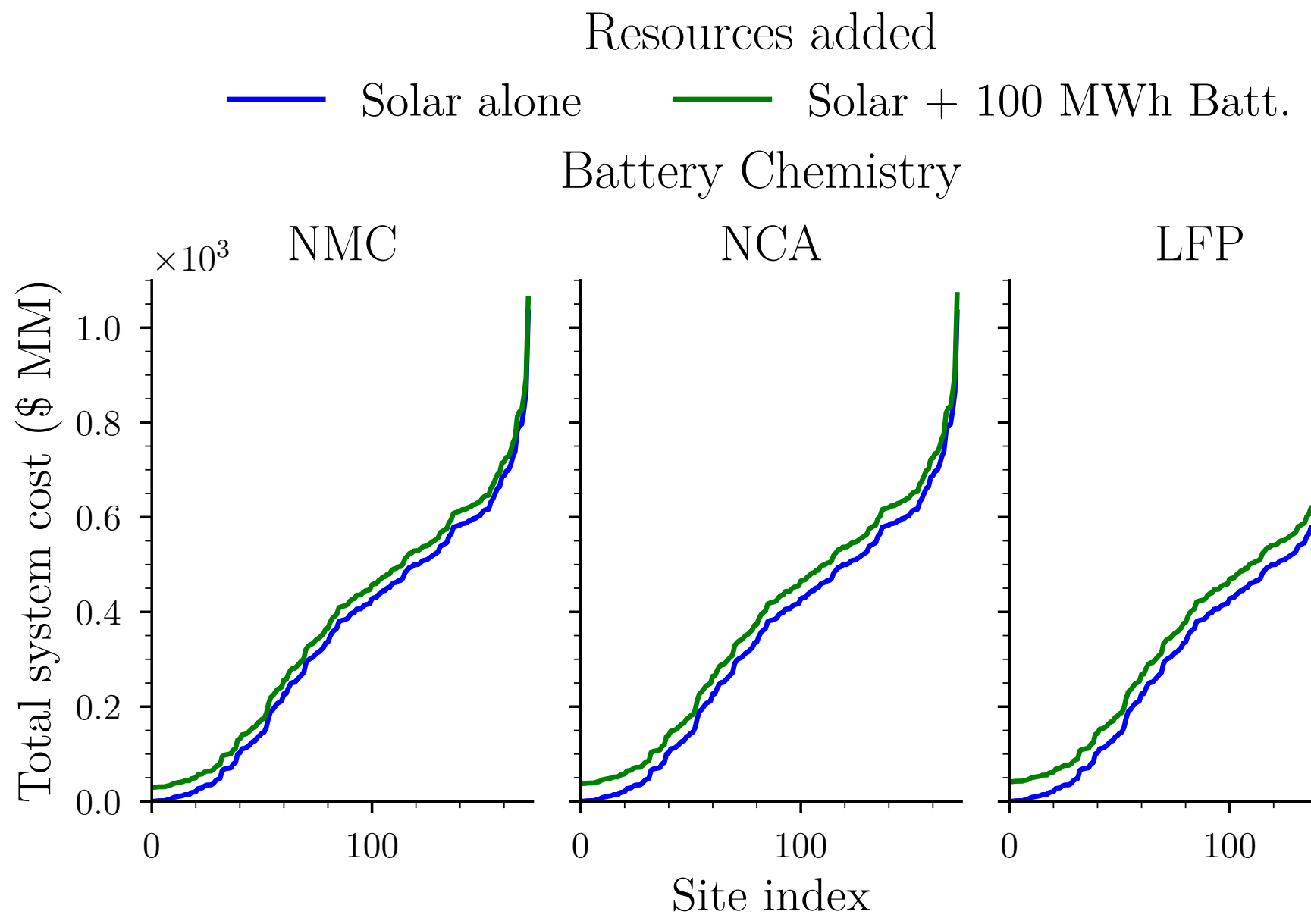

*Figure S15: Total system cost of deploying a microgrid with solar farm alone (blue curve) and a microgrid with a solar farm and a 100 MWh battery system (green curve) for each of the viable sites within the region.*

*5.3 Site-specific LCOPR*

As we are considering the excess load demand for the full fleet, we consider the utility to be responsible in managing the costs of deployment of solar and battery resources. The fleet operator, then uses electricity from a grid that is augmented with these resources. We give estimates of the costs for these two different stakeholders as quantified in Section 4 for every identified viable site in the Port of Savannah region.

The LCOPR for the viable sites in the Port of Savannah region are shown in *Figure S16*. We see a relatively stable value for the LCOPR of ~26.40 $/MWh across the viable sites in the region, with only the lowest 5 nameplate capacity sites having a significantly higher LCOPR. The stable value, roughly speaking, results from the cost of larger nameplate solar capacity farms increasing proportionally to the energy output of the sites.

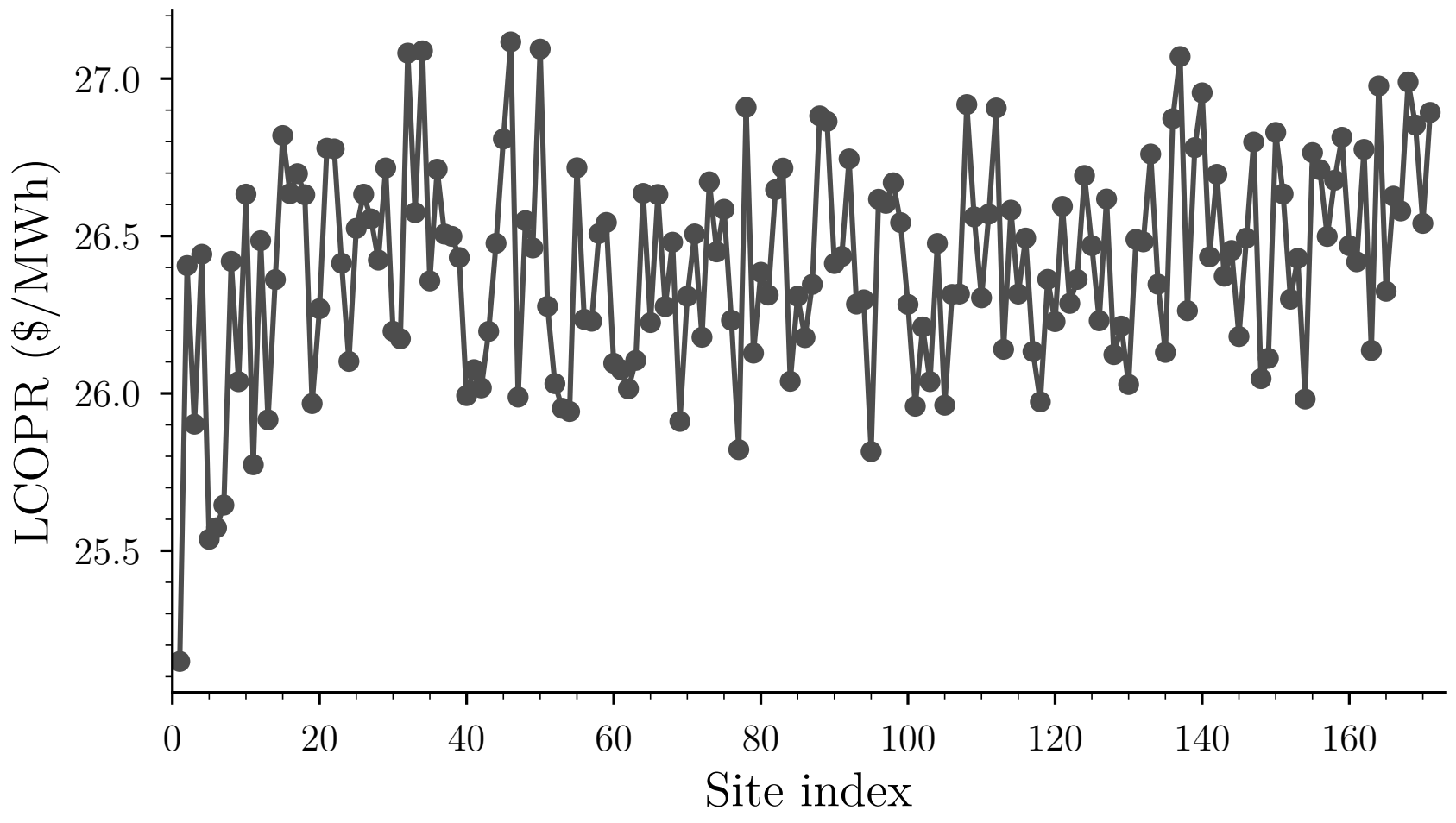

*Figure S16: Levelized cost of photovoltaic recharge (LCOPR) for each of the viable sites in the region.*

**6. Capabilities of a solar site adjacent to the Port of Savannah**

To systematically explore the effects of adding solar and energy storage into the energy network for a given site location, we explore the capabilities of the hypothetical microgrid deployment, incorporating solar and battery power, geographically adjacent to the Port of Savannah. The nameplate capacity of the solar farm and the battery size are assumed to be arbitrary. This hypothetical scenario thus neglects the area requirements for deploying a microgrid and the land-use restrictions that would otherwise prevent a deployment adjacent to the port.

*6.1 Available solar capacity estimation*

To obtain an estimate of the available solar capacity $\boldsymbol{p_s(t)}$ for the hypothetical site, we utilize the tool PVWatts by NREL[22]. This web application estimates the hourly available solar capacity, of a grid-scale photovoltaic system at a given location of the U.S. in a year. To do this, PVWatts cross-references the National Solar Radiation Database to determine which cell in the database corresponds to the desired location and uses that as an input to its energy generation prediction sub-models. To determine the hourly available solar capacity, we use PVWatts assuming a solar farm with a 1 MW nameplate solar capacity, where the solar panels are fixed, open-rack, with a 20-degree tilt and 180-degree azimuth angle. The default standard modules, corresponding to crystalline silicone cell material, is used. The system losses are estimated to be 14.08%. The available solar capacity profile is shown in *Figure S17*. To obtain the capacities for a larger nameplate capacity site, we linearly scale the 1 MW available solar capacity profile by the desired nameplate solar capacity.

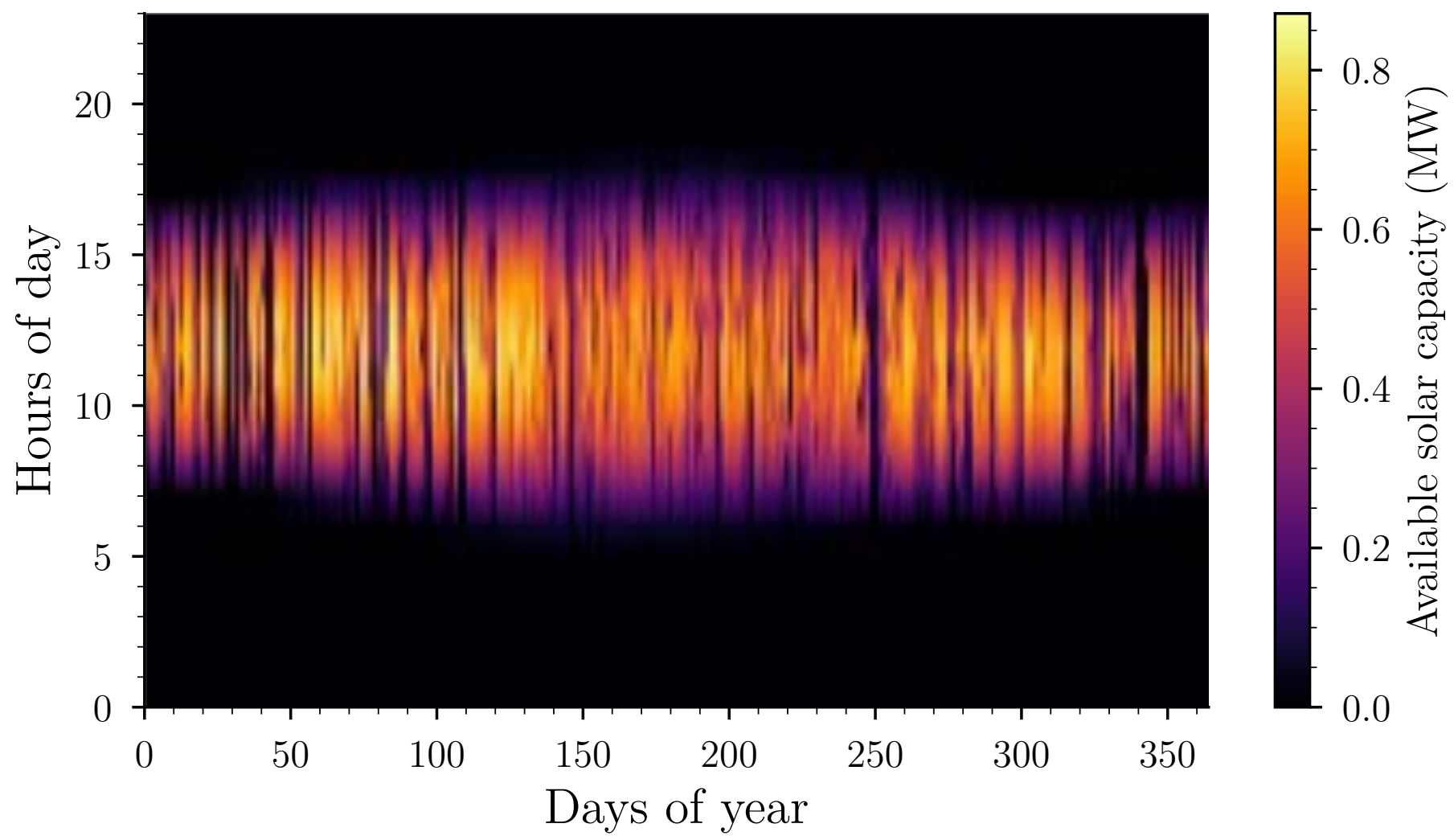

*Figure S17: Hourly available solar capacity for a hypothetical 1 MW nameplate solar capacity site, deployed adjacent to the Port of Savannah.*

*6.2 Optimal excess CO2 emissions reduction under different scenarios*

Using the available solar capacity estimate from PVWatts, we perform the dispatch optimization procedure described in Section 3 for each combination of nameplate solar capacity up to 100 MW and battery sizes up to 100 MWh and *Figure S18* shows the maximal total excess CO2 emissions reduction. In general, we find that the total excess CO2 emissions reduction increases monotonically both for increasing site nameplate solar capacity and battery size; however, it is much more sensitive to the nameplate solar capacity. We observe that, for nameplate solar capacities of up to 100 MW and battery sizes up to 100 MWh, a maximum of 77.8% total excess CO2 emissions reduction with the renewable network is achievable, corresponding to a site with 100 MW nameplate solar capacity and a 100 MWh battery system.

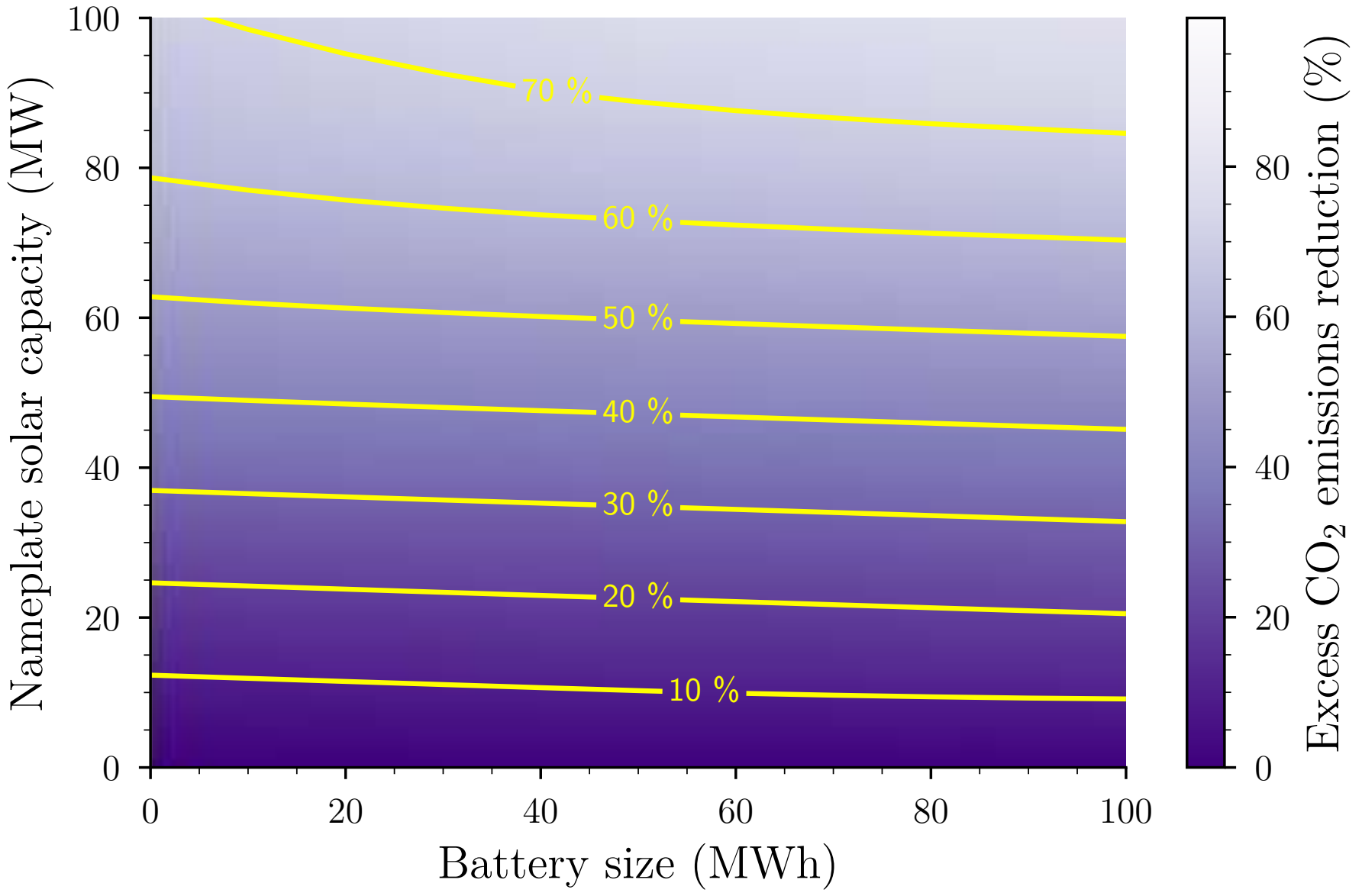

*Figure S18: Maximum excess CO2 emissions reduction for a solar farm & battery system built just outside of the Port of Savannah.*

Notably, the contour lines in *Figure S18* are relatively shallow as a function of battery size, indicating that the total excess CO2 emissions reduction is relatively insensitive to the battery size. This arises due to the alignment of peak solar availability and excess load demand, as seen in *Figure S19* for the first five days of the year. The port uses the most electricity near noon time and attains the highest excess load demand at that time; coincidentally, noon time is also when solar availability is the highest.

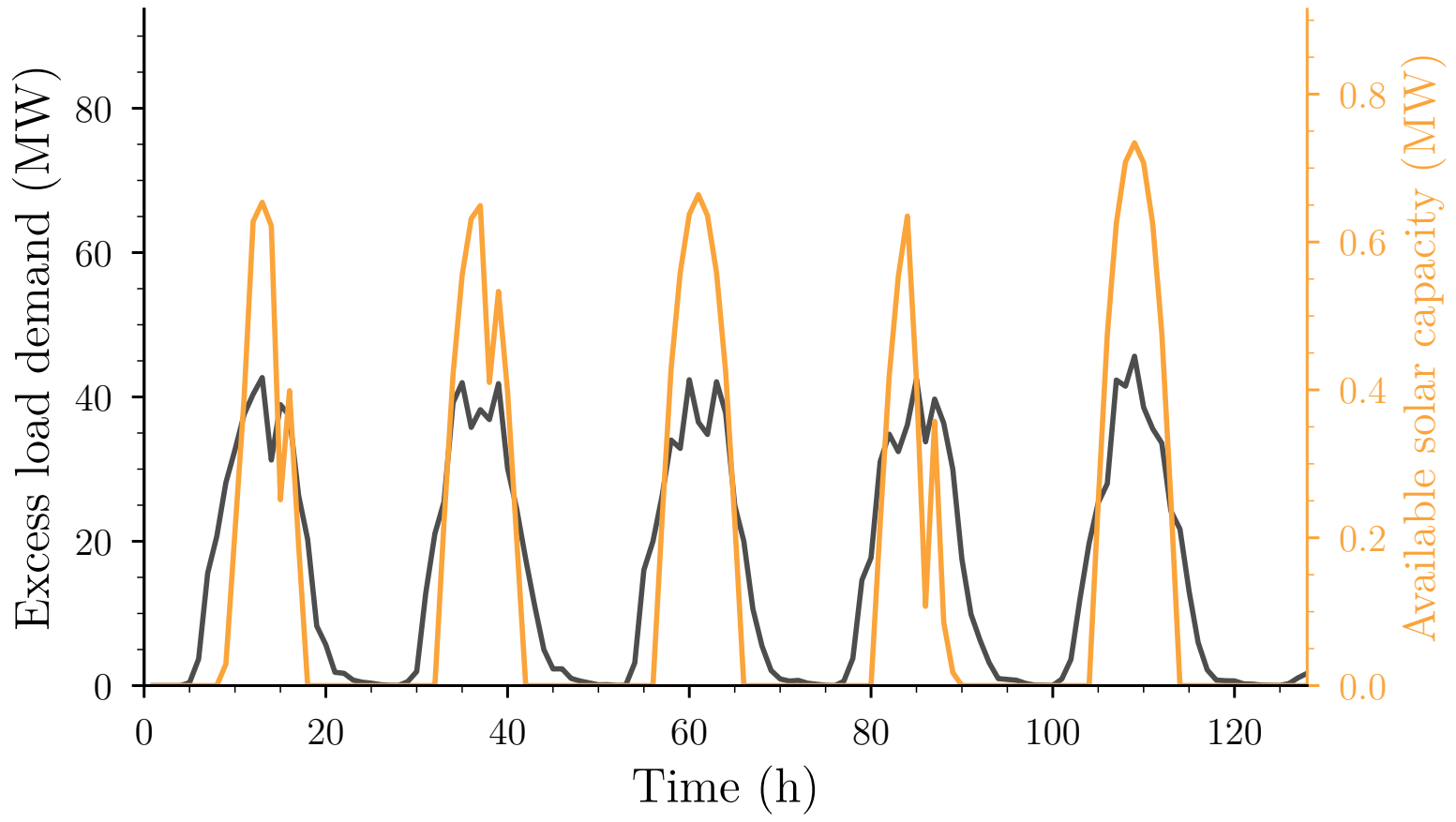

*Figure S19: The peak of solar availability at the hypothetical site is generally aligned with the peak of the excess load demand*

The battery is expected to have a greater impact when the peak of the excess load demand is shifted with respect to the peak of solar availability. This scenario allows the battery to serve as an energy re-distribution mechanism, taking in energy at times when demand is low but solar availability is high, storing it, and discharging at times where the demand is higher. To explore this, we artificially shift the excess load demand signal ahead or delayed with respect to the peak of solar by 4 hours, as illustrated in *Figure S20*.

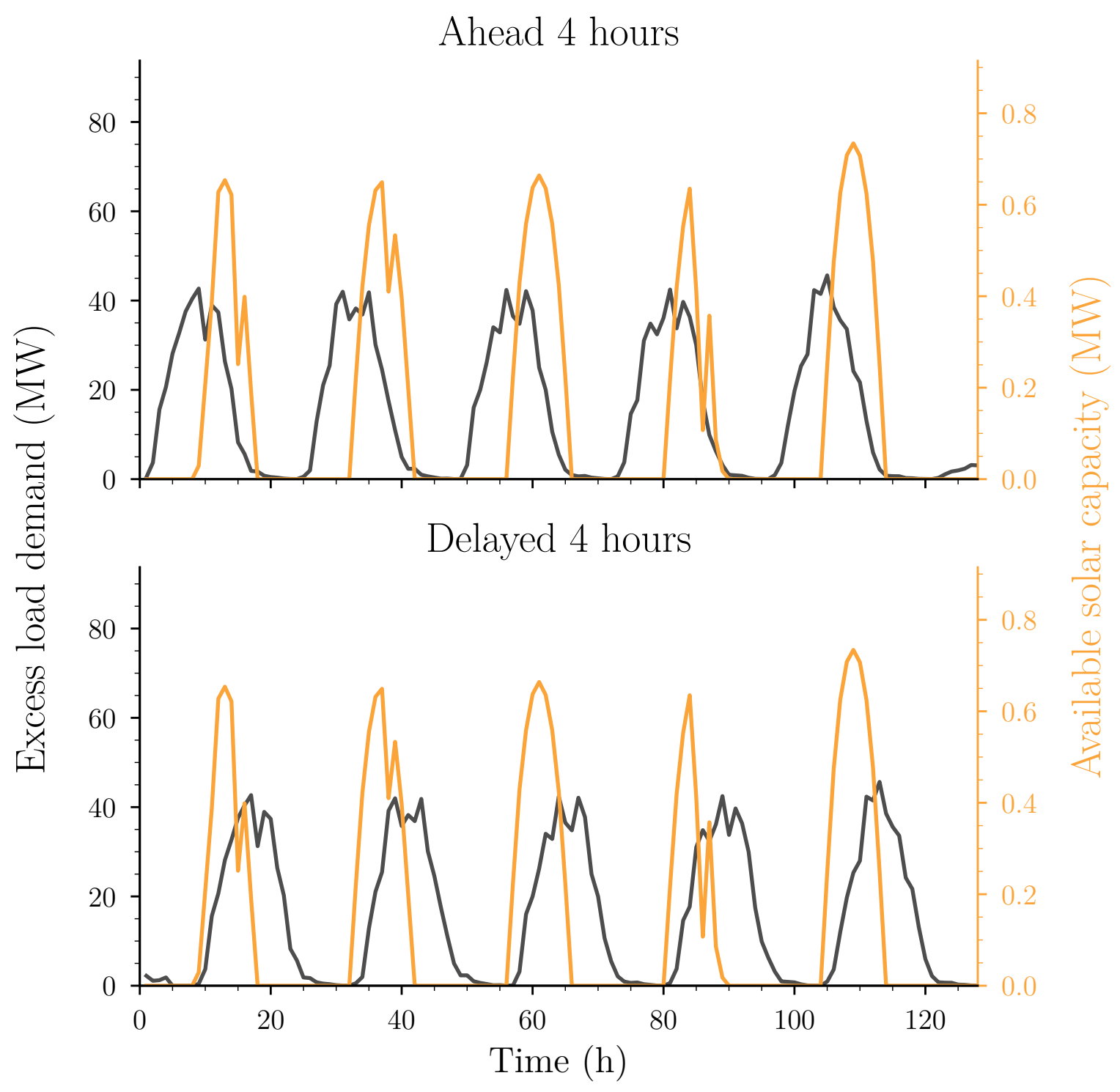

*Figure S20: Artificially shifting the peak of excess load demand with respect to the peak of solar availability. (Top plot) Excess load demand is shifted 4 hours ahead of the peak of solar availability. (Bottom plot) Excess load demand is delayed 4 hours relative to the peak of solar availability.*

We find in the left and right subfigures of *Figure S21* that artificially shifting the excess load demand signal results in an increase in the sensitivity of the total excess CO2 emissions reduction to the battery size, evidenced by the steeper slopes of the isolines with respect to the battery size. However, a demand shift also reduces the maximum total excess CO2 emissions reduction attainable. For the maximum size of solar farm and battery system, the excess CO2 emissions can only be reduced by ~60% when there is a demand shift present.

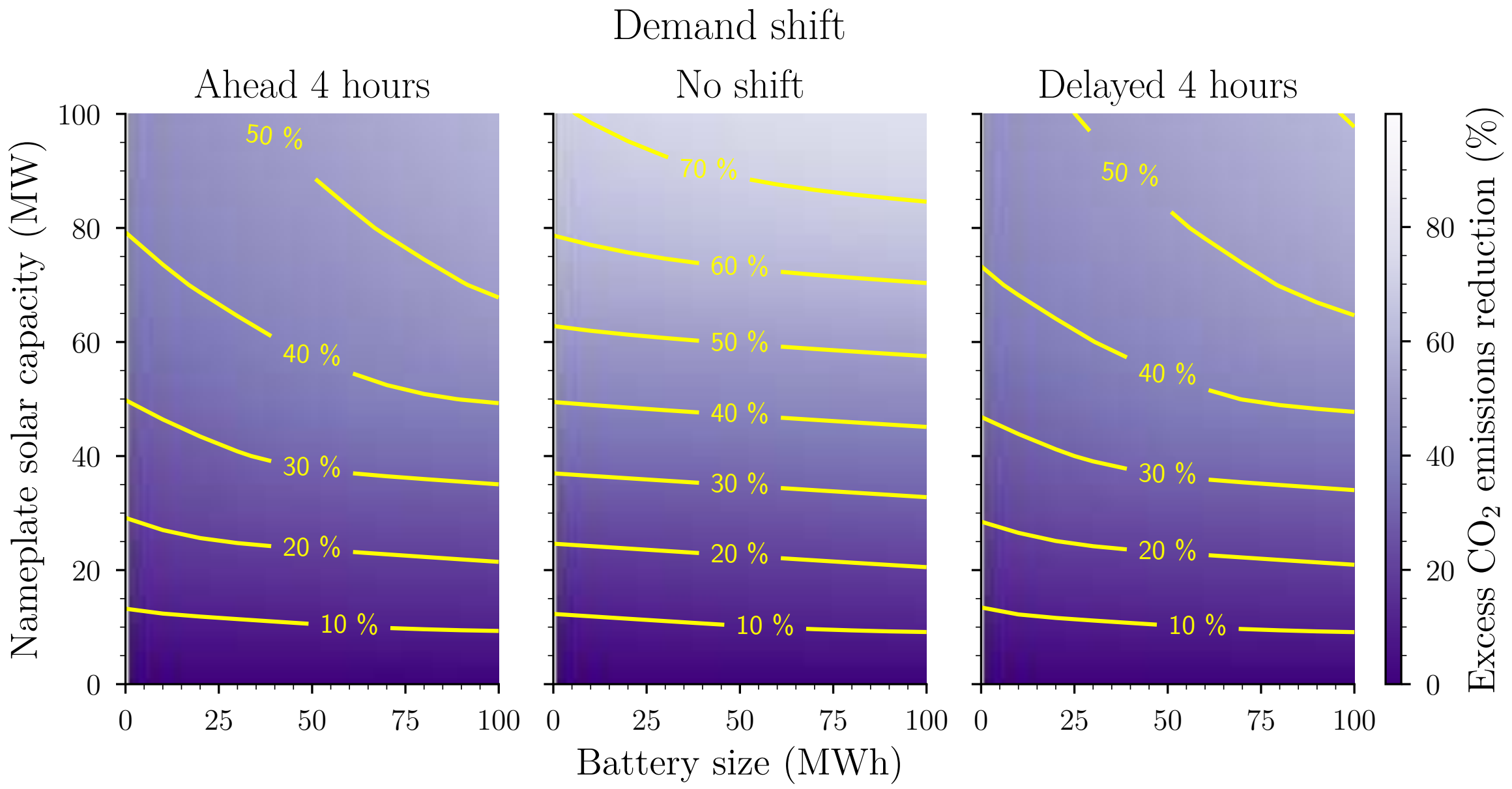

*Figure S21: Maximum total excess CO2 emissions reduction under demand shifted scenarios. (Left plot) Excess load demand peak shifted 4 hours ahead of solar availability peak. (Middle plot) No demand shift. Same plot as Figure 13. (Right plot) Excess load demand peak delayed 4 hours relative to solar availability peak.*

### *6.3 Resource utilization*

To understand the limitations on the total excess $CO_2$ emissions reduction, we investigate the how the energy used meet the excess load demand is partitioned amongst the different energy sources. The average utilization of each resource is shown in *Figure S22* for different combinations of nameplate solar capacity and battery size, under different demand shift scenarios. Aligning with intuition, the average utilization of the grid decreases as sites with higher nameplate solar capacity are considered. Interestingly, for a given demand shift scenario, the average utilization of solar remains constant irrespective of the size of battery system added, as seen from the horizontal isolines in the solar utilization. As the battery system size is increased, for a fixed nameplate solar capacity, the average utilization of the grid decreases.

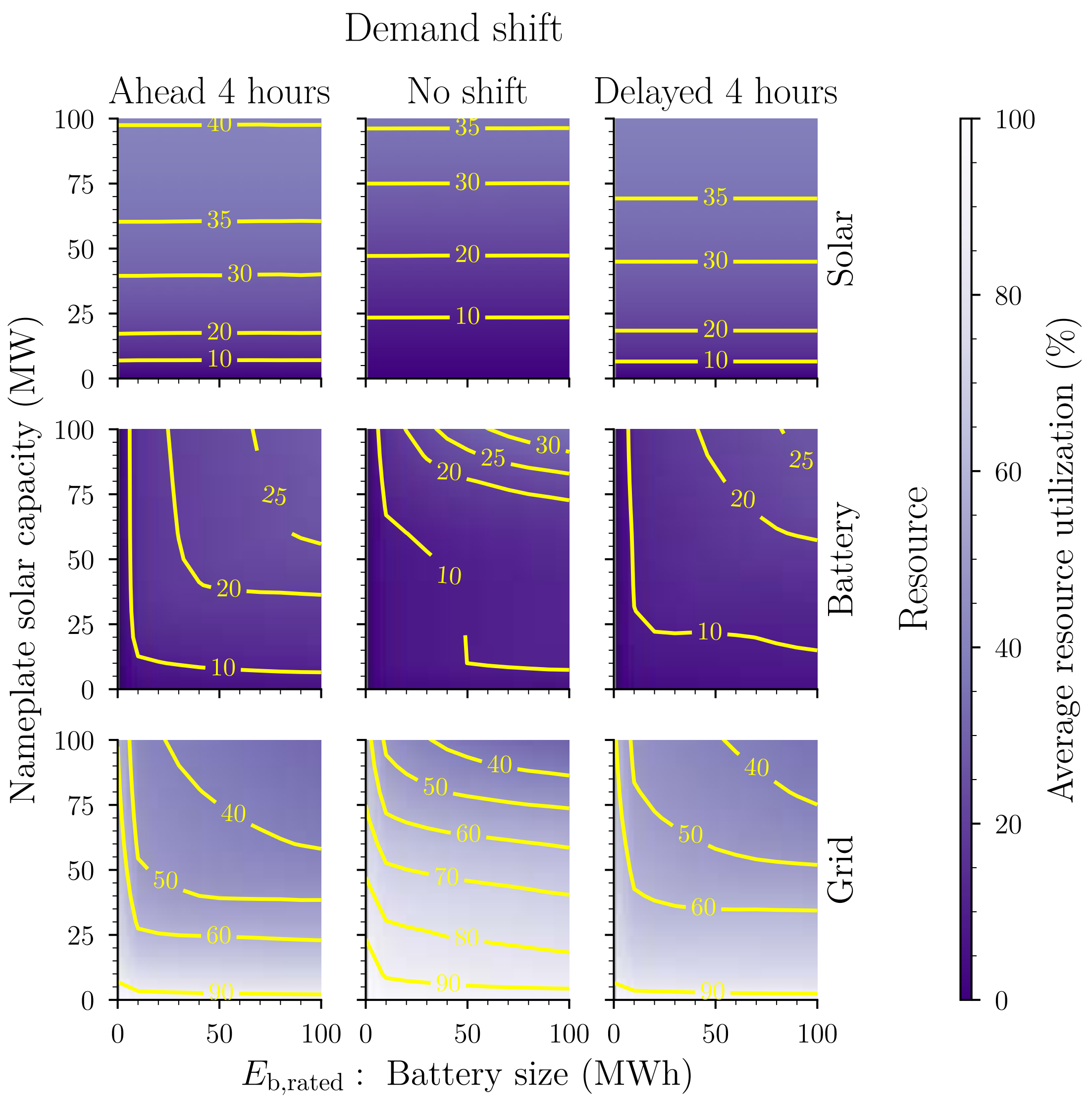

*Figure S22: Resource utilization for different combinations of nameplate solar capacity and battery size for a hypothetical renewable energy site adjacent to Port of Savannah. Each column: different demand shift scenarios. Each row: different battery chemistry.*

Interestingly, we observe that, even for sites with nameplate solar capacities of up to 100 MW and battery sizes of up to 100 MWh, it is impossible to meet the excess load demand purely with solar and battery resources alone. The limited efficiency of solar resources imposes that grid resources must be used, leading to a limitation on the maximum total excess CO2 emissions reduction.

Counterintuitively, due to the power limits on the battery, introducing a demand shift lowers the utilization of the battery for a given combination of nameplate solar capacity and battery size, relative to the no shift scenario. The power limitation of the battery is compensated for by grid power leading to the lower excess CO2 emissions reduction observed for the demand shifted scenarios.

*6.4 Cost estimates of deployment*

Having estimated the capabilities of a renewable energy network, where solar and battery system is deployed adjacent to the Port, we now estimate the cost of deployment for this system. We first focus on estimating the TEC, as it is the central quantity for the cost estimation for either of the stakeholders we consider in this work. To do this, we compute the LCOPR from the estimate of the hourly available solar capacity. We find that the LCOPR is constant with respect to nameplate solar capacity with a value of ~25 $/MWh. This indicates that the overnight capital costs scales linearly with the nameplate solar capacity of the hypothetical solar farm. Notably, the LCOPR is significantly lower than that of the grid electricity price of the Savannah region.

Given the average resource utilization computed from the optimal dispatch profiles of each network element, *Figure S23* shows the TEC under different demand shift scenarios and for different choices of LIB chemistry. For the NMC chemistry, we observe that the TEC isolines are relatively flat as a function of battery size. Instead, the TEC level for this chemistry seems most sensitive to the nameplate solar capacity. Additionally, due to the LCOPR being slightly lower than the grid electricity price, introducing a demand shift lowers the overall TEC as more solar power is used to meet the load demand. Using an LFP system can attain the lowest TEC out of all the other chemistries, particularly for larger nameplate solar capacities and battery system sizes. Due to its LCOS being much lower than the grid electricity price, a small increase in the battery system utilization for this chemistry can significantly decrease TEC. In contrast, as an NCA storage system has a significantly greater cost than the grid electricity price, an increase in the usage of the battery system causes an increase in the TEC. As such, the most cost-efficient system is one that where solar is not augmented by an NCA battery system regardless of the demand shift scenario. Computing the TEC allows us to subsequently compute the cost metrics for both the utility and the fleet operator. We will now examine each, in turn.

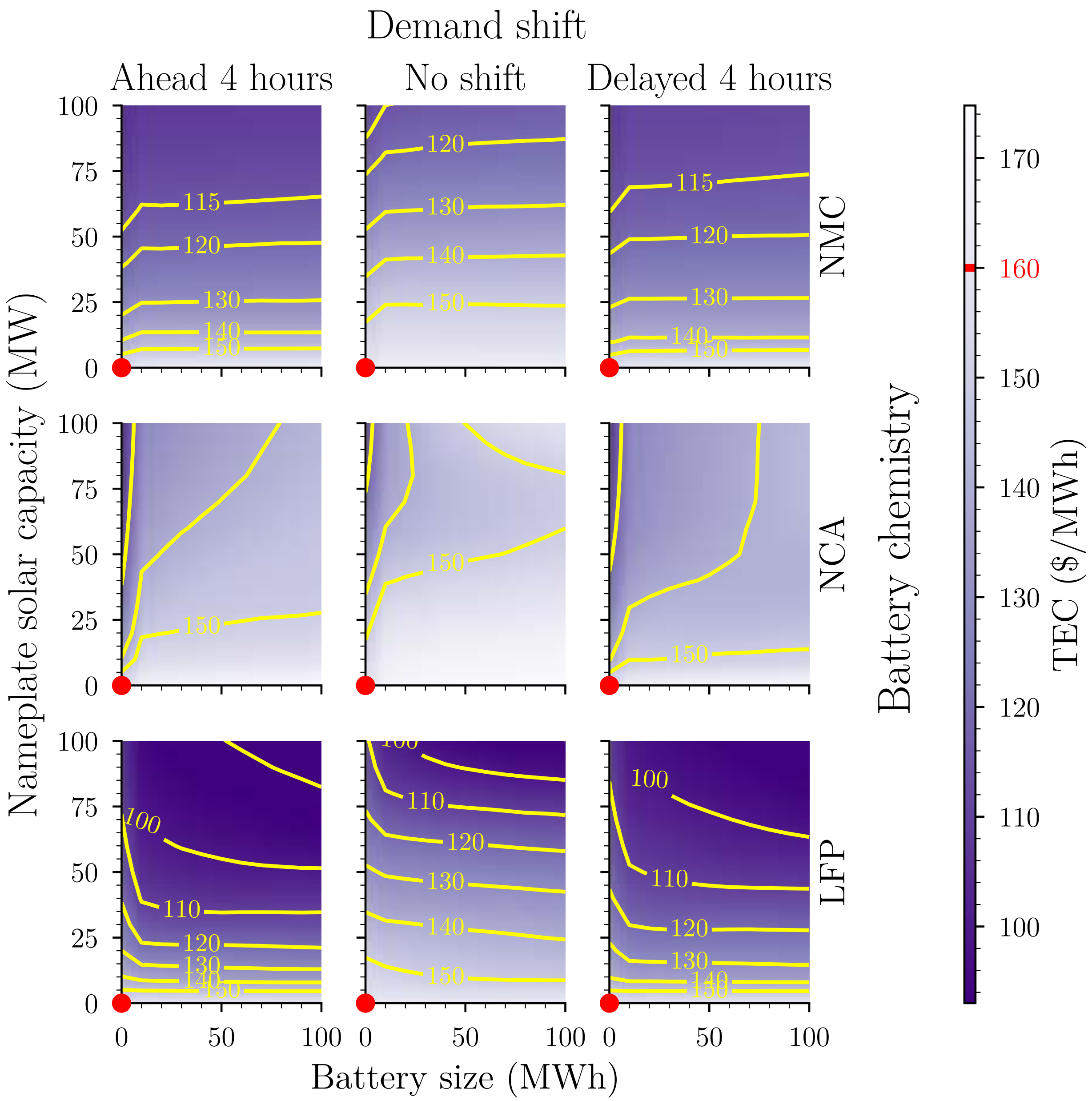

*Figure S23: Total electricity cost (TEC) for different nameplate solar capacity and battery size combinations deployed at a hypothetical renewable energy site adjacent to the Port of Savannah. Each column: different demand shift scenarios. Each row: different battery chemistry. Cost for a system with no solar and battery size is the grid electricity price of 160 $/MWh.*

6.4.1 Cost to the utility

In *Figure S24*, the cost metric for the utility is shown for different choices of battery chemistry and different demand shift scenarios. We observe that for NMC- and NCA-based battery systems, the lowest cost solution is to build only a solar system without a battery.

In contrast, for an LFP-based system, a range of battery sizes can achieve a low-cost metric provided that the nameplate capacity of the solar system is sufficiently large.

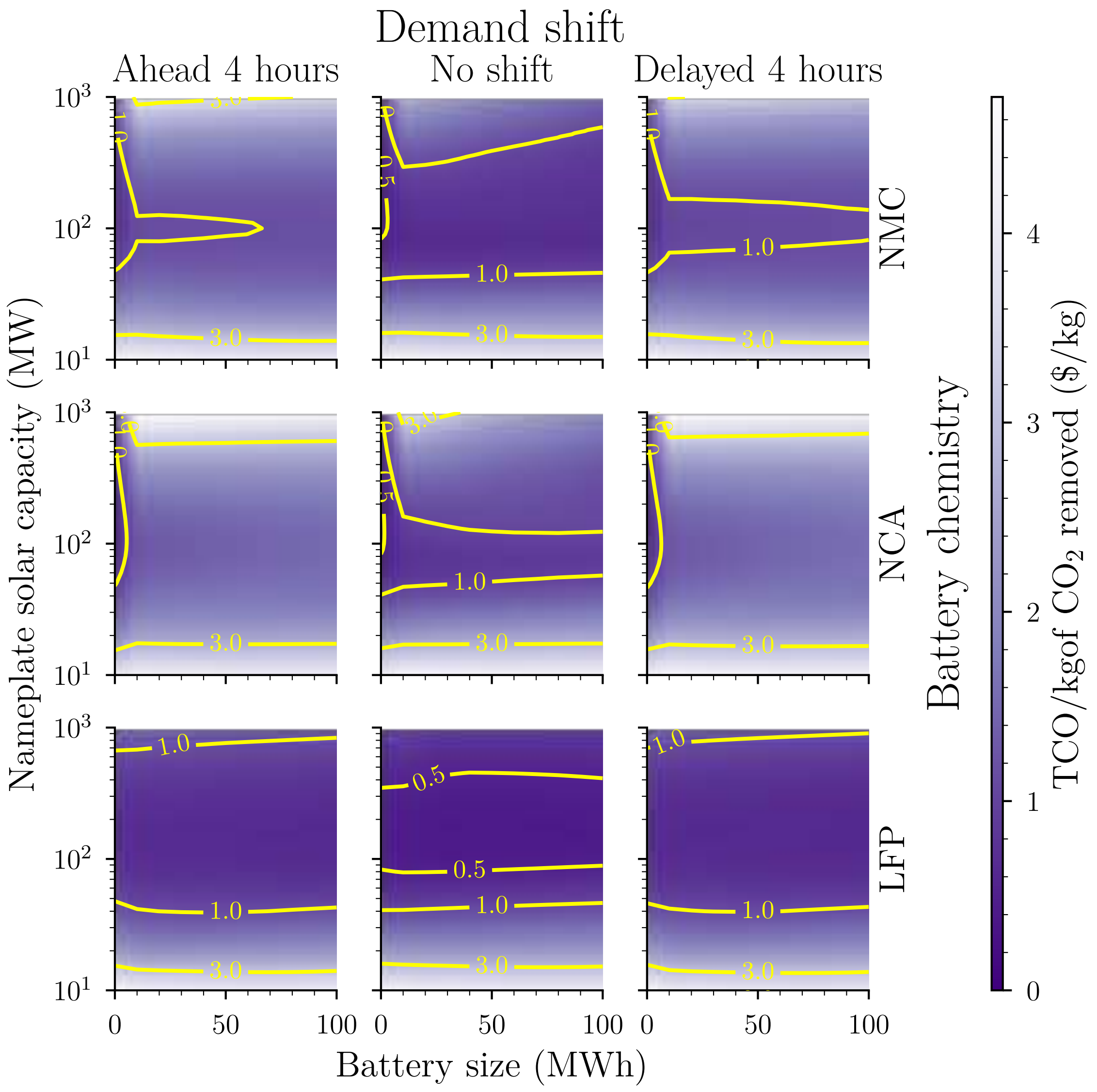

*Figure S24: TCO/kg of CO2 removed for different combinations of nameplate capacity and battery size deployed at a hypothetical site adjacent to the Port of Savannah. Each column: different demand shift scenarios. Each row: different battery chemistry.*

6.4.2 Cost to the fleet operator

The fleet operator cost metric is computed in *Figure S25* for different choices of battery chemistry and different demand shift scenarios. We observe very similar trends in the TCO/mile as we observed in the TEC. In particular, the lowest cost battery chemistry is identified to be LFP, across all three demand shift scenarios. The NMC-based battery system is relatively insensitive to the battery system size and the level of cost is largely determined by the nameplate capacity of the solar system. An NCA-based battery system remains the most expensive and the lowest cost solution for this chemistry is to deploy only solar without an accompanying battery system.

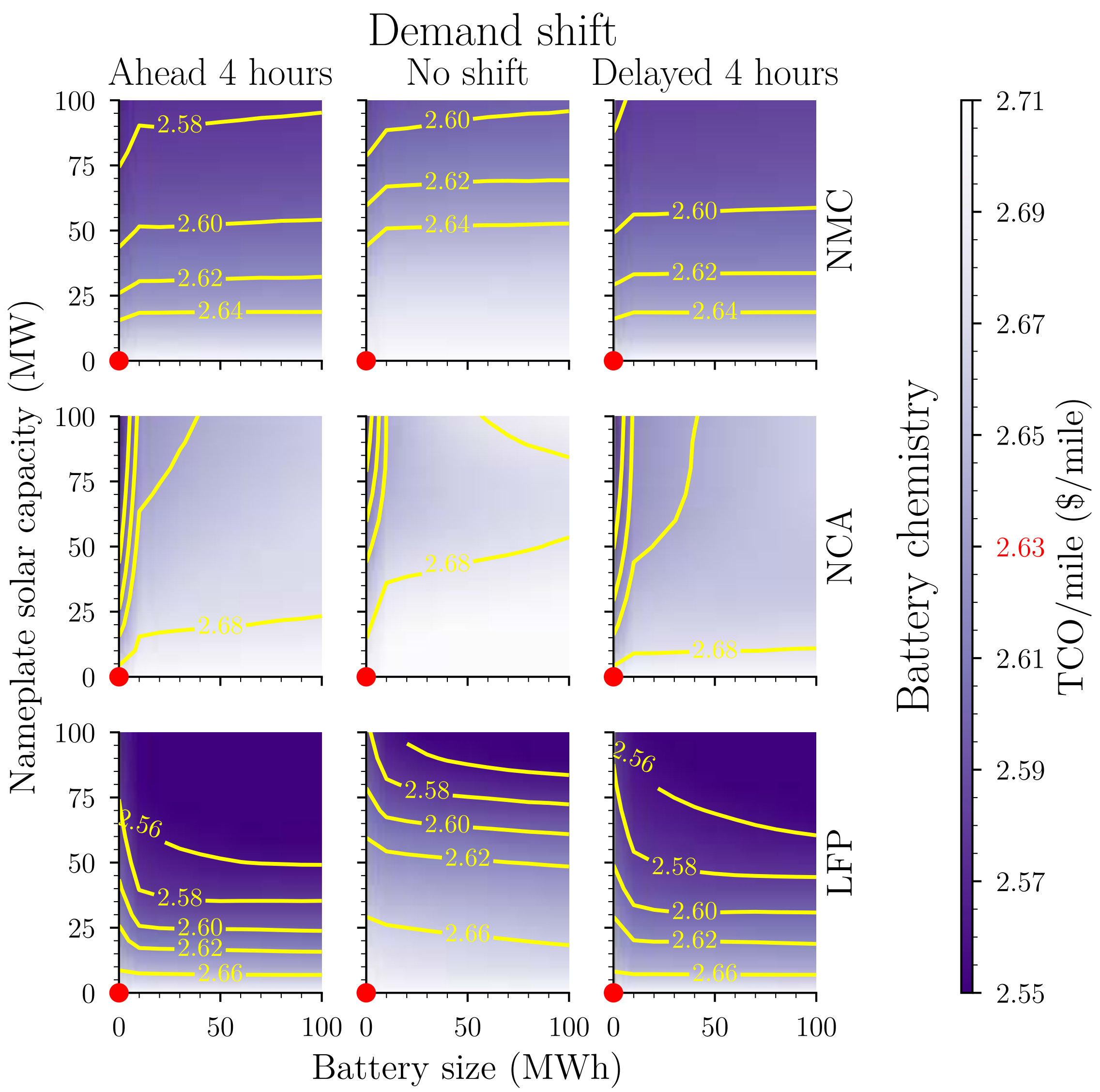

*Figure S25: TCO/mile of the HDCV fleet for different combinations of nameplate capacity and battery size deployed at a hypothetical site adjacent to the Port of Savannah. Each column: different demand shift scenarios. Each row: different battery chemistry. Baseline cost using only the grid to power the excess load demand is 2.63 $/mile.*

We emphasize that the shape of the contours and the effect of the demand shift observed in the results of this section are specific to the Port of Savannah. As the hourly available solar capacity and excess CO2 emissions are location dependent, if we chose a different location for our analysis the exact trends we would observe would vary. This highlights the importance of doing a region-based analysis to determine where a solar farm of a given size can viably be installed.